\documentclass[prd,amsmath,aps,floats,amssymb,floatfix,superscriptaddress,nofootinbib,twocolumn,preprintnumbers]{revtex4-2}
\UseRawInputEncoding
\usepackage{tabularx}
\usepackage{amsmath,amssymb,natbib,latexsym,times}
\usepackage[T1]{fontenc}
\usepackage{aecompl}

\usepackage{multirow}
\usepackage{rotating}
\usepackage{booktabs}
\usepackage{soul}
\usepackage{enumitem}

\usepackage{verbatim}
\usepackage[usenames,dvipsnames]{xcolor}
\usepackage{tabulary}
\usepackage{tabularx}
\usepackage{todonotes}
\usepackage{hyperref}
\hypersetup{colorlinks=true, linkcolor=blue, citecolor=blue, filecolor=blue, urlcolor=blue}
\usepackage{ulem}
\newcommand{\maglimpp}{\textsc{MagLim++}}

\newcommand{\photoz}{photo-$z$}

\newcommand\lcdm{$\Lambda$CDM\,}
\newcommand\wcdm{$w$CDM}

\newcommand*{\diff}{\ensuremath{{\rm d}}}

\newcommand\mcal{{\textsc{metacalibration}}} %{{\tt metacal}}
\newcommand\mdet{{\textsc{metadetection}}}

\usepackage[dvipsnames]{xcolor}

\newcommand\be{\begin{equation}}
\newcommand\ee{\end{equation}}
\def\bea{\begin{eqnarray}}
\def\eea{\end{eqnarray}}

\newcommand\red[1]{\textcolor{Black}{#1}} %% Change this back to RED to see new text

\newcommand{\rcwr}[1]{\textcolor{black}{#1}}
\newcommand{\journal}[2]{\st{#1}\textcolor{black}{#2}}

\allowdisplaybreaks

\defcitealias{y6-gold}{\textsc{Y6Gold}}
\defcitealias{y6-piff}{\textsc{Y6PSF}}
\defcitealias{y6-mask}{\textsc{Y6Mask}}
\defcitealias{y6-balrog}{\textsc{Y6Balrog}}
\defcitealias{y6-metadetect}{\textsc{Y6Shear}}
\defcitealias{y6-maglim}{\textsc{Y6Pos}}
\defcitealias{y6-sourcepz}{\textsc{Y6ShearPZ}}
\defcitealias{y6-lenspz}{\textsc{Y6PosPZ}}
\defcitealias{y6-wz}{\textsc{Y6WZ}}
\defcitealias{y6-nzmodes}{\textsc{Y6Modes}}
\defcitealias{y6-magnification}{\textsc{Y6Mag}}
\defcitealias{y6-gglens}{\textsc{Y6GGL}}
\defcitealias{y6-methods}{\textsc{Y6Model}}
\defcitealias{y6-ppd}{\textsc{Y6PPD}}
\defcitealias{y6-imagesims}{\textsc{Y6Imsim}}
\defcitealias{y6-cardinal}{\textsc{Y6Mock}}
\defcitealias{y6-1x2pt}{\textsc{Y6-1x2pt}}
\defcitealias{y6-2x2pt}{\textsc{Y6-2x2pt}}
\defcitealias{y6-extensions}{\textsc{Y6Extend}}

\begin{document}
\title{Dark Energy Survey Year 6 Results: \\Cosmological Constraints from Galaxy Clustering and Weak Lensing}
% Author list file generated with: mkauthlist 1.4.0 
% mkauthlist --force --sort --journal prd DES-2025-0929_author_list.csv authors.tex 

\author{T.~M.~C.~Abbott}
\affiliation{Cerro Tololo Inter-American Observatory, NSF's National Optical-Infrared Astronomy Research Laboratory, Casilla 603, La Serena, Chile}

\author{M.~Adamow}
\affiliation{Center for Astrophysical Surveys, National Center for Supercomputing Applications, 1205 West Clark St., Urbana, IL 61801, USA}

\author{M.~Aguena}
\affiliation{INAF-Osservatorio Astronomico di Trieste, via G. B. Tiepolo 11, I-34143 Trieste, Italy}
\affiliation{Laborat\'orio Interinstitucional de e-Astronomia - LIneA, Av. Pastor Martin Luther King Jr, 126 Del Castilho, Nova Am\'erica Offices, Torre 3000/sala 817 CEP: 20765-000, Brazil}

\author{A.~Alarcon}
\affiliation{Institute of Space Sciences (ICE, CSIC),  Campus UAB, Carrer de Can Magrans, s/n,  08193 Barcelona, Spain}

\author{S.~S.~Allam}
\affiliation{Fermi National Accelerator Laboratory, P. O. Box 500, Batavia, IL 60510, USA}

\author{O.~Alves}
\affiliation{Department of Physics, University of Michigan, Ann Arbor, MI 48109, USA}

\author{A.~Amon}
\affiliation{Department of Astrophysical Sciences, Princeton University, Peyton Hall, Princeton, NJ 08544, USA}

\author{D.~Anbajagane}
\affiliation{Kavli Institute for Cosmological Physics, University of Chicago, Chicago, IL 60637, USA}

\author{F.~Andrade-Oliveira}
\affiliation{Physik-Institut, University of Z\"{u}rich, Winterthurerstrasse 190, CH-8057 Z\"{u}rich, Switzerland}

\author{S.~Avila}
\affiliation{Centro de Investigaciones Energ\'eticas, Medioambientales y Tecnol\'ogicas (CIEMAT), Madrid, Spain}

% ...

\collaboration{DES Collaboration}

\date{\today}
%\pubyear{2026}
\label{firstpage}

\begin{abstract}
We present cosmology results combining galaxy clustering and weak gravitational lensing measured in the full six years (Y6) of observations by the Dark Energy Survey (DES) covering $\sim$5000\,deg$^2$. We perform a large-scale structure analysis with galaxy samples defined from the final data extending to redshift range $z \lesssim 2$, using three two-point correlation functions (3$\times$2pt): (i) cosmic shear measuring correlations among the shapes of 140 million source galaxies, (ii) auto-correlations of the spatial distribution of 9 million lens galaxies, and (iii) galaxy galaxy lensing from the cross-correlation between lens positions and source shapes.
We perform the analysis under a rigorous blinding protocol to prevent confirmation biases. We model the data in flat \lcdm and $w$CDM cosmologies.  We find consistent cosmological results from subsets of the three two-point correlation functions. Their combined analysis yields $S_8\equiv \sigma_8 (\Omega_{\rm m}/0.3)^{0.5} = 0.789^{+0.012}_{-0.012}$ and matter density $\Omega_{\rm m} = 0.333^{+0.023}_{-0.028}$ in $\Lambda$CDM (68\% CL), where $\sigma_8$ is the clustering amplitude. The factor of two improvement in constraining power in the $\Omega_{\rm m}$--$\sigma_8$ plane relative to DES Year 3 is due to higher source number density, extended redshift range, and improved modeling. These constraints show a (full-space) parameter difference of 1.8$\sigma$ from a combination of cosmic microwave background (CMB) primary anisotropy datasets from {\it Planck} 2018, ACT-DR6, and SPT-3G DR1. Projected only into $S_8$ the difference is $2.6\sigma$. In $w$CDM the Y6 3$\times$2pt results yield $S_8 = 0.782^{+0.021}_{-0.020}$, $\Omega_{\rm m} = 0.325^{+0.032}_{-0.035}$, and dark energy equation-of-state parameter $w = -1.12^{+0.26}_{-0.20}$. For the first time, we combine all DES dark-energy probes: 3$\times$2pt, SNe Ia, BAO and Clusters. In \lcdm this combination yields a $2.8\sigma$ parameter difference from the CMB. When combining DES 3$\times$2pt with other most constraining low-redshift datasets (DESI DR2 BAO, DES SNe Ia, SPT clusters), we find a 2.3$\sigma$ parameter difference with CMB. A joint fit of Y6 3$\times$2pt, CMB, and those low-$z$ datasets produces the tightest $\Lambda$CDM constraints to date: $S_8 = 0.806^{+0.006}_{-0.007}$, $\Omega_{\rm m} = 0.302^{+0.003}_{-0.003}$, $h = 0.683^{+0.003}_{-0.002}$, and $\sum m_\nu < 0.14$ eV (95\% CL). In $w$CDM, this dataset combination yields $w = -0.981^{+0.021}_{-0.022}$, with no significant preference over $\Lambda$CDM.
\end{abstract}

%\begin{keywords}
%\end{keywords}

\preprint{DES-2025-0929}
\preprint{FERMILAB-PUB-26-0026-PPD}
\maketitle

\section{Introduction}\label{sec:intro}

The concordance model of cosmology posits that the universe is well described by
General Relativity and \rcwr{is composed of} $\sim5\%$ baryons, $\sim$25\% cold dark matter (CDM) and $\sim$70\% dark energy. \rcwr{The nature of dark matter is largely unknown aside from the fact that it interacts gravitationally, but it is canonical to assume that it is non-relativistic and only weakly interacting (CDM). The nature of dark energy is even less constrained---typically, it is assumed to be either a cosmological constant ($\Lambda$) or describable as an effective fluid with equation of state parameter $w$. In this paper, we analyze and compare results in the context of two cosmological models, the standard $\Lambda$CDM paradigm and $w$CDM, where in the latter case $w$ is assumed to be constant in time but can differ from the $\Lambda$CDM value of $w=-1$. Most previous analyses of data in the context of these two models \citep{Planck2020_cosmo, Lodha2025, Louis2025, Camphuis2025, Brout2022, TDCOSMO2025, Wright2025, Dalal2023, Unions2025, Euclid2025} have found results consistent with $w=-1$, i.e., with $\Lambda$CDM. \journal{}{A separate paper \cite{y6-evolvingDE} considers DES constraints in the context of models with time-varying $w$}, which have garnered considerable recent interest \citep{DESY5SN2024, desi-dr1, Popovic2025b}.}

The Dark Energy Survey (DES; \citep[][]{DES2005}) was developed to \rcwr{test the $\Lambda$CDM paradigm and probe the nature of dark energy} through a combination of cosmological probes. Between 2013 and 2019, DES imaged $\sim$1/8th of the sky to leverage information encoded in the large-scale structure (LSS) of galaxies and \rcwr{matter}, as well as a large number of other science goals \citep{DES2016}.
This paper presents the LSS analysis from the full six years of DES data (DES Y6), combining measurements of the galaxy angular overdensity field $\delta_{\rm g}$ and the weak gravitational lensing shear field $\gamma$, which measures the coherent distortion of galaxy shapes caused by deflection of light by the cosmic mass distribution. In particular, we perform a 3$\times$2pt analysis, which uses two distinct galaxy samples to measure three complementary 2-point correlation functions. The \textit{source sample} consists of galaxies selected for shape measurements to quantify weak lensing shear, whereas the \textit{lens sample} comprises galaxies whose positions trace the large-scale matter distribution. The three correlation functions are: the auto-correlation of galaxy shear, $\xi_\pm = \langle \gamma \gamma \rangle$, often referred to as cosmic shear; the correlation between galaxy density and shear, $\gamma_\mathrm{t}=\langle \delta_{\rm g} \gamma \rangle$, often referred to as galaxy-galaxy lensing; and the {auto-correlation of galaxy density}, $w=\langle \delta_{\rm g}  \delta_{\rm g} \rangle$, or galaxy clustering. 

\begin{table}
    \centering
    \caption{Reference table for supporting papers to this work.} 
    \label{tab:alias}
        \begin{tabular}{lp{1.5cm}p{5cm}}
            \hline 
            \hline
            Shorthand  & Reference & Description \\ 
        \hline
             \textsc{Y6Gold} &   \cite{y6-gold}   & Coadd object catalog and associated   \\
                             &                    & products \\
             \textsc{Y6Balrog} &   \cite{y6-balrog}   & Synthetic source injection  \\
             \textsc{Y6PSF} &   \cite{y6-piff}   & Point spread function (PSF) models  \\
             \textsc{Y6Mask} &   \cite{y6-mask}   & Survey mask  \\
             \textsc{Y6Shear} &   \cite{y6-metadetect}   & \textsc{MetaDetect} shear catalog  \\
             \textsc{Y6Pos} &   \cite{y6-maglim}   & \textsc{MagLim++} lens selection and galaxy  \\
             & & clustering \\
             \textsc{Y6ShearPZ} &   \cite{y6-sourcepz}   & Source photo-$z$ calibration  \\
             \textsc{Y6PosPZ} &   \cite{y6-lenspz}   & Lens photo-$z$ calibration  \\
             \textsc{Y6WZ} &   \cite{y6-wz}   & Clustering redshift  \\
             \textsc{Y6Mode} &   \cite{y6-nzmodes}   & Method for marginalizing redshift   \\
             &  &  uncertainty \\
             \textsc{Y6Imsim} &   \cite{y6-imagesims}   & Image simulation  \\
             \textsc{Y6Mock} &   \cite{y6-cardinal}   & Mock galaxy catalog  \\
             \textsc{Y6Model} &   \cite{y6-methods}   & Modeling and analysis choices  \\
             \textsc{Y6Mag} &   \cite{y6-magnification}   & Characterising lens magnification  \\
             \textsc{Y6PPD} &   \cite{y6-ppd}   & Quantifying internal consistency with    \\
             & & posterior predictive distribution \\
             \textsc{Y6GGL} &   \cite{y6-gglens}   & Galaxy-galaxy lensing  \\
             \textsc{Y6-1x2pt} &   \cite{y6-1x2pt}   & Cosmology from cosmic shear  \\
             \textsc{Y6-2x2pt} &   \cite{y6-2x2pt}   & Cosmology from galaxy-galaxy lensing \\
                                & & and galaxy clustering \\
             \hline
             \hline
        \end{tabular}
\end{table}

The combination of galaxy clustering with \rcwr{weak gravitational lensing} was proposed by \citep{HuJain} in 2004 \rcwr{as a way to increase sensitivity relative to weak lensing alone}.
Jointly analyzing these three probes allows the data to self-calibrate astrophysical and nuisance parameters, resulting in tighter and more robust cosmological constraints. This technique was first applied to data in the first-year DES cosmology analysis (DES Y1) \cite{y1-keypaper}, and was later built upon in the DES three-year analysis (DES Y3) \citep{y3-3x2ptkp}. 
Many improvements in data quality, data quantity, and methodology have been implemented for successive DES analyses. The major methodological changes from Y3 to the present Y6 analysis are summarized in Appendix~\ref{app:changes_since_y3}.

Other collaborations have conducted 3$\times$2pt analyses on multi-survey data, relying on overlapping spectroscopic lens samples. The Kilo-Degree Survey\footnote{\url{http://kids.strw.leidenuniv.nl/}} \citep[KiDS][]{Kuijken2015,Heymans2020} presented in \citep{Heymans2020} a 3$\times$2pt analysis using a combination of the KiDS-1000 shear catalog and lenses from the Baryon Oscillation Spectroscopic Survey (BOSS) and spectroscopic 2-degree Field Lensing Survey (2dFLenS). {The most up-to-date KiDS shear analysis was recently presented in \cite{Wright2025}, and its extension to 3$\times$2pt is currently ongoing.} The Hyper Suprime-Cam Subaru Strategic Program\footnote{\url{https://www.naoj.org/Projects/HSC/}} \citep[HSC-SSP,][]{Aihara2017,Hikage2018,Hamana2019} carried out a 3$\times$2pt analysis in \citep{Sugiyama2023, Miyatake2023, Zhang2025} using the HSC Year 3 shear catalog and the SDSS DR11 spectroscopic galaxies. 

This paper presents constraints \rcwr{from} DES Y6 3$\times$2pt statistics on a baseline 6-parameter {flat} $\Lambda$CDM model and its extension to $w\ne-1$ ($w$CDM).  These results build on two papers that present subsets of the 3$\times$2pt probes: \citepalias{y6-1x2pt} carries out a detailed analysis using only cosmic shear (1$\times$2pt), while \citepalias{y6-2x2pt} investigates the combination of galaxy-galaxy lensing and galaxy clustering probes (2$\times$2pt). We also present cosmological constraints \rcwr{by combining DES Y6 3$\times$2pt results with those from other DES cosmological probes}. In particular, we present results from the combination of four dark energy probes using the DES dataset: Y6 3$\times$2pt (this work) and previous measurements of supernovae type Ia (SNe Ia) \citep{DESY5SN2024}, galaxy clusters \citep{DESY3CL2025}, and baryon acoustic oscillation (BAO) \citep{y6-BAOkp}. A \rcwr{follow-up} paper \citep{y6-clusters} will update the analysis of galaxy clusters using DES Y6 data, completing the final internal combination that was envisioned in the design of DES.  

In this paper we also analyze combinations of the DES Y6 3$\times$2pt results with state-of-the-art external probes and assess the consistency between them. Future work in \citep{y6-extensions} will use the DES Y6 data discussed in this paper to explore alternative cosmological models such as modified gravity and evolving dark energy. Further followup papers will present cosmological constraints using different lens and/or source samples and differing sets of summary statistics. 

This paper is organized as follows. In Section~\ref{sec:data} we describe the DES Y6 data used in this work. In Section~\ref{sec:twopoint} we present the measurements of the data vector. In Section~\ref{sec:method} we describe the modeling framework and analysis choices. We present our main 3$\times$2pt results in Section~\ref{sec:results}, combine with other DES probes in Section~\ref{sec:des_probes}, and combine with external probes in Section~\ref{sec:ext}. We conclude in Section~\ref{sec:conclusion}.
The DES Y6 cosmological constraints build on a series of \rcwr{18} supporting papers\footnote{\rcwr{Some of these papers are still in preparation, but are expected to be published in the near future.}}\rcwr{, and references within,} that develop the individual components that go into the final cosmological inferences. To ease the reader's navigation, this paper refers to the supporting papers with the labels listed in Table~\ref{tab:alias}.

\begin{figure*}
\includegraphics[width=\textwidth]{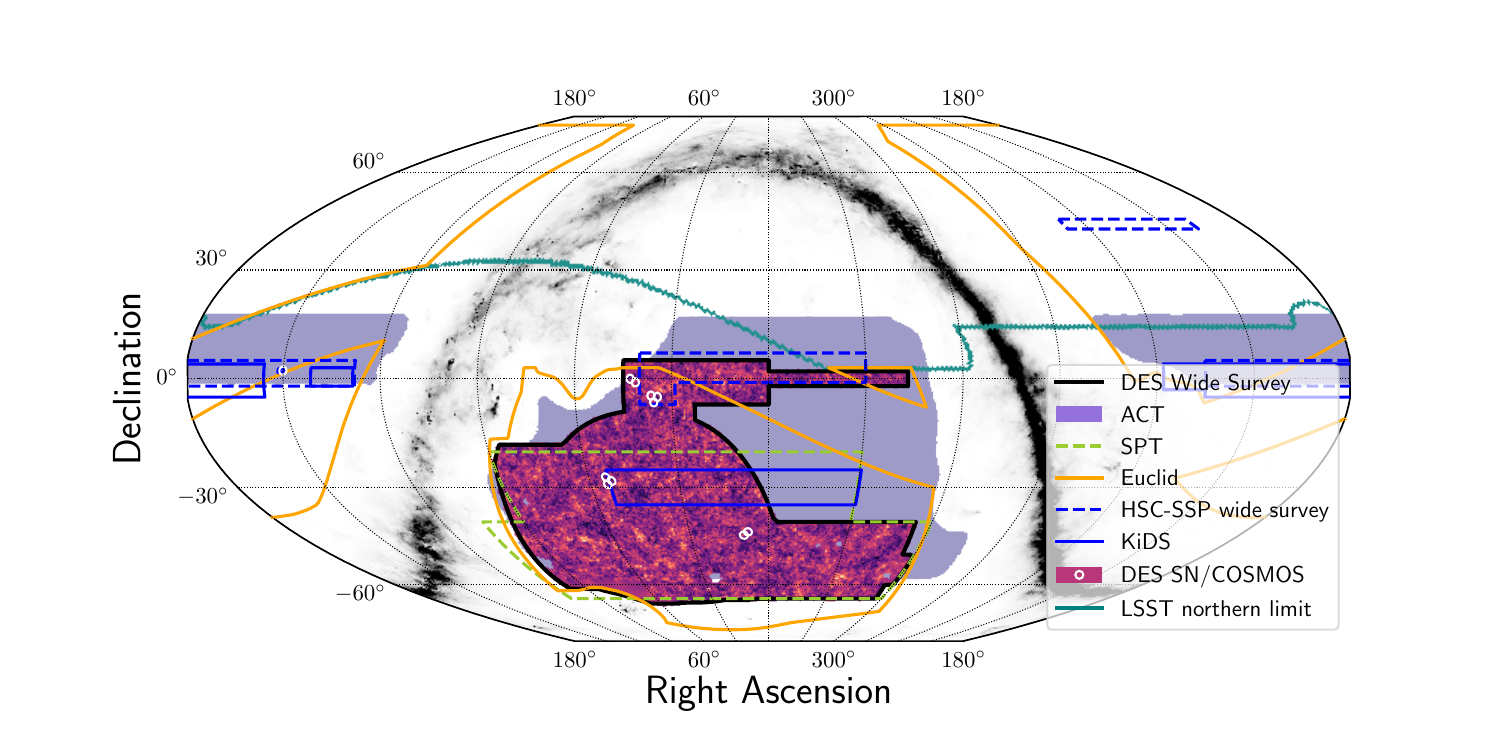}
\caption{\label{fig:footprint}
DES footprint in equatorial coordinates. The $\sim 5000 \deg^2$ \rcwr{wide-field} survey footprint is shown as a black outline, with overplotted convergence map using the Wiener filter reconstruction method \citep{y3-massmapping}. The white circles, scaled to approximately one full DECam field-of-view, show the supernovae field locations. Other current and planned surveys are shown as well. The South Pole Telescope footprint includes the SPTpol and SPT-SZ surveys described in \citep{Bocquet2024}. }
\end{figure*}
\section{The Dark Energy Survey Year 6 Data Products}\label{sec:data}

\subsection{Y6 Gold Catalog}

DES was allocated 760 distinct full- or half-nights of observations between 2013 August 15 and 2019 January 9 \citep{Diehl2020} to use the Dark Energy Camera (DECam; \citep[][]{Honscheid2008,DECam}), a 570-megapixel camera with a 3 deg$^2$ field-of-view installed at the prime focus of the 4-meter V\'{i}ctor M.\ Blanco Telescope at Cerro Tololo Inter-American Observatory (CTIO) in Chile.  
The DES wide-field survey observed in five broad bands, $grizY$, spanning $400\,\text{nm}\lesssim \lambda \lesssim 1060\,\text{nm}$.
The final DES data release, DES Data Release 2 (DES DR2; \citep[][]{DESDR2}), includes 76,217 DECam exposures from the wide-field survey in the $grizY$ bands that passed baseline survey quality criteria.  The image reduction, detrending, and processing for DES DR2 were performed using the DES Data Management system at NCSA \citep{imageproc}, with the configuration described in \cite{DESDR2}. 

DES DR2 is used to produce the DES Y6 Gold data collection, which serves as the basis for most cosmological analyses with the DES wide-field survey data \citepalias{y6-gold}. Y6 Gold augments DES DR2 with value-added information including survey systematics maps, improved multi-epoch photometry, and catalog quality flags.
The DES Y6 Gold catalog contains 669 million objects spanning 4,923\,deg$^2$ of the sky with a median coadd depth of $i = 23.8$\,mag, defined as the magnitude corresponding to a $S/N = 10$ in 1.95 arcsec diameter apertures \citep{DESDR2}. Each galaxy in the Y6 Gold sample is assigned a photometric redshift estimate, $z_{\rm DNF}$, via the Directional Neighbourhood Fitting (DNF; \citep[][]{DeVicente2016}) method.
The cosmology analyses also make use of deeper catalogs by assembling coadded images from exposures in the DES supernovae (SN) fields \citep{y3-deepfields}, which were then used to seed synthetic source injection studies using the \textsc{Balrog} software \citep{Suchyta2016, y3-balrog, y6-balrog}. 
Figure~\ref{fig:footprint} shows the sky footprint of the DES wide survey and the locations of the DES deep fields (DES SN/COSMOS).

\begin{table*}
    \centering
    \caption{\label{tab:shapes} Characteristics of the source and lens samples. For each tomographic source bin and their union, we list the mean redshift, number of objects, effective number density, shape noise, shear response and weighted residual mean shear. For lens bins we give the mean redshift, number of objects and number density for the lens sample.}
     \begin{tabular}{c|ccccccc|ccc}
\hline\hline
        & \multicolumn{7}{c}{Source sample} & \multicolumn{3}{c}{Lens sample}  \\
     Bin & $\left< z_\mathrm{s}\right>$ & num.  & $n_{\rm eff}$  & $\sigma_{\rm e}$ & $\langle R \rangle$ & $\langle e_1 \rangle $ & $\langle e_2 \rangle $ & $\left< z_\mathrm{l}\right>$ & num. & $n_{\rm gal}$\\ 
     
        & & of objects &   [gals/arcmin$^2$] & && $ \times 10^{-4}$ & $ \times 10^{-4}$ & & of objects & [gals/arcmin$^2$]\\     
     \hline 
      %\vspace{0.5cm}
        1 & 0.414 & 33,707,071 &  2.05 & 0.265 & 0.856 & +1.6 & +0.62 & 0.306 & 1,852,538 & 0.128 \\
        2 & 0.538 & 34,580,888 &  2.10 & 0.287 & 0.869 & +0.45 & -0.58 & 0.435 & 1,335,294 & 0.092 \\
        3 & 0.846 & 34,646,416 &   2.14 & 0.282 & 0.819 & +2.4 & -1.7 & 0.624  & 1,413,738 & 0.097 \\
        4 & 1.157 & 36,727,798 & 2.32 & 0.347 & 0.692 & +3.1 & +1.1 & 0.778 & 1,783,834 & 0.123 \\
        5 & - & - & - & - & - & - & - & 0.903 & 1,391,521 & 0.096 \\
        6 & - & - & - & - & - & - & - & 1.011 & 1,409,280 & 0.097 \\
        All & & 139,662,173  & 8.29 & 0.289 & 0.819 & +1.4 & -0.15 & & 9,186,205 & 0.633 \\
\hline\hline
    \end{tabular}
\end{table*}

\subsection{Source galaxies}
\label{sec:sources}
The DES Y6 shear catalog \citepalias{y6-metadetect} is constructed by applying \mdet\ \citep{Sheldon2020} to cell-based coadded images (e.g., \cite{Armstrong2024}). The principle of \mdet\ is to produce five shear catalogs for each coadd cell (of size $200\times200$ pixels): one from the unsheared image and two with positive and negative synthetic shears applied to the image in each of the spin-2 axes; $\pm e_1$, $\pm e_2$.  A nominal shear estimator is applied to each variant image; the variation of the measurement under synthetic shear is used to calculate the response factor $\langle R \rangle$ of the nominal estimator to $\gamma$ in each tomographic bin.  

Due to detecting objects on cell-based coadds, the \mdet\ catalog does not map onto the Gold catalog. We make similar quality selections as when constructing the Gold catalog, however, to ensure the quality of photometric redshifts (see Section 3.1 in \citepalias{y6-sourcepz}), and to make a common selection of regions between lens and source samples \citepalias[see][]{y6-metadetect,y6-mask}. This leaves 139,881,005 galaxies with the effective number density $n_{\rm eff} = 8.29$ galaxies/arcmin$^2$ and shape noise (which includes the intrinsic ellipticity dispersion) $\sigma_\mathrm{e} = 0.289$. Table~\ref{tab:shapes} summarizes shear responses, mean densities, and shear noise levels for each tomographic bin.

\mdet \, shear estimations make use of PSF models created with the \textsc{PiFF} software \citep{y3-piff}. The PSF models for Y6 newly incorporate the chromatic dependence of the PSF, significantly reducing modeling errors in all $griz$ bands \citepalias{y6-piff}. However, due to residual modeling errors in $g$-band, shapes measured on $g$-band images are not suitable for cosmology, but the fluxes are less affected. We therefore use the $g$-band fluxes for better photometric redshift estimates in addition to $riz$ fluxes. 

The cosmological inferences include nuisance parameters to marginalize over multiplicative biases $m$ in the \mdet\ shear estimators. Residual additive biases $c$ (measured as the weighted mean shear, see Table \ref{tab:shapes}) are substracted empirically.
Priors on $m$ 
in each source bin are estimated by building a suite of image simulations that mimic our DES images, then processing them through the same measurement pipeline as the data. The simulations are particularly important for quantifying the effect of ``blended'' (i.e., overlapping) galaxies in the images, for which the apparent shape of the observed light profile is subject to the weak lensing shear applied at two different redshifts. The Y6 image simulations \citepalias{y6-imagesims} improve upon those from \cite{y3-imagesims} in several ways: the simulations include a more realistic distribution of stars, the galaxies include a redshift-dependent clustering signal informed by an $N$-body simulation \citepalias{y6-cardinal}, and the determination of the calibration accounts for the full covariance between shear and redshift distribution calibration parameters. This last improvement is especially important in Y6, with its higher rate of blending relative to Y3.

\subsection{Lens galaxies}
\label{sec:lenses}
The lens sample used in this work is the \maglimpp{} sample described in \citepalias{y6-maglim}.   \maglimpp\ is designed to balance shot-noise against redshift resolution; we do this by selecting galaxies with with $i$-band magnitudes restricted to \citep{y3-2x2maglimforecast}:
\begin{equation} \label{eq:ilim_maglim}
    17.5 < i < 18 + 4 \times z_\mathrm{DNF}.
\end{equation}
Two additional cuts reduce systematic contamination from stars and other interlopers. First, a redshift-bin optimized star-galaxy classifier \cite{stargalsep} leverages near-infrared \texttt{unWISE} \cite{Lang_2014, Meisner_2017} data to identify optimal color cuts in $(r-z,\ z-W1)$ space for each bin, complementing the GOLD morphological classifier \citepalias{y6-gold} and removing [1.1, 0.6, 2.0, 3.3, 3.2, 1.3]\%
of objects as residual stellar contamination across redshift bins.
Second, a selection in the space of a self-organized map (SOM) of the $griz$ color space removes objects occupying compact regions with large photo-$z$ dispersion, due to either color-redshift degeneracies or training data limitations. This removes an additional $1.66\%$ of objects, primarily QSOs.
The final number density of lens galaxies in each bin is listed in Table~\ref{tab:shapes}. These number densities are significantly smaller than those of the sources due to the need for a higher photometric redshift accuracy for galaxy clustering. 

The observed galaxy density is modulated by spatially-varying observing conditions or survey properties, which modify the detection probability. For example, regions of high dust extinction often have fewer detected galaxies in them; other examples of these varying survey properties include variations due to changing observing conditions such as seeing and sky noise. To mitigate the impact of these, each galaxy is assigned a weight constructed from template maps of the spatially-varying survey properties.
Two different weights methods are used---(\texttt{ENET} \cite{Weaverdyck:2020mff} and \texttt{ISD} \cite{y3-galaxyclustering})---and the covariance matrix for the $w(\theta)$ measurements incorporates analytic marginalization over the results of these methods.

\subsection{Tomographic bin assignment}
\label{sec:tomoz}

We select lens galaxies in six disjoint $z_{\rm DNF}$ ranges to form bins in which we measure their $\delta_{\rm g}$ fields. Note that we use the $z_{\rm DNF}$ values only for bin assignment; the redshift distributions $n(z)$ of each bin's members are determined by fully distinct methods, described in Section~\ref{sec:pzs}. We select four disjoint bins of source galaxies to use for tomographic estimation of the weak lensing shear $\gamma.$ We select galaxies into these bins by using a SOM to compress their $griz$ fluxes into a two-dimensional grid of cells. This grid is further partitioned into four sets of cells according to the mean redshift estimated for each cell (see Section~\ref{sec:pzs}) to form roughly equipopulated bins (see Table \ref{tab:shapes}). \journal{}{We note that the different approaches used for tomographic binning in the lens and the source sample (using $z_{\rm DNF}$ vs. SOM) results in different number of nuisance parameters for the two samples (Table~\ref{tab:params}).}

\subsection{Photometric redshift distributions}
\label{sec:pzs}

Estimates of the redshift distribution $n^i(z)$ for each tomographic bin $i$ are required to model the 3$\times$2pt signal (see Section~\ref{sec:model}). The $n(z)$ estimations for source bins are described in \citepalias{y6-sourcepz}, and for lens bins in \citepalias{y6-lenspz}. Each uses a hybrid framework combining photometry information (SOMPZ) and clustering information (WZ) \citepalias{y6-wz}.  

The SOMPZ process begins with photometry from the DES deep fields \citep{y3-deepfields}, a sample of $1.68$ million galaxies spanning $5.88\,{\rm deg}^2$ over four different fields (\journal{}{white circles in} Figure~\ref{fig:footprint}). 
This deep data combines (i) $10\times$ the depth of the wide-field coadded images with (ii) extended wavelength coverage from DES $u$-band and near-infrared $JHK$ imaging provided by overlapping VIDEO \citep{video} or UltraVISTA \citep{ultravista} observations and (iii) highly accurate redshift information, primarily from spectroscopic and multi-band photometric data in the COSMOS field. Using the 8-band fluxes of deep-field galaxies that are bright enough to be detected in the wide-field survey, we construct a 2D SOM. The redshift distribution of each
cell is constrained by its associated deep-field galaxies with high-quality spectroscopy \citep{Gschwend2018, Lilly09zcosmos, Masters2017, C3R2_DR2, C3R2_DR3, vvds, scodeggio2018, desiedr} or many-band photo-$z$ from COSMOS \citep{cosmos2020} or PAUS \citep{paus, PAUSW1}.

A distinct two-dimensional SOM is constructed for the wide-field galaxies, using 4 bands ($griz$) instead of 8. We want the ``transfer function'' giving the probability that members of a given deep-SOM cell will be assigned to each wide-SOM cell when observed in the DES wide-field survey. This is measured using the synthetic-source catalog created by the \textsc{Balrog} process \citepalias{y6-balrog}: each galaxy in the \textsc{Balrog} catalog is a model of a true deep field galaxy that has been added into the wide-field $griz$ images and measured with the wide-field pipeline. The \textsc{Balrog} transfer function then allows mapping the redshift distributions of the deep-field SOM cells into an $n(z)$ estimate for the collection of wide-field SOM cells that comprise a given source or lens galaxy bin.

The SOMPZ derivation of the $n(z)$s is realized many times by sampling over uncertainties in COSMOS and PAUS photometric redshift assignments, over the uncertainties in the relative photometric calibrations of the deep fields, and over sample variance in the limited  area of both the deep and redshift samples. For the source galaxies, the dominant uncertainty contribution comes from imperfections in the redshift sample, primarily due to the lack of representative spectra of faint galaxies. For the lens galaxies there is no single source of uncertainty clearly dominating across all tomographic bins. As described in \citepalias{y6-wz}, the SOMPZ realizations are importance-sampled by the likelihood that each $n(z)$ realization would reproduce the observed correlations $w(\theta,z)$ against spectroscopic galaxies in several intervals of $z$ \citepalias{y6-wz}.
The correlations used in the clustering redshift method, WZ, are limited to angular scales that are smaller than those used in the fiducial $3\times2$pt cosmological analysis.  The entire process yields samples of $n(z)$ that are drawn from the joint SOMPZ+WZ likelihood. 

Multiplicative shear calibration $m$ is coupled with redshift calibration through blending, and both are incorporated at the level of the source redshift distributions.
This is represented by an effective sheared redshift distribution, $n^i_\gamma:=R^i(z)\,n^i(z)$, where the redshift dependence of the response matrix is evaluated using image simulations \citepalias{y6-imagesims}.
Since $n^i_\gamma(z)$ are not, in general, normalized, we introduce  multiplicative shear bias coefficients $m^i=\int n^i_\gamma(z)\,\mathrm{d}z-1 $, such that 
\begin{equation}
    n_{\gamma}^i=(1+m^i)\,n_{\gamma,{\rm \,norm}}^i(z).
\end{equation}
The new normalized effective sheared distributions can differ significantly from the original $n(z)$ distributions with mean-$z$ shift for each bin: $-0.013,\, -0.021,\, -0.002,\, -0.034$ \citepalias{y6-imagesims}. In the rest of this work, we use the normalized effective sheared source redshift distributions, but we omit the $\gamma$ notation.

\begin{figure}
    \centering
    \includegraphics[width=\columnwidth]{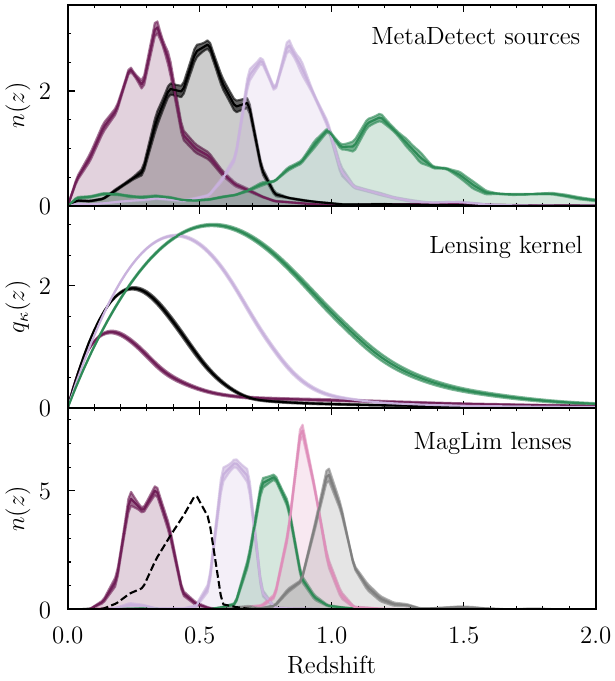}
    \caption{Estimated redshift distributions of the Y6 source (top) and lens (bottom) samples. Different colors represent different tomographic bins. The middle panel shows the lensing efficiency (Equation~\ref{eq:lensing_eff}) for each source bin. In each case, the solid lines and band represent the mean and standard distribution of 100 $n(z)$ reconstructions, drawn from the posterior of our fiducial $3\times2$pt linear bias analysis. Note that lens bin 2 is unshaded because it is excluded from our fiducial analysis (see Appendix~\ref{sec:unblinding_details} for details).}
    \label{fig:nofz}
\end{figure}

To efficiently marginalize over redshift uncertainties in the cosmological inference, we apply a mode projection approach \citepalias{y6-nzmodes} to these samples. The ensemble of SOMPZ+WZ redshift realizations are decomposed into linear sums of components $U^{i,\beta}(z)$ that capture the directions of largest cosmological impact:
\begin{equation}
  n^i(z) = \bar n^i(z) + \sum_\beta u^{\beta}\, U^{i,\beta}(z),
  \label{eq:nzmodes}
\end{equation}
{where $u^\beta$ are the mode amplitudes that are being marginalized over.}
We find that a total of seven $\beta$ modes are sufficient to describe all variations seen in the SOMPZ+WZ samples of the four source bins' $n^i(z)$ that have the capacity to produce detectable variations in the 3$\times$2pt statistics \citepalias{y6-sourcepz}.  For each lens bin, three modes are found sufficient to specify any detectable influence on the observables \citepalias{y6-lenspz}. In total, 25 nuisance coefficients $u^\beta$ specify the uncertainties in the SOMPZ+WZ estimations of $n^i(z)$ for the 10 source and lens bins.  The modes are constructed such that each $u^\beta$ has a prior with zero mean and unit variance. In most cases the priors are normal distributions; for a few of the $u^\beta,$ these are detectably non-Gaussian, but a simple transformation of variables renders the prior Gaussian again \citepalias{y6-nzmodes}.

We make a final joint adjustment to the priors on the multiplicative shear errors $m^i$ and the redshift mode coefficients $u^\beta$ to make the shear measurement model consistent with the effects of galaxy blending measured in the image simulations. This correction, as described in  \citepalias{y6-imagesims} and \citepalias{y6-sourcepz}, results in a prior with small non-zero covariances among the $m^i$ and $u^\beta$. 

We further validate our redshift calibration using the shear ratio (SR) test \citep{y3-shearratio}, by taking the ratio of galaxy-galaxy lensing using the same lens and different source bins. As described in \citepalias{y6-gglens}, we use non-overlapping lens-source pairs to ensure the approximation of geometric ratios holds, where the evolution of lens properties within the lens bins is ignored. This results in using the two lowest lens bins and the two highest source bins. This ensures SR is sensitive only to redshift and shear calibration, while being complementary at higher redshift to the information in the galaxy clustering constraint. We find that our SR result is consistent with SOMPZ + WZ + blending.

The final redshift distributions for the source and lens samples reconstructed by the modes are shown in the top and bottom panel of Figure~\ref{fig:nofz}, respectively.

\section{Two-point Measurements}\label{sec:twopoint}

From the galaxy positions in the lens sample and the shear and position of galaxies in the source sample, we measure three projected two-point correlation functions.
\begin{enumerate}
    \item The angular two-point function $w^{ij}(\theta)$ between positions of lens galaxies in redshift bins $i$ and $j$, measured by comparing counts of galaxy pairs separated by angular separation $\theta$ to the number of random-point pairs. We retain only auto-correlations, discarding measurements where $i \neq j$. The $w^{ij}(\theta)$ estimator, its measurement, and validation process are described in detail in \citepalias{y6-maglim}. 
    \item The galaxy-galaxy lensing signal $\gamma_\mathrm{t}^{ij}(\theta)$ captures the shear $\gamma$ of background (source) galaxies caused by the mass distribution traced by foreground (lens) galaxy positions. We measure $\gamma_\mathrm{t}^{ij}(\theta)$ by averaging the tangential shear of source galaxies in bin $j$ around lens galaxies in bin $i$ as a function of the angular separation between lens-source pairs. The $\gamma_\mathrm{t}^{ij}(\theta)$ estimator, its measurement, and validation process are described in detail in \citepalias{y6-gglens}. 
    \item The cosmic shear functions $\xi_\pm^{ij}(\theta)$ capture the coherent distortion of galaxy shapes. We measure $\xi^{ij}_\pm(\theta)$ as the correlation of the shear of source galaxy pairs from source bins $i$ and $j$ as a function of their angular separation $\theta$. The $\xi_\pm^{ij}(\theta)$ estimators are defined as the sum and difference of the products of the tangential and cross-components of the projected shear. More details on the estimator, the cosmic shear measurement, and validation process can be found in \citepalias{y6-1x2pt}. 
\end{enumerate}

\begin{figure*}
    \centering
    \includegraphics[width=\textwidth]{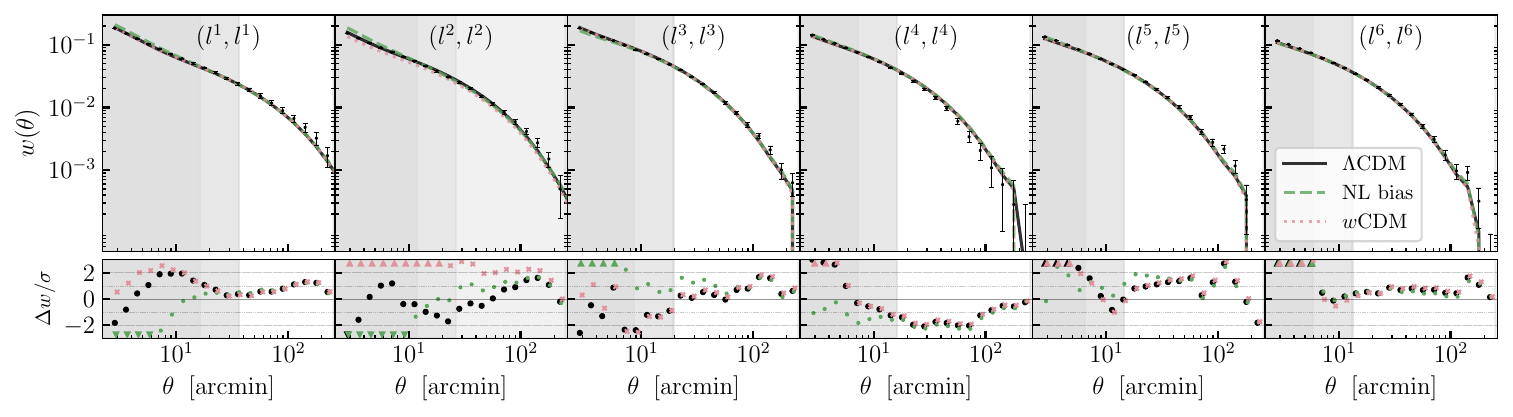}
    \caption{Angular galaxy clustering auto-correlation function as a function of angular separation $\theta$ for different lens redshift bins. Upper panels show measurements (black points) with best-fit models: $\Lambda$CDM linear galaxy bias (black solid), $\Lambda$CDM nonlinear bias (green dashed), and $w$CDM linear bias (pink dotted). Lower panels show residuals in units of the expected standard deviation. Gray shaded regions indicate excluded scales: lighter gray for linear bias cuts, darker gray for nonlinear bias cuts. Lens bin 2 is entirely shaded because it is excluded from the fiducial analysis (see Appendix~\ref{sec:unblinding_details}). Triangle markers indicate the residuals are larger than 3$\sigma$.} 
    \label{fig:wtheta-dv}
\end{figure*}

\begin{figure*}
    \centering
    \includegraphics[width=\textwidth]{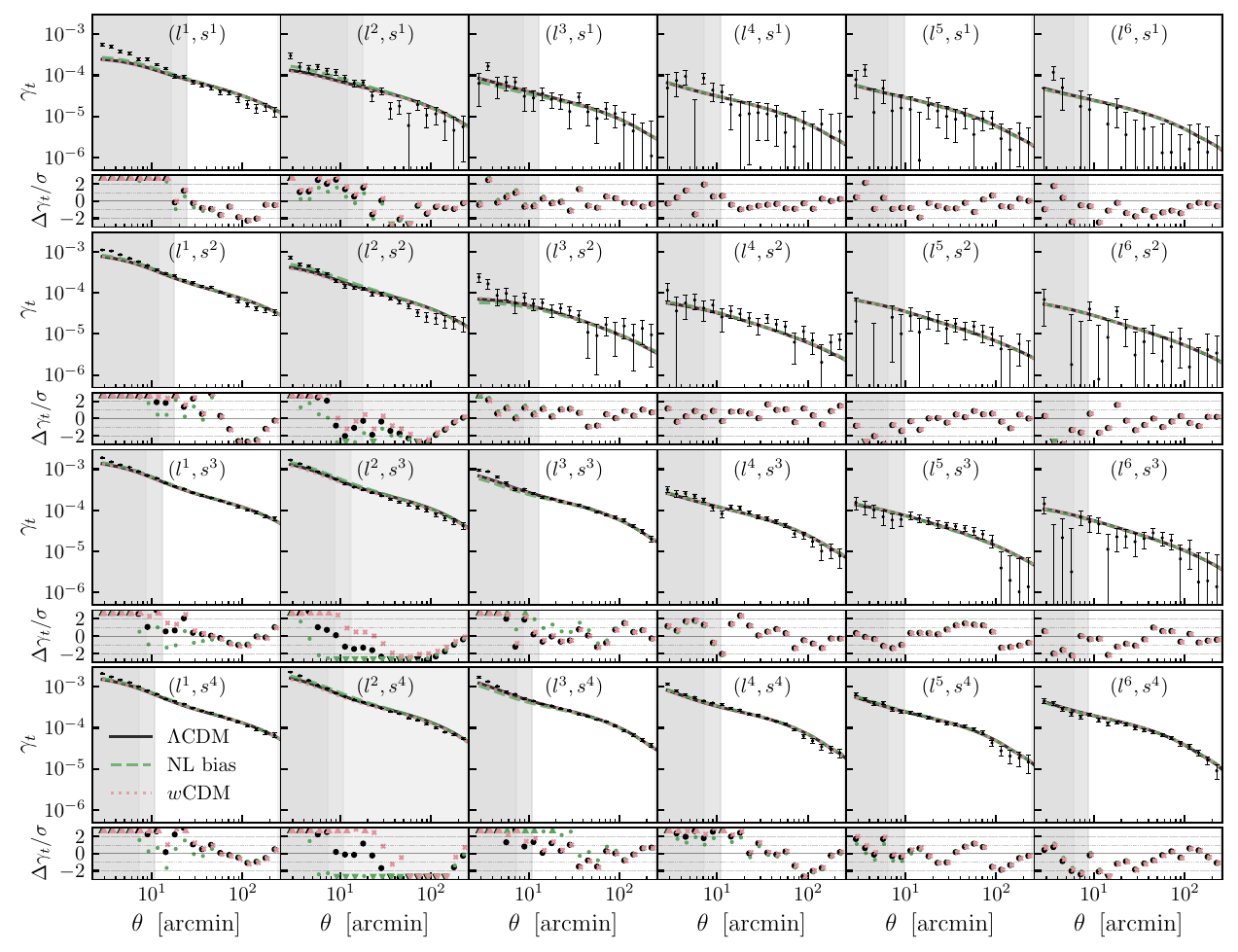}
    \caption{Galaxy-galaxy lensing correlation function $\gamma_\mathrm{t}$ as a function of angular separation $\theta$ for different lens-source bin combinations $(l^i, s^j)$. Upper panels show measurements (black points) with best-fit models: $\Lambda$CDM linear galaxy bias (black solid), $\Lambda$CDM nonlinear bias (green dashed), and $w$CDM linear bias (pink dotted). Lower panels show residuals.  Gray shaded regions indicate excluded scales: {lighter gray for linear bias cuts, darker gray for nonlinear bias cuts}. Lens bin 2 is entirely shaded because it is excluded from the fiducial analysis (see Appendix~\ref{sec:unblinding_details}). Note that this figure does not reflect the point-mass marginalization procedure, which is applied directly to the inverse covariance matrix during likelihood evaluation and  cannot be visualized through the covariance matrix elements shown here.}
    \label{fig:gammat-dv}
\end{figure*}

\begin{figure*}
    \centering
    \includegraphics[width=\textwidth]{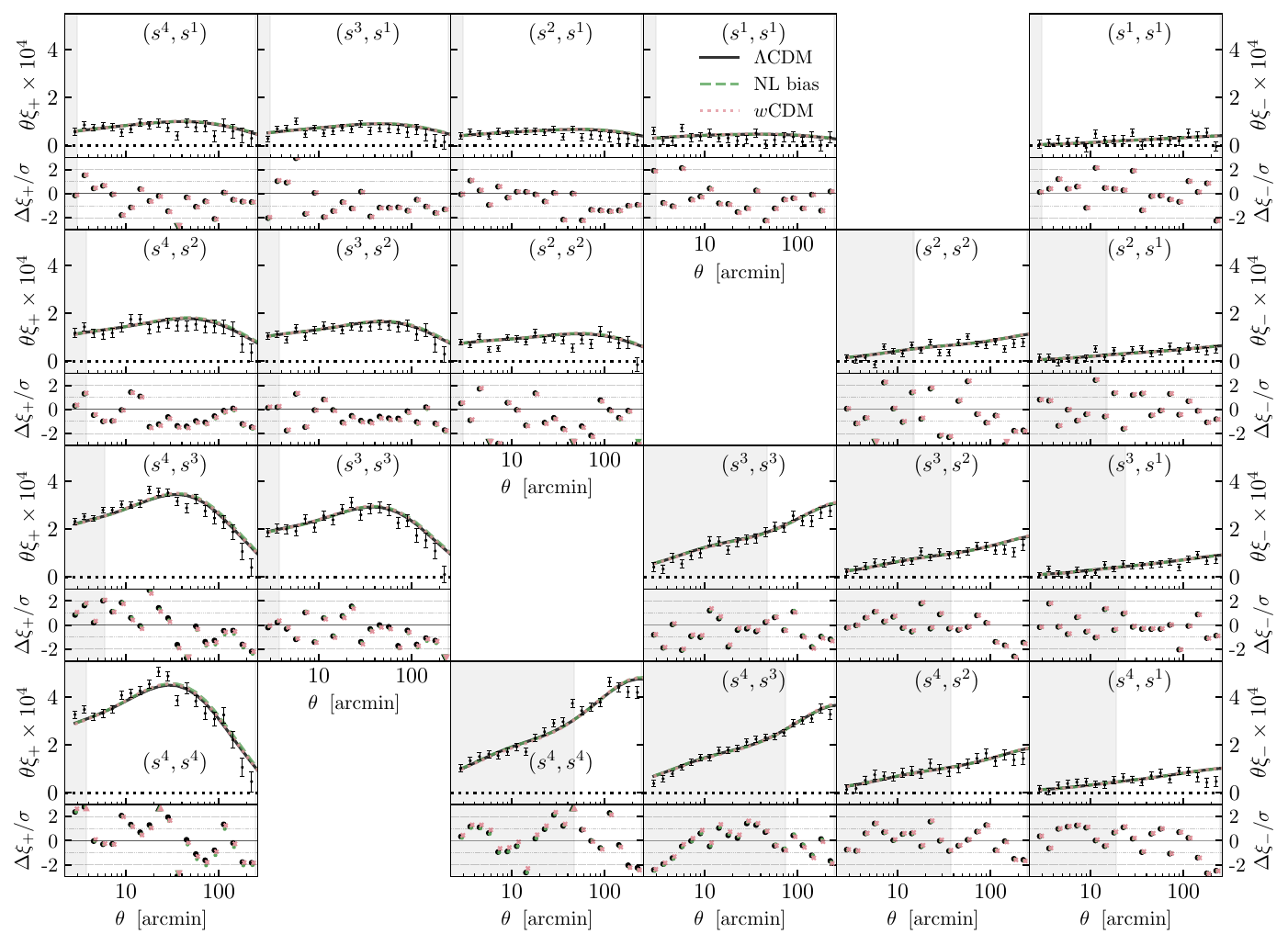}
    \caption{Cosmic shear correlation functions $\xi_+(\theta)$ (left panels) and $\xi_-(\theta)$ (right panels) for different source redshift bin combinations $(i,j)$.  Upper panels show measurements (black points) with best-fit models: $\Lambda$CDM linear galaxy bias (black solid), $\Lambda$CDM nonlinear bias (green dashed), and $w$CDM linear bias (pink dotted). Lower panels show residuals in units of the expected standard deviation. Gray shaded regions indicate excluded scales.} 
    \label{fig:xipm-dv}
\end{figure*}

To compute these data vectors, we use the fast tree code \textsc{TreeCorr} \texttt{v5.1.1} \citep{Treecorr}.  We measure these correlation functions in 26 logarithmically-spaced angular bins spanning $2.5^\prime < \theta \lesssim 1000^\prime$. Only the 20 bins between $2.5^\prime < \theta < 250^\prime$ are  used for the main analysis, as larger separations carry minimal additional information for the cosmological models we test here \journal{}{(see Appendix G of \citepalias{y6-methods})}. We have 6 lens bins and 4 source bins, leading to a total of 1000 data points, excluding galaxy clustering cross-correlations. After scale cuts (see Section~\ref{sec:scale_cuts}) and the exclusion of lens bin 2 (see Section~\ref{sec:unblinding}), the final 3$\times$2pt data vector consists of 602 (652) data points for the linear (nonlinear) galaxy bias model. We plot all of the data in Figures~\ref{fig:wtheta-dv} through \ref{fig:xipm-dv}. 

The total signal-to-noise of the 3$\times$2pt data vector  is ${\rm S}/{\rm N} =$ 110 (95) after fiducial linear galaxy bias scale cuts without (with) point-mass marginalization (see Section~\ref{sec:model}), where ${\rm S}/{\rm N} = \xi_\text{data}\mathbf{Cov}^{-1}  \xi_{\text{model}}/\sqrt{\xi_\text{model} \mathbf{Cov}^{-1} \xi_\text{model}}$, with covariance matrix $\mathbf{Cov}$ and best-fit model $\xi_\text{model}$. This is a 26\% improvement over the DES Y3 3$\times$2pt ${\rm S}/{\rm N} .$ We obtain \rcwr{S/N = 160 (147)} for the nonlinear galaxy bias case. 

\section{Analysis}
\label{sec:method}

We use the \textsc{CosmoSIS} \citep{Zuntz2015} package to perform the inference of parameters $\mathbf{p}$ from the observed 3$\times$2pt data $\mathbf{D}.$ Throughout we assume a Gaussian likelihood of the form, 
\begin{align}
  \mathcal{L}(\mathbf{D}|\mathbf{p}) \propto \exp\left(-\frac{1}{2}\left[\mathbf{D} - \mathbf{M}(\mathbf{p})\right]^\top \mathbf{Cov}^{-1} \left[\mathbf{D} - \mathbf{M}(\mathbf{p})\right]\right),
  \label{eq:likeli}
\end{align}
where $\mathbf{M}(\mathbf{p})$ is the model for the observables, and $\mathbf{Cov}$ is a predicted covariance matrix for the data.  We now describe the model, its parameters, their priors, and the covariance matrix, and then the tools of the inference process.

\subsection{Model}
\label{sec:model}

The 3$\times$2pt data vector used in this work concatenates the real-space galaxy clustering $w(\theta)$, galaxy-galaxy lensing $\gamma_\mathrm{t}(\theta)$ and cosmic shear $\xi_{\pm}(\theta)$. \rcwr{All of these statistics are Fourier transforms} of their respective angular (cross-) power spectra $C_\mathcal{AB}(\ell)$ which under the full-sky prescription take the form:

\begin{align}
    w^{ij}(\theta) &=  \sum_{\ell} \frac{2\ell+1}{4\pi} P_\ell(\cos\theta)\, C^{ij}_{\delta_{\rm g,o} \delta_{\rm g,o}}(\ell),\\
    \gamma_\mathrm{t}^{ij}(\theta) &= (1 + m^j) \sum_\ell \frac{2\ell+1}{4\pi} \frac{P_\ell^2(\cos\theta)}{\ell(\ell+1)} \,C^{ij}_{\delta_{\rm g,o} E}(\ell),\\
    \xi^{ij}_{\pm}(\theta) &= (1 + m^i)(1 + m^j) \times \\ 
    &\sum_{\ell} \frac{2\ell + 1}{4\pi} \frac{2(G^+_{\ell,2}(x) \pm G^-_{\ell,2}(x))}{\ell^2(\ell+1)^2}  
          [C^{ij}_{EE}(\ell) \pm C^{ij}_{BB}(\ell)].\nonumber
\end{align}
{Here, $P_\ell$ and $P^2_\ell$ are the Legendre and associated Legendre polynomials of degree $\ell$,
$G^{\pm}_{\ell,2}$ are given by Equation (4.19) in \cite{Stebbins1996}}, $i$ and $j$ indicate the tomographic redshift bins, $\delta_{\rm{g,o}}$ refers to the observed galaxy density contrast, $E/B$ refer to the weak lensing $E/B$-mode signals, and the set of $m^i$ corresponds to the multiplicative shear bias parameters \citepalias{y6-imagesims},  which account for biases in the shear calibration. The model is integrated over each specific angular bin range to ensure proper comparison with measurements.

The observed angular power spectra between the observed galaxy density contrast and the shear field are
\begin{equation}
    \begin{aligned}
         &C^{ii}_{\delta_{\rm g,o} \delta_{\rm g,o}}(\ell) = C^{ii}_{\delta_{\rm g} \delta_{\rm g}}(\ell) + C^{ii}_{\delta_\mu \delta_\mu}(\ell) + C^{ii}_{\delta_{\rm RSD} \delta_{\rm RSD}}(\ell)  \\
          &\hspace{2cm}+ 2 \, C^{ii}_{\delta_{\rm g} \delta_\mu}(\ell) +  2\, C^{ii}_{\delta_{\rm g} \delta_{\rm RSD}}(\ell) + 2\, C^{ii}_{\delta_{\rm RSD} \delta_\mu}(\ell), \\
        &C^{ij}_{\delta_{\rm g,o} E}(\ell) =   C^{ij}_{\delta_{\rm g} \kappa} + C^{ij}_{\delta_{\rm g} I_E} + C^{ij}_{\delta_\mu \kappa} + C^{ij}_{\delta_\mu I_E}, \\
        &C^{ij}_{EE}(\ell) =  C^{ij}_{\kappa \kappa}(\ell) + C^{ij}_{\kappa I_E}(\ell) + C^{ji}_{\kappa I_E}(\ell) + C^{ij}_{I_E I_E}(\ell) , \\
        &C^{ij}_{BB}(\ell) =  C^{ij}_{I_B I_B}(\ell), 
    \end{aligned}
\end{equation}
where the observed galaxy density receives contributions from lens magnification ($\delta_\mu$), affecting the observed number of galaxies and modeled as described in \citepalias{y6-magnification}, and from redshift-space distortions ($\delta_{\rm RSD}$) arising from the peculiar velocities of galaxies. The shear field receives contributions from the convergence field $\kappa$, including shear calibration effects, as well as from the intrinsic-alignment (IA) E/B modes ($I_{E/B}$), discussed in detail in \citepalias{y6-1x2pt}, which mimic the cosmic shear signal and are produced due to the alignment of galaxies with their local environment.

For $C^{ij}_{\delta_{\rm g} \kappa}(\ell)$ and $C^{ij}_{\kappa \kappa}(\ell)$, we compute the angular power spectra  with the Limber approximation, \red{which for a flat Universe takes the form:} 
\begin{equation}
    C^{ij}_{\mathcal{AB}}(\ell) = \int \diff \chi\, \frac{q^i_\mathcal{A}(\chi)\,q^j_\mathcal{B}(\chi)}{\chi^2} P_{\rm m}\left( k=\frac{\ell + 1/2}{\chi}, z(\chi)\right),
\end{equation}
where \(\mathcal{A}\) and \(\mathcal{B}\) represent the fields being correlated, which in our case correspond to the galaxy overdensity field \(\delta_{\rm g}\) and the convergence field \(\kappa\). The function \(P_{\rm m}(k,z)\) represents the three-dimensional matter power spectrum, which we evaluate at a given redshift \(z\) and wavenumber \(k\). The linear matter power spectrum is computed with CAMB \citep{Lewis2000, Lewis2002}, while the nonlinear power spectrum is based on \textsc{HMCode2020} \citep{Mead2021}. {Due to the narrow width of the lens tomography bins, the Limber approximation is not sufficient for evaluating the angular galaxy clustering signal. Instead we carry out the non-Limber integrals using the linear/non-linear growth split described in \cite{Fang_2020}}. {The kernel or radial weight function for the galaxy overdensity field is}
\begin{equation}
    q^i_{\delta_{\rm g}}(\chi) = b^i_{\rm l} \, n^i_{\rm l}(z(\chi)) \, \frac{\diff z}{\diff \chi} \, ,
\end{equation}
and for the convergence field, the lensing efficiency is
\begin{equation}
    q^i_\kappa(\chi) = \frac{3\,H_0^2\, \Omega_{\rm m}\, \chi}{2\,a(\chi)} \int^{\chi_H}_{\chi} \diff\chi'  \frac{\diff z}{\diff \chi'}\frac{\chi' - \chi}{\chi'}\, n^i_{\rm s}(z(\chi'))\,  \, ,
    \label{eq:lensing_eff}
\end{equation}
where \(b^i_{\rm l}\) are the linear galaxy bias coefficients, \(n_{\rm l/s}\) are the normalized redshift distribution of the lens/source galaxies, \(H_0\) is the Hubble constant, and \(a(\chi)\) is the scale factor for a given comoving distance \(\chi\). The lensing efficiency for the source sample is shown in the middle panel of Figure~\ref{fig:nofz}.

The matter power spectrum $P_{\rm m}(k,z)$ and the distances $\chi(z)$ that appear in these equations are functions of the cosmological model. Our nominal $\Lambda$CDM model has the 6 free parameters $A_s, \Omega_{\rm m}, h, \Omega_b, n_s$ and $\sum m_\nu$. For a given set of parameters, we usually translate $A_s$ to alternative measures of clustering more relevant at low-redshift, replacing it with either $\sigma_8$ (the RMS of linear fluctuations at scales of 8 Mpc/$h$) or $S_8=\sigma_8 (\Omega_{\rm m}/0.3)^{0.5}$. We also explore the $w$CDM model, with an additional free parameter, the dark energy equation of state $w$. 

The models contain several classes of nuisance parameters.  Section~\ref{sec:sources} introduced the multiplicative shear bias parameters $m^i,$ with $i\in[1,2,3,4]$ indexing source bins, with priors constrained by the shear image simulations. Section ~\ref{sec:pzs} described the priors for the $n(z)$ perturbative mode amplitudes $u_\mathrm{s}^\beta,$ with $\beta \in [1,\ldots,7]$, and the 18 total lens-redshift mode amplitudes $u_\mathrm{l}^{i,\beta}$ ($\beta \in [1,2,3]$).\footnote{The full covariance among the $m^i$ and $u^\beta_\mathrm{s}$ parameters derived in \citepalias{y6-imagesims} that accounts for source blending will be available with our ancillary data products.}  We also use magnification coefficients $\alpha^i$ for each lens bin to model $\delta_\mu,$ with \red{calibration} of their priors from image simulations detailed in \citepalias{y6-magnification}.

The remaining nuisance parameters in our model are required to account for uncertainties in astrophysical theory and observing conditions, as described below.\\

\paragraph*{Nonlinear power spectrum:} \rcwr{We model the impact of nonlinear structure formation on the matter power spectrum using \textsc{HMCode2020} \citep{Mead2021}. A significant}  uncertainty here is how much baryonic feedback, mainly from AGN, suppresses matter clustering.\textsc{HMCode2020} introduces the effect of baryons via the sub-grid heating parameter $\Theta_{\rm AGN} = {\rm log_{10}} (T_{\rm AGN}/{\rm K})$, indicating the strength of the AGN. \rcwr{We employ angular scale cuts, to be described below, to mitigate potential biases from this uncertainty while fixing $\Theta_{\rm AGN}=7.7$ in our fiducial analysis (see \citepalias{y6-methods} for more details). This value} corresponds to the one employed in the fiducial \textsc{Bahamas} hydrodynamical simulations \citep{McCarthy2017}. \rcwr{We examine the cosmological impact of other choices of parameters in Appendix~\ref{sec:robustness}. We find them to cause $<0.3\sigma$ changes in the inferred cosmology.}

\paragraph*{Nonlinear galaxy bias:} Our simplest model assumes linear scale-independent galaxy bias, leading to a single free bias parameter $b_1^i$ for each lens bin. We also explore a more complex nonlinear bias model with an additional free parameter $b^i_2$  for local quadratic bias in each lens bin.  \rcwr{For more details see} \citepalias{y6-methods}. 

\paragraph*{Point-mass marginalization:} Because the tangential shear signal $\gamma_\mathrm{t}(\theta)$ is an integral of the mass distribution at radii $\le\theta,$ modeling uncertainty in the small-scale mass distribution around galaxies can propagate to larger radii. We account for this by marginalizing over a mean central point mass (PM) in each lens \cite{MacCrann2020}, which introduces a term with $1/\theta^2$ scaling,
\journal{}{
\begin{equation}
    \gamma_\mathrm{t}^{ij}(\theta) \;\rightarrow\; \gamma_{\mathrm{t},\mathrm{model}}^{ij}(\theta) + \frac{A^{i} \beta^{ij}}{\theta^2},
  \end{equation}
where, 
\begin{equation}
\beta^{ij}\equiv\frac{1}{\rho_{\rm crit}\Omega_m}\int d\chi\,
  \frac{n^i_{\rm l}(z(\chi))\,\frac{dz}{d\chi}\,q^j_\kappa(\chi)}{\chi^2}
\label{eq:betaij}
\end{equation}
with $\rho_{\rm crit}$ the present-day critical density and $q^j_\kappa$ the lensing efficiency of Eq.~(\ref{eq:lensing_eff}). The $\beta^{ij}$ are purely geometric and are computed at a fixed fiducial cosmology. The PM amplitudes $A^i$ are assumed to have negligible redshift evolution within the corresponding tomographic lens bins. We then marginalize analytically over one PM amplitude per lens bin, assuming Gaussian priors}, by adding a term to the inverse covariance.

In practice this removes sensitivity to the mass distribution within our smallest retained $\theta$ value. \journal{}{The implementation of the PM term has been thoroughly tested in \citep{y3-generalmethods, y3-2x2ptbiasmodelling, Prat2023}, including a discussion of the assumed prior for $A^i$.}

\paragraph*{Intrinsic alignments (IA):} The fiducial model for IA of the source galaxies is the tidal alignment and tidal torquing model \citep[TATT,][]{blazek19} described in \citepalias{y6-methods}. \rcwr{Free parameters are required for IA scaling linearly with the tidal field ($A_1$), for IA quadratic in the tidal field ($A_2$), and for the dependence of each on source redshift $(\eta_1,\eta_2)$. A fifth parameter, the coefficient $b_{\rm TA}$ for the IA term that scales as the product of the density field and its tidal field, is held fixed.} The pivot redshift $z_0$ is chosen to match the peak sensitivity of DES Y6 shear data. In Appendix~\ref{sec:robustness}, we investigate the consequences of freeing $b_{\rm TA},$ and of fixing $A_2=0$ to reduce to the simpler nonlinear alignment \citep[NLA,][]{hirata04, bridle07} model. 
\paragraph*{Observing conditions marginalization:} We analytically marginalize over uncertainty in computing the  weights that remove correlations of the density field with observing conditions. See \citepalias{y6-maglim} for a description of the weights methods and marginalization and \citepalias{y6-mask} for details on the survey properties used as weights templates.

\subsection{Covariance}

The covariance of the $3 \times 2$pt signal is computed analytically using \textsc{CosmoCov} \cite{Krause2017, Fang2020}, which incorporates both Gaussian and non-Gaussian contributions. The covariance estimation code was thoroughly validated for the DES Y3 $3 \times 2$pt analysis \cite{y3-covariances}. Furthermore, we marginalize analytically over \rcwr{uncertainties in the weights of observational systematics} that modulate $w(\theta),$ as well as point mass contributions to the galaxy-galaxy lensing signal.

There are several changes in the Y6 covariance \rcwr{and modeling choices} \citepalias{y6-methods} compared to Y3, which have \rcwr{a well-understood impact on the error estimates}: (i) we incorporate non-zero central values for multiplicative shear bias, requiring adjusted effective source galaxy densities; (ii) we employ \textsc{HMCode2020} for signal modeling but \rcwr{keep} \textsc{Halofit} \citep{Takahashi2012} for covariance calculation; this leads to minor discrepancies in individual covariance elements ($\lesssim 2.5\%$) and negligible impact in our error estimates (see Appendix \ref{sec:cov-missmatch}); (iii) we compute survey geometry corrections to shape/shot noise \cite{Troxel2018} directly in configuration space from the Y6 random catalogs. The latter is necessitated by the \rcwr{fact that the Y6 mask contained more small-scale structures compared to Y3.} 

\subsection{Scale cuts}
\label{sec:scale_cuts}

Reference \citepalias{y6-methods} derives lower bounds on $\theta$ for the DES Y6 $\xi_{\pm} (\theta), \gamma_\mathrm{t} (\theta)$, and $w(\theta)$ that reduce the influence of uncertainties in the theoretical model to well below the other uncertainties in the inference. These scale cuts were tested for both $\Lambda$CDM and $w$CDM. For linear galaxy bias, the scale cuts are 9 Mpc/$h$ for $w(\theta)$, 6 Mpc/$h$ for $\gamma_\mathrm{t}(\theta)$, and \rcwr{$[\Delta \chi^2]_{\rm max}=5$} for $\xi_{\pm}(\theta)$. For $w(\theta)$ and $\gamma_\mathrm{t}(\theta)$, we convert the projected physical scales to an angular scale assuming the fiducial cosmology in \citepalias{y6-methods} and the mean redshift of the lens bins. For $\Delta \chi^2$, we calculate the $\chi^2$ difference between the theoretical cosmic shear data vector computed with our fiducial model and two versions generated using an alternate model for nonlinear dark matter clustering: one without baryonic feedback effects and another one with more extreme baryonic feedback contamination than our fiducial.
We adjust the scale cut until $\Delta \chi^2< [\Delta \chi^2]_{\rm max}$. As such, $[\Delta \chi^2]_{\rm max}=5 $ maps into concrete $\theta_{\rm min}$ values per bin-pair (with smallest $\theta_{\rm min} \sim 3'$ in our case). These cuts are similar to those used in DES Y3 \citep{y3-generalmethods}, with $\xi_{\pm}(\theta)$ \rcwr{including slightly more scales}. The nonlinear bias model enables use of smaller-scale data---down to 4 Mpc/$h$ for both $w(\theta)$ and $\gamma_\mathrm{t}(\theta)$ as seen in Figures~\ref{fig:wtheta-dv} and~\ref{fig:gammat-dv}. This changes comes at the cost of having to marginalize over additional nuisance parameters for bias, resulting in nearly equal final precision~\citep{y3-2x2ptaltlensresults, y3-2x2ptbiasmodelling}.

\begin{table}
    \centering
\caption{Fiducial values and priors for the cosmological, astrophysical, and calibration parameters. Parameters are sampled using either flat priors, denoted by [min, max], or Gaussian priors, \journal{}{following the notation $\mathcal{N}(\mu, \, \sigma)$ for a normal distribution with mean $\mu$ and variance $\sigma$}. The priors on the calibration parameters of the source galaxy samples, namely the multiplicative shear bias $m$ and the source redshift distribution modes, are correlated, see Sec. \ref{sec:model}. The width of their Gaussian priors is determined by the covariance matrix of the parameters.}
    \label{tab:params}
        \begin{tabular}{ll}
            \hline \midrule
            Parameter  & Prior \\ 
            \midrule
            \multicolumn{2}{l}{\textbf{Cosmology}} \\
            $\Omega_{ \rm m}$ & [0.1, 0.6] \\ 
            $A_{\rm s} \times 10^9$  & [0.5, 5] \\ 
            $h$  & [0.58, 0.8] \\ 
            $\Omega_{\rm b}$  & [0.03, 0.07] \\ 
            $n_{\rm s}$  & [0.93, 1.00] \\ 
            $w$  & [-2, -1/3] \\
            $m_\nu$ [eV]  & [0.06, 0.6] \\ 
            \midrule
            \multicolumn{2}{l}{\textbf{Intrinsic alignment}} \\
            $A_1$  & [-1, 3]  \\
            $A_2$  & [-3, 3] \\
            $\eta_1$  & $\mathcal{N}(0.0,3.0) \in [-5,5]$ \\
            $\eta_2$  &  $\mathcal{N}(0.0,3.0) \in [-5,5]$ \\
            $b_{\rm TA}$  & 1, fixed \\
            $z_0$ & 0.3, fixed \\
            \midrule
            \multicolumn{2}{l}{\textbf{Lens galaxy bias}} \\
            $b_1^i (i \in [1, 6])$  & [0.8, 3]\\
            $b_2^i (i \in [1, 6])$  & [-3, 3]\\
            \midrule
            \multicolumn{2}{l}{\textbf{Lens magnification}} \\
            $\alpha^1$  & $\mathcal{N}(1.58, 0.04)$\\
            $\alpha^2$  & $\mathcal{N}(1.38, 0.11)$\\
            $\alpha^3$  & $\mathcal{N}(2.04, 0.08)$\\
            $\alpha^4$  & $\mathcal{N}(2.21, 0.08)$\\
            $\alpha^5$ & $\mathcal{N}(2.45, 0.14)$\\
            $\alpha^6$  & $\mathcal{N}(2.42, 0.13)$\\
            \midrule
            \multicolumn{2}{l}{\textbf{Lens n($z$) modes}} \\
            $u_{\rm l}^{i, \beta} \; (i \in [1, ..., 6], \; \beta \in [1,2, 3])$  & $\mathcal{N}(0, 1) \in [-3,3]$ \\
            \midrule
            \multicolumn{2}{l}{\textbf{Source n(z) modes}} \\
            $u_{\rm s}^{\beta} \; (\beta \in [1,..., 7])$  &  $\mathcal{N}(0, \, 1) \in [-3,3]$ \\
            \midrule
            \multicolumn{2}{l}{\textbf{Shear calibration}} \\
            $m^1$ & $\mathcal{N}(-0.0034,0.0058)$ \\
            $m^2$ & $\mathcal{N}(0.0065,0.0066)$ \\
            $m^3$ & $\mathcal{N}(0.0159,0.0059)$ \\
            $m^4$ & $\mathcal{N}(0.0017,0.0122)$ \\

            \midrule
            \multicolumn{2}{l}{\textbf{External CMB data}} \\
            $A_{\rm CMB}$ (ACT \& \textit{Planck}) &  $\mathcal{N}(1.0,0.0025) \in [0.9,1.1]$ \\
            $T_{\rm cal}$ (SPT-3G) &  $\mathcal{N}(1.0,0.0036) \in [0.8,1.2]$ \\
            $P_{\rm ACT}$ (ACT)  &  $[0.9,1.1]$ \\
            $E_{\rm cal}$ (SPT-3G)  &  $[0.8,1.2]$ \\
            \midrule \hline
        \end{tabular}

\end{table}

\subsection{Sampling}
\label{sec:sampling}

Table~\ref{tab:params} lists all the parameters used in the model $\mathbf{M}(\mathbf{p})$ for the likelihood in Equation~\ref{eq:likeli} of our 3$\times$2pt data. The cosmological parameters of interest ($A_s$, $\Omega_{\rm m}$) and galaxy bias parameters take wide, flat priors, while tight, informative Gaussian priors are placed on nuisance parameters that we are able to constrain in simulations of our data (lens magnification, redshift and shear calibration parameters). Priors on the IA parameters are motivated by our current understanding of IA from observations (see \citepalias{y6-methods}).

We sample $\mathcal{L}$ in Equation~\ref{eq:likeli} using the \textsc{Nautilus} sampler \citep{Lange2023}.\footnote{All runs with \textsc{Nautilus} use the following settings: $n_{\rm live} = 10,000$, $\mathrm{discard\_exploration} = \mathrm{T}$, $n_{\rm networks} = 16$.} The sampling results in Monte Carlo chains for 3$\times$2pt as well as other individual probes. For 3$\times$2pt chains, we also specifically test in Appendix~\ref{sec:projection} for potential prior-volume effects, whereby the posterior density is strongly influenced by nonlinear projection of the prior volume into the space of cosmological parameters (also known as ``projection effects''). For our fiducial results, we only quote parameters that are not shifted significantly by prior-volume effects.

For some purposes, e.g.\rcwr{,}\ combining the likelihoods or posteriors of different probes across a common parameter space, it is necessary to produce a density estimator of the sampler output.  
We do this using normalizing flows \cite{Raveri:2024dph, Raveri:2021wfz, gatti2024}. Normalizing flows are machine-learning models that transform a simple base distribution (typically a multivariate Gaussian) into a complex target via a sequence of invertible, differentiable mappings. Once the mapping and its Jacobian are learned, one can both evaluate the probability density of the target distribution and draw samples from it. This enables combining \rcwr{independent probes} without rerunning chains for the full joint likelihood.
The normalizing flow is implemented in the
 {\tt tensiometer} package~\cite{Raveri:2021wfz, RaveriHu}. 

\journal{}{Once we obtain the posterior samples, we can calculate the Figure-of-Merit (FoM) in specific parameter spaces to facilitate comparison between chains,
\begin{equation}
    \text{FoM}{p_1, p_2} \equiv \left( \det{\text{Cov}(p_1, p_2)} \right)^{-1/2}.    
\end{equation}
}

\subsection{Goodness of fit and internal consistency} 
\label{sec:goodness_of_fit}

We assess the ability of our model to fit the data using two approaches: a posterior predictive distribution (PPD) metric that quantifies the internal consistency of measurements; and point estimates of the goodness of fit for the maximum \textit{a posteriori} (MAP) parameters. 

The PPD is the probability distribution for realizations of a set of test observables  $\mathbf{D}_{t}$ conditioned on a measurement of a (potentially different) set of observables, $\mathbf{D}_c^{\rm obs}.$ Given model parameters $\mathbf{p}$ and an estimate of the posterior $P(\mathbf{p}|\mathbf{D}_{c}^{\rm obs})$, 
\begin{equation}
    P(\mathbf{D}_{t}|\mathbf{D}_{c}^{\rm obs}) = \int P(\mathbf{D}_{t}|\mathbf{D}_{c}^{\rm obs},\mathbf{p})\,P(\mathbf{p}|\mathbf{D}_{c}^{\rm obs})\,\diff\mathbf{p}.
\end{equation}
Comparing an actual measurement $\mathbf{D}_t^{\rm obs}$ to the PPD assesses the consistency of measurements $\mathbf{D}_t$ and $\mathbf{D}_c$ under the assumed model. We can compress a high-dimensional PPD distribution to a single summary statistic giving the fraction of $\mathbf{D}_t$ with lower PPD than the observed $\mathbf{D}_t^{\rm obs}$:
\begin{equation}
    \Delta_{\rm PPD} = \int_{P(\mathbf{D}_{t}|\mathbf{D}_{c}^{\rm obs})<P(\mathbf{D}_{t}^{\rm obs}|\mathbf{D}_{c}^{\rm obs})}P(\mathbf{D}_{t}|\mathbf{D}_{c}^{\rm obs})\,\diff\mathbf{D}_{t}.
\end{equation}
Small values of $\Delta_{\rm PPD}$ indicate a lack of consistency, while values close to one signify over-fitting. If $\mathbf{D}_t=\mathbf{D}_{c}$, or if the test observables are a subset of  $\mathbf{D}_{c}$, this serves as an assessment of the goodness of fit. We can also evaluate the PPD for cases where $\mathbf{D}_t$ and $\mathbf{D}_c$ are disjoint but correlated sets of observables in order to assess the internal consistency between those measurements.

For Y6, PPDs are estimated using Gaussian mixture models.  The Y6 PPD methods and their improvements over Y3 methods \citep{y3-inttensions} are detailed in \citepalias{y6-ppd}.

Another statistic used to gauge goodness of fit is the \rcwr{value of} $\chi^2$  at the best-fit parameter values. The standard way to compute this would be $\chi^2 = -2\ln{\mathcal{L}}_{\rm max}$, evaluated at the parameter values maximizing the likelihood given in Equation~\ref{eq:likeli}.  In this work, we employ a slightly different definition that incorporates agreement with Gaussian priors as part of our goodness-of-fit assessment, following \cite{y3-3x2pt-ext-kp}. If the posterior for data $\bf{D}$ at a location $\mathbf{p}$ in parameter space is $P(\bf p|\bf{D}) \propto \mathcal{L}(\bf{D}|\bf{p})\pi_G(\bf{p})\pi_F(\bf{p})$, where $\pi_G$ is the product of all Gaussian priors and $\pi_F\equiv \prod_i(\Delta\theta_i)^{-1}$ is the product of all flat priors (where $\Delta \theta$ is the prior width for each parameter $i$), then we define $\chi^2 = -2\ln{\left[\mathcal{L}\pi_G\right]_{\rm max}}$. Since the same parameters that maximize the posterior will also maximize the product $\mathcal{L}\pi_G$, we can simply evaluate this from the MAP by subtracting the contribution from $\pi_F$.

If $P_{\mathrm{MAP}}$ is the value of the posterior associated with the MAP parameters, we \red{thus} compute 
\begin{equation}\label{eq:chi2}
    \chi^2 = -2 \ln{P_{\mathrm{MAP}}} - 2\sum_i \ln{\Delta\theta_i}.
\end{equation}
While in this paper we only report $\chi^2$ statistics in the context of model comparison, we did evaluate $\chi^2$ per degree of freedom as a cross-check on PPD statistics during pre-unblinding validation checks (see Table~\ref{tab:chi2} from Appendix~\ref{sec:unblinding_details}). For that evaluation, the number of degrees of freedom $\nu=N_{\rm data} - N_{\rm eff}$ is evaluated by adjusting the number of sampled parameters to
\begin{equation}
    N_{\rm eff} = N_{\rm sampled} - {\rm Tr}(\mathcal{C}_P\,\mathcal{C}_{\pi}^{-1}),
\end{equation}
where $\mathcal{C}_P$ and $\mathcal{C}_{\pi}^{-1}$ are parameter covariances estimated from the posterior and prior, respectively. 

We estimate MAP parameters using a two-step optimization procedure, where the $N_{\rm guess}=20$ highest posterior samples from a chain are used as starting guesses for optimizer searches.\rcwr{\footnote{We use \texttt{scipy}'s Powell optimizer in our analysis, though we find similar results for noise levels with \texttt{minuit} and \texttt{BOBYQA} algorithms. See Appendix~E of \citepalias{y6-methods} for more details.}} We select $\mathbf{p}_{\rm MAP}$ to be the parameters associated with the highest posterior found among those $N_{\rm guess}$ searches. 
To indicate when marginalized 1D posteriors may be dominated by volume effects, we also compute and show the projected joint highest posterior density (PJ-HPD) of parameters. The PJ-HPD indicates the highest posterior density region that contains 68\% of a parameter's marginalized posterior mass \cite{joachimi21}. By definition it includes the MAP and large differences between the marginalized 1D posteriors and the PJ-HPD indicate where projection or volume effects may be significant.

\subsection{Tension metrics and external consistency}
\label{sec:tension_metrics}

We require a measure of the consistency of probability distributions over the same parameters $\mathbf{p}$ conditioned on two different experiments, $A$ and $B$---for example, we must do so before ascribing any meaning to their joint constraint $P(\mathbf{p} | A,B).$
We measure consistency from the probability of a parameter difference, following the method described in ~\cite{Raveri:2021wfz, RaveriHu}, as employed in the recent DES analyses~\cite{gatti2024, DES-BAO-SN-cosmo}. The parameter difference probability density, $P(\Delta \mathbf{p})$, is given by the cross-correlation integral:
\begin{equation} \label{Eq:ParameterDifferencePDF}
P(\Delta \mathbf{p}) = \int P(\mathbf{p} | A)\, P(\mathbf{p} -\Delta \mathbf{p}|B) \,\diff\mathbf{p}.
\end{equation}
A scalar consistency estimator is the $\Delta\mathbf{p}$ probability above the iso-density contour corresponding to no parameter shift ($\Delta \mathbf{p} = 0$):
\begin{equation} 
\Delta = \frac{1}{\sum_{i=1}^{n} w_i} \sum_{i=1}^{n} w_i\, S( \hat{P}(\Delta \mathbf{p}_i)-\hat{P}(0) ),
\label{Eq:NFMCMCParamShiftProbability}
\end{equation}
where $S(x)$ is the Heaviside step function (unity for $x > 0$, zero otherwise), $n$ is the number of 
samples $\Delta \mathbf{p}_i$ from $P(\Delta \mathbf{p})$ with weights $w_i$ and $\hat{P}$ represents their probability as given by the normalizing flow model for the parameter difference distribution.

Tension results are reported in terms of the effective number of standard deviations, i.e.\ the number of standard deviations that an event with the same probability would have had if it had been drawn from a Gaussian distribution \citep{RaveriHu}:
\begin{equation} 
n_\sigma \equiv \sqrt{2}\,{\rm erf}^{-1}(\Delta)\,,
\label{Eq:EffectiveSigmas}
\end{equation}
where ${\rm erf}^{-1}$ is the inverse error function. We set $n_\sigma < 3$ as the criterion for different data sets agreeing well enough that it makes sense to examine their combined constraint.

While the integrations in Equations~\ref{Eq:ParameterDifferencePDF} and ~\ref{Eq:NFMCMCParamShiftProbability} can be done if we are interested in consistency for a single dimension of parameter space, more generally it involves a high-dimensional integral further complicated by knowing $P(\mathbf{p} | A)$ and $P(\mathbf{p} | B)$ only through a set of incommensurate samples.  In practice, the estimation of $\Delta$ is done by constructing density estimators with normalizing flows, and executing the integral in Equation~\ref{Eq:NFMCMCParamShiftProbability} via Monte Carlo methods, as described in \citep{Raveri:2021wfz} and executed by the {\tt tensiometer} code. In this paper, most of the discussion will center around full-parameter tensions as listed in Table~\ref{tab:consistency}, but we will also quote 1D or 2D numbers when needed.

\subsection{Model comparison}\label{sec:model_comparison}
We employ two metrics for deciding whether the data prefer $\Lambda$CDM or $w$CDM: 
\begin{itemize}
\item \textbf{Posterior differences:} Taking advantage of the fact that $\Lambda$CDM is a nested model of $w$CDM, we require  $|{-}1-w| \gg \sigma(w)$ to have significant preference for $w$CDM over $\Lambda$CDM, where $w$ and $\sigma(w)$ are the mean and standard deviation of the marginalized posterior of the $w$CDM model. 
\item \textbf{Goodness-of-fit differences:} We compute goodness-of-fit differences as:
\begin{equation}\label{eq:deltachi2}
\Delta \chi^2 \equiv \chi^2_{w\mathrm{CDM}} - \chi^2_{\Lambda \mathrm{CDM}} , 
\end{equation}
where each $\chi^2$ is computed with Equation~\ref{eq:chi2}. We note that the fact that the $\Lambda$CDM parameter space is a subset of $w$CDM guarantees $\Delta \chi^2\leq 0$. To assess the significance of the resulting value, we conduct a likelihood ratio test: we evaluate the probability $p$ of finding a value higher than $|\Delta\chi^2|$ when drawing from a $\chi^2$ distribution with one degree of freedom. We then convert this to the effective number of standard deviations for a $\chi^2$ variable\footnote{\journal{}{After submission we found that the $\beta_{ij}$ factors in Equation \ref{eq:betaij} had been initialized and fixed to different input nuisance parameters for each \textsc{Nautilus} worker, spanning the priors. We checked that this produces a negligible impact on the posteriors (shifts of less than 0.02$\sigma$ in the mean and less than 1\% in the width of the marginalized constraints on cosmological and nuisance parameters). The maximum a posteriori (MAP) estimates are more sensitive, since each minimizer run used a slightly different likelihood normalization and point-mass template, shifting the best-fit $\chi^2$ by $\mathcal{O}(1)$. However, re-evaluating and re-optimizing the MAP searches for 3$\times$2pt with a common fixed $\beta_{ij}$ does not change any of the conclusions of this article based on $\Delta \chi^2$. Hence, we decide to keep the originally quoted posterior means, MAP values and $\Delta\chi^2$.}}.
\end{itemize}

\subsection{Blinding and unblinding}
\label{sec:unblinding}

To avoid confirmation bias, it is necessary for decisions about the analysis to be made without knowledge of whether they will drive the inferred cosmological parameters closer to or further from particular values, such as previous experiments' results. This is particularly true when previous results have shown marginal evidence for failure of the $\Lambda$CDM model.  A blinding framework also forces a large collaboration to codify the criteria that will be used to detect internal inconsistencies in their data and resolve them before knowing the results.
The DES Y6 3$\times$2pt analysis has three levels of blinding \journal{}{in place from the beginning}. \journal{}{We unblind each level consecutively after certain pre-defined tests are passed as described below.}

\begin{itemize}
\item \textbf{Catalog level:} The galaxy shapes $e$---and the inferred shears $\gamma$---in the \mdet\  catalog are rescaled by some blind factor  $0.9<f<1.1$ (more precisely, $\eta = \tanh^{-1}e$ is rescaled).  All work proceeds with the blinded shear catalog until a suite of null tests on the catalog, described in \citepalias{y6-metadetect}, are successfully passed. For example, the inferred shear $\gamma$ should have no correlation with the properties of the PSF.

\item \textbf{Data-vector level:} The summary statistics derived from the catalogs are transformed using the technique from \citep{y3-blinding}, whereby they are changed by factors that would transform data from a nominal cosmology to some (blind) other cosmology.  This preserves the ability to check for consistency within and between the summary statistics under the cosmological models being considered.  The null tests performed are described in \citepalias{y6-1x2pt} (for $\xi_\pm$), \citepalias{y6-gglens} (for $\gamma_\mathrm{t}$) and \citepalias{y6-maglim} (for $w$).  All choices of measurement and astrophysical models are complete before the data vectors are unblinded. Once all tests pass in this stage, we unblind the data vector. Note that we do not run chains on the blinded data vectors in this framework. 

\item \textbf{Cosmology level:} Cosmological inference chains {for $3\times2$pt, $2\times2$pt ($w + \gamma_{\rm t}$) and $1\times2$pt ($\xi_\pm$)} are run without revealing the values of cosmological parameters they yield.  The covariance matrix for these runs assumes a fiducial cosmology.  The $\Delta_{\rm PPD}$ values from these chains must exceed the predetermined threshold $\Delta_{\rm PPD}>0.01$, and internal consistency between the probes is also tested and required to have $\Delta_{\rm PPD}>0.01$. We also examine the $\chi^2/\nu$ values of the MAP fits (without revealing the MAP values).  We check whether the posterior distributions of any of the redshift, shear and magnification nuisance parameters push strongly against the priors. Once these tests are passed, we recalculate the covariance matrix using the (still blind) MAP cosmology from the $3\times2$pt case, then rerun the chains and recheck the $\Delta_{\rm PPD}$ values.  Only after these pass do we unblind the  cosmological parameters.   
\end{itemize}

During the last stage of unblinding,  several PPD tests failed, and the posterior for one of the $n(z)$ modes of lens bin 2 was pushing strongly against its prior. After a thorough investigation, we were unable to determine the exact cause of this behavior of lens bin 2. To be conservative, we decided to complete the analysis without using any data involving lens redshift bin 2.  After this change, all blinding tests passed, and the results were unblinded. 

We emphasize that the aforementioned investigation involved reviewing most of the pipeline and three iterations of blind inference. The decision-making process was outlined in advance by a detailed analysis plan, which prevented any post hoc decisions. In itself, this serves as an important lesson for future Stage IV missions.

The complete unblinding criteria, and the investigation of lens bin 2, are described in Appendix~\ref{sec:unblinding_details}. We also show that in the post-unblinding results, neither the values nor the constraining power of the inferred cosmology are meaningfully altered by the omission of lens bin 2.

\begin{figure*}
    \centering
    \includegraphics[width=\columnwidth]{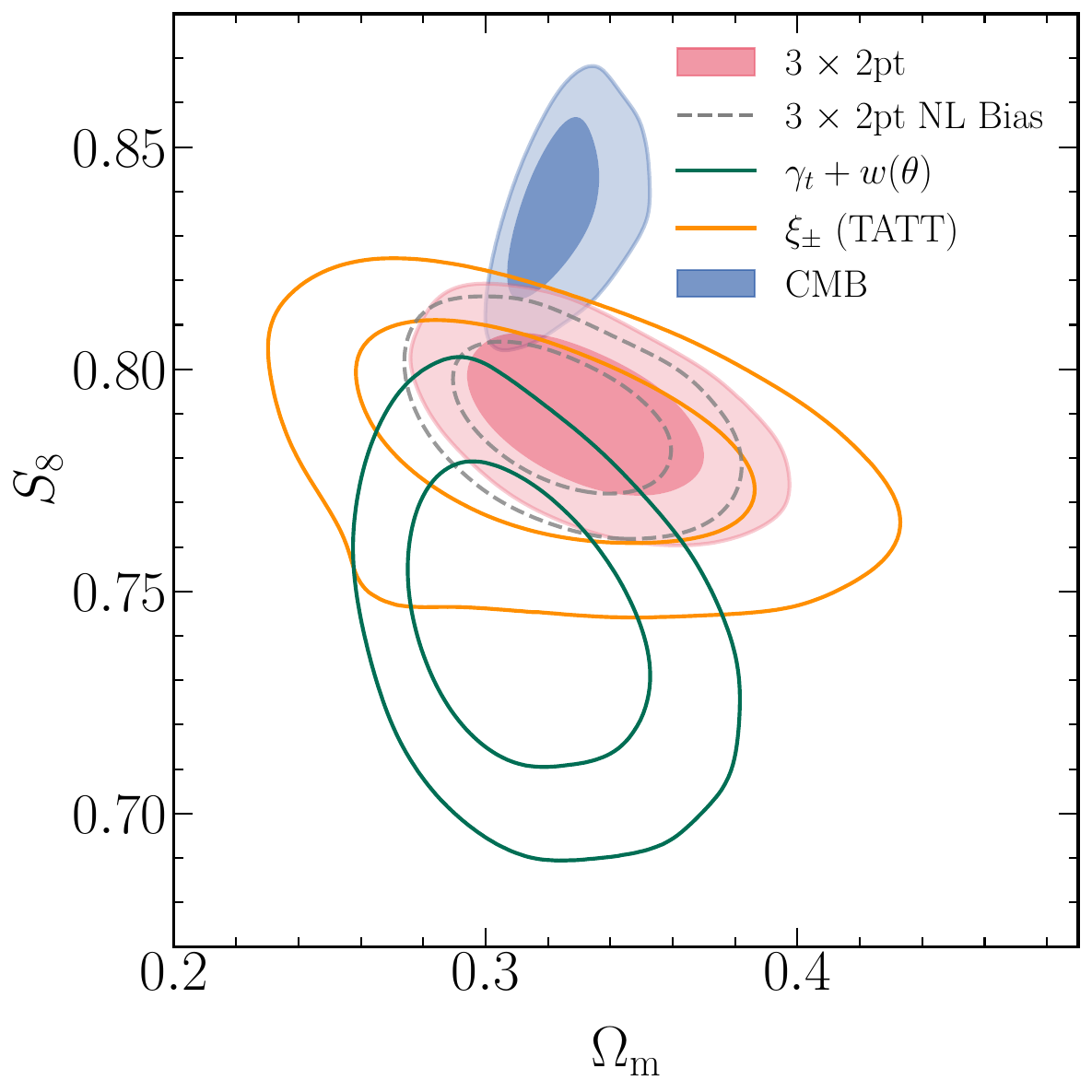}~
    \includegraphics[width=\columnwidth]{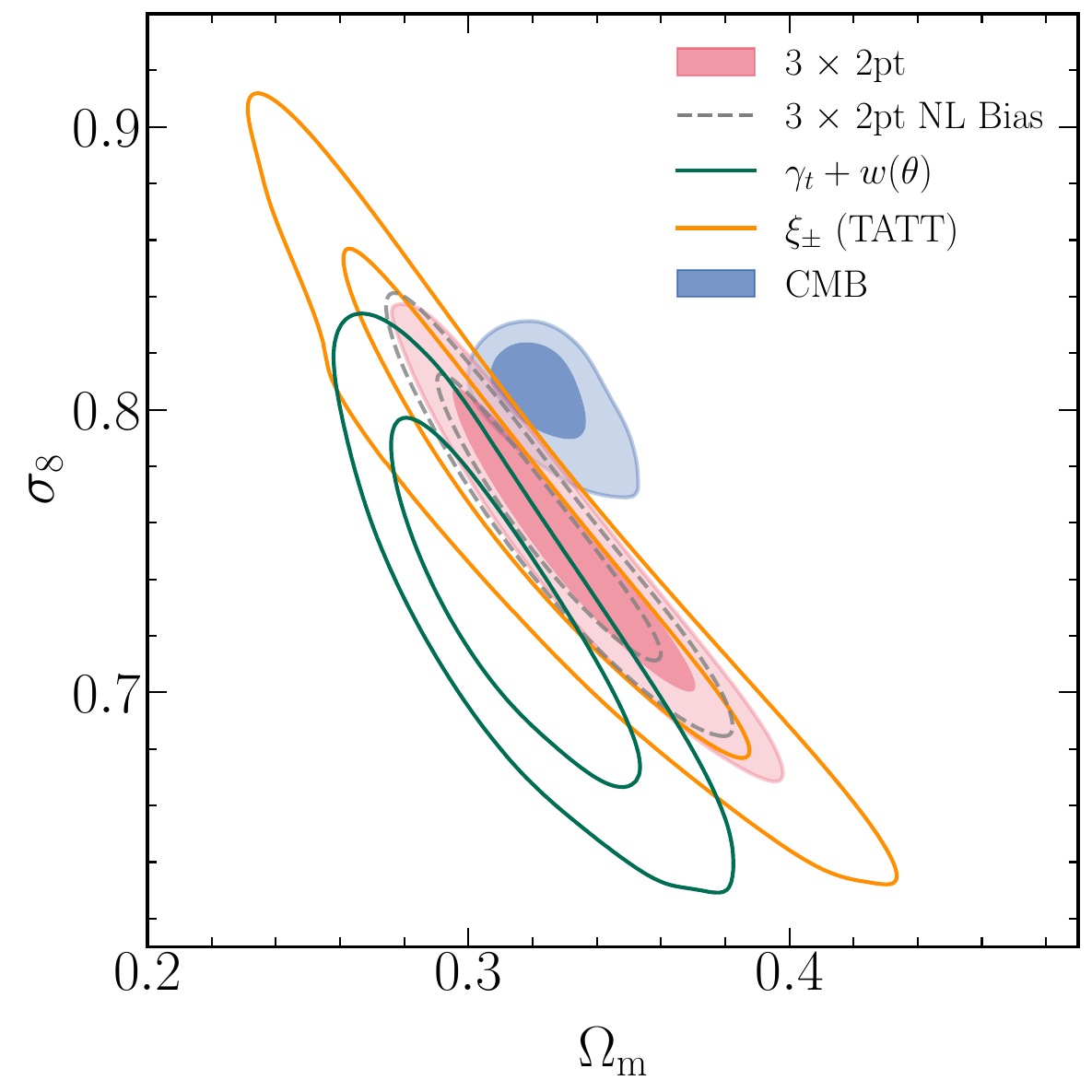}
    \caption{\label{fig:lcdm_lin}Marginalized constraints in $\Omega_{\rm m}$, $\sigma_8$, and $S_8 \equiv \sigma_8 \sqrt{\Omega_{\rm m}/0.3}$ in the $\Lambda$CDM analysis of cosmic shear ($\xi_{+\-}$, orange), galaxy clustering plus galaxy-galaxy lensing ($w(\theta)+\gamma_\mathrm{t}$, green), and their combination ($3 \times 2$pt, pink). Additionally, we show the result from the nonlinear galaxy bias ($3 \times 2$pt NL Bias, dashed grey line) \rcwr{and the CMB (solid blue)}. \journal{}{CMB here stands for the CMB combination described in Section~\ref{sec:ext}: \textit{Planck} 2018, ACT DR6, SPT-3G DR1}. Depicted contours represent the 68\% and 95\% credible levels.}
\end{figure*}

\section{Cosmology from DES Y6 3$\times$2pt}
\label{sec:results}

\begin{figure*}
    \centering
    \includegraphics[width=0.8\linewidth]{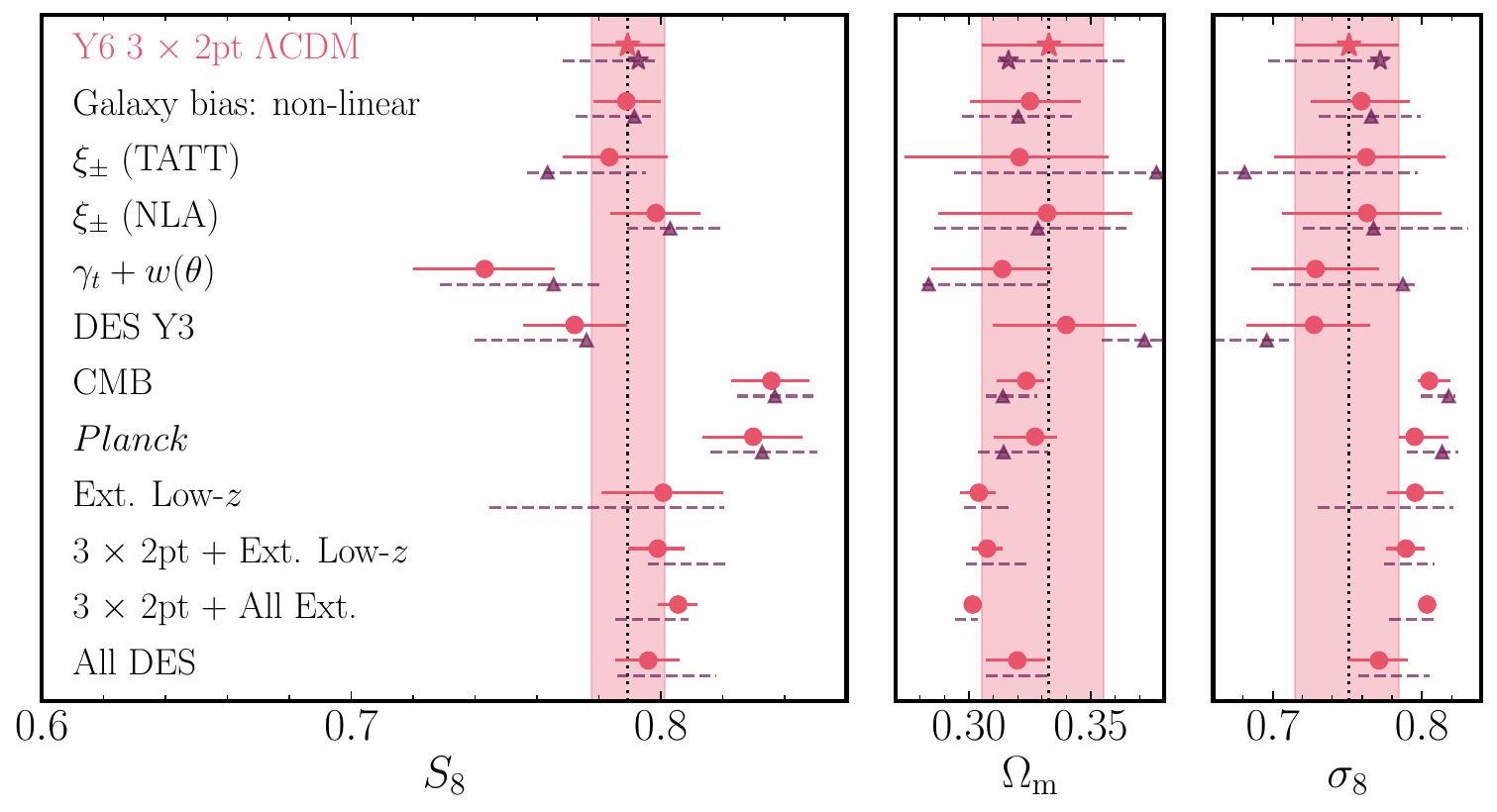}
    \caption{\label{fig:s8bar_results} Constraints on $S_8$, $\Omega_{\rm m}$ and $\sigma_8$ in $\Lambda$CDM -- all constraints are shown as marginalized 1$\sigma$ constraints. In the first block, we show the DES Y6 3$\times$2pt results and its individual components, $\xi_{\pm}$ and 2$\times$2pt. Second block shows the previous 3$\times$2pt results, from DES Y3. Third block shows the external constraints of CMB (both \textit{Planck} and \textit{Planck}+ACT+SPT). Final block shows the various combined results that include Ext. Low-$z$, DES 3$\times$2pt + Ext. Low-$z$, , 3$\times$2pt + All Ext and lastly all DES probes. The pink circles and lines represent the marginalized mean and 68\% uncertainty. {The purple triangle and dashed lines for each row indicate the MAP with \rcwr{the projected joint highest posterior density (PJ-HPD), which is less sensitive to volume effects. \cite{joachimi21}}. For the filled triangles, the MAP estimate is obtained by running a minimizer. For the last four rows, corresponding to data combinations obtained from normalizing flows whose MAP estimates cannot be reliably determined, we only show the PJ-HPD.}}
\end{figure*}

In this section, we present the cosmological constraints derived from the DES Y6 3$\times$2pt probes. We present our constraints for $\Lambda$CDM in Section~\ref{sec:lcdm} and compare them to those from the CMB in Section~\ref{sec:3x2vsCMB}. Constraints for $w$CDM are presented in Section~\ref{sec:wcdm}. For all constraints, we report the mean in each parameter, along with the 68\% credible level (CL) of posterior volume around the mean and the MAP in parentheses. Also, as described in Appendix~\ref{sec:projection}, we omit quoting quantities that suffer from severe prior volume effects as tested in simulations. Table~\ref{tab:lcdm_param_constraint} and Table~\ref{tab:postw} summarize all the numerical results from this paper, {while Figure \ref{fig:s8bar_results} provides a visual representation of $S_8$, $\Omega_{\rm m}$ and $\sigma_8$ constraints in the $\Lambda$CDM model}. 

\subsection{$\Lambda$CDM}
\label{sec:lcdm}

\subsubsection{Linear galaxy bias}
\label{sec:linear_gal}

The main 3$\times$2pt constraints assuming $\Lambda$CDM and linear galaxy bias are shown in Figure~\ref{fig:lcdm_lin}, overlaid by the individual $\xi_{\pm}$ and 2$\times$2pt constraints.  We show in Appendix~\ref{sec:full_param} the posterior in the full parameter space.  From Figure~\ref{fig:lcdm_lin} we observe that $\xi_{\pm}$ and 2$\times$2pt individually have similar constraining power in the $\Omega_{\rm m}$--$S_8$ \rcwr{two-dimensional} plane, with $\xi_{\pm}$ having better precision in $S_8$ and 2$\times$2pt in $\Omega_{\rm m}$. The calculated {figures of merit FoM$_{\sigma_8,\Omega_{\mathrm{m}}}$ in the $\sigma_8,\Omega_{\mathrm{m}}$ plane} (Table~\ref{tab:lcdm_param_constraint}) are similar for  $\xi_{\pm}$ and 2$\times$2pt, and their combination into 3$\times$2pt increases \rcwr{the} FoM$_{\sigma_8,\Omega_{\mathrm{m}}}$ by more than a factor of two due to degeneracy breaking in the high-dimensional space. 

$\xi_\pm$ and $2\times2$pt are consistent \rcwr{with each other, having} $\Delta_{\rm PPD} = 0.049$, even though the marginalized \rcwr{two-dimensional} contours appear somewhat offset from each other. For two highly covariant data sets, there is no a priori expectation or requirement that the projection of their joint posterior in a small subset of the parameter space should be centered on the mean of the two individual contours. 

For the main cosmological parameters from 3$\times$2pt, we find
\begin{align}
S_8 = 0.789^{+0.012}_{-0.012} \quad (0.793), \notag \\
\Omega_{\rm m} = 0.333^{+0.023}_{-0.028} \quad (0.316), \notag \\
\sigma_{8} =0.751^{+0.034}_{-0.036} \quad (0.772),
 \end{align}

We are also able to independently constrain $\Omega_b \times 10^2 = 4.58^{+0.75}_{-0.93}$ using DES $3\times 2$pt. We show in Appendix~\ref{sec:robustness} that these results are robust to many variations of the analysis choices. In Appendix~\ref{sec:breakdown}, we break down the relative contributions to the uncertainties from different sources by fixing the nuisance parameters and looking at the resulting change in the cosmological constraints. 

\begin{figure}
    \centering
    \includegraphics[width=\columnwidth]{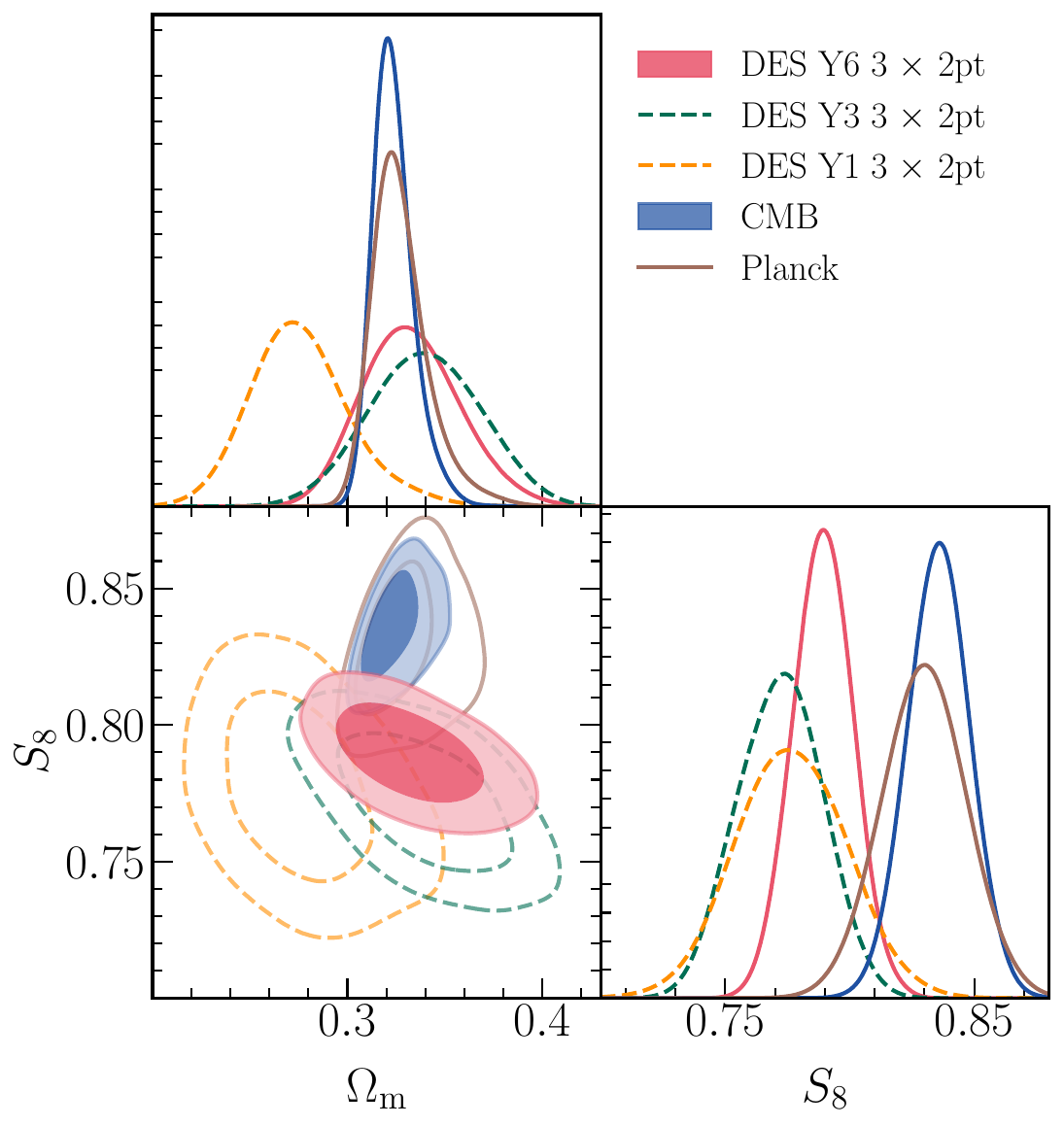}
    \caption{$S_8$ and $\Omega_{\rm m}$ constraints in $\Lambda$CDM from DES Y6 3$\times$2pt compared to the combined primary CMB constraints \journal{}{(\textit{Planck} 2018 + ACT DR6 + SPT-3G DR1)}. We also include the results from the DES Y3 3$\times$2pt analysis, in which we compared with the primary \textit{Planck} 2018 constraints, also overlaid in the figure. When comparing with \textit{Planck} only, the difference in the $S_8-\Omega_{\rm m}$ plane is reduced from 1.9$\sigma$ (Y3) to 1.4$\sigma$ (Y6). This difference increases to 2.2$\sigma$ in Y6 when comparing to the combined CMB likelihood. We also overlay the published DES Y1 3$\times$2pt constraints. }
    \label{fig:lcdm_lin_desy136_cmb}
\end{figure}

The DES Y6 3$\times$2pt constraints are compared to previous DES 3$\times$2pt results in Figure~\ref{fig:lcdm_lin_desy136_cmb}. The Y1 constraint is from \citep{y1-keypaper} and the Y3 constraint from \citep{y3-3x2ptkp}. We do not attempt to homogenize the analysis choices here between the three results, as there has been significant evolution of both the data samples and the modeling approach since Y1. Overall, we see a sustained gain in constraining power from Y1 to Y3 to Y6. Table~\ref{tab:lcdm_param_constraint} shows FoM$_{\sigma_8,\Omega_{\mathrm{m}}}$ gains as $1085\rightarrow2068\rightarrow3907.$ \rcwr{The three sets of contours are consistent, with Y1 having slightly lower $\Omega_{\rm m}$ central value.} 

\subsubsection{Nonlinear galaxy bias}
\label{sec:nonlinear_gal}

We also analyze our 3$\times$2pt data vector assuming a nonlinear galaxy bias model, going to smaller scales with the 2$\times$2pt data vector while allowing for one additional free parameter per lens bin as well as two additional bias parameters fixed by a relationship to the linear bias \citepalias{y6-methods}. 
The constraints are shown as dashed lines in Figure~\ref{fig:lcdm_lin}. We find 
\begin{align}
S_8 = 0.789^{+0.011}_{-0.011} \quad (0.788), \notag \\
\Omega_{\rm m} = 0.325^{+0.021}_{-0.025} \quad (0.330), \notag \\
\sigma_{8} =0.759^{+0.032}_{-0.034} \quad (0.752).  
\end{align}
 These constraints are consistent with the linear bias analysis, and only slightly tighter than those assuming linear galaxy bias, with a less than 10\% improvement on $S_8$ and an increase of about 14\% in FoM$_{\sigma_8,\Omega_{\mathrm{m}}}$ (see Table \ref{tab:lcdm_param_constraint}). These gains are very similar to the ones found in DES Y3 \cite{y3-3x2ptkp}. These results indicate that our fiducial $\Lambda$CDM 3$\times$2pt constraints are robust to galaxy bias modeling. Due to the similarity in constraints, and because the gain in using nonlinear galaxy modeling is not sufficient to justify the added computational complexity when exploring analysis variants, in the remainder of the paper we primarily present results using linear galaxy bias (unless otherwise specified). We also find slightly larger prior-volume effects for the nonlinear bias model ($\sim0.5\sigma$ for the main parameters of interest, compared to $\sim0.2\sigma$ for linear bias), as detailed in Appendix~\ref{sec:projection}.

\subsection{Comparison with the CMB in $\Lambda$CDM}
\label{sec:3x2vsCMB}

Next, we compare our results with the primary CMB constraints. As we will describe in more detail in Section~\ref{sec:ext}, the particular combination of CMB likelihoods used here is the combined likelihood of \textit{Planck} 2018 \citep{Planck2020_cosmo}, ACT-DR6 \citep{Louis2025}, and SPT-3G DR1 \citep{Camphuis2025}, for primary temperature and polarization data, but without CMB lensing. Figure~\ref{fig:lcdm_lin} gives a visual impression of the two sets of constraints. DES Y6 3$\times$2pt (pink) and CMB (blue) overlap at the 2$\sigma$ contour in this parameter subspace. The $S_8$ constraining power of the CMB is comparable to that of DES 3$\times$2pt, while the CMB constrains $\Omega_{\rm m}$ about a factor 2-3 more tightly than DES 3$\times$2pt. We calculate the statistical consistency between these two constraints in Table~\ref{tab:consistency}. We compare the two constraints in the one-dimensional ($S_8$) and two-dimensional ($S_8-\Omega_{\rm m}$) parameter spaces most constrained by 3$\times$2pt, and in the full $N$-dimensional parameter space based on Equation~\ref{Eq:EffectiveSigmas}. We find the difference to be 2.6$\sigma$, 2.2$\sigma$ and 1.8$\sigma$ respectively\footnote{We also examined $S_8^{\alpha}\equiv\sigma_8\, (\Omega_{\rm m}/0.3)^\alpha$ with optimized $\alpha=0.617$, our best-constrained direction, which yields $S_8^{\alpha=0.617} = 0.799^{+0.011}_{-0.010}$ (MAP: 0.797). This remains in $2.5\sigma$ tension with CMB, similar to the $2.6\sigma$ found for standard $S_8$.}. 
$3\times 2$pt is most sensitive to directions in parameter space like $S_8$, so we are more likely to detect at higher significance any separations between the two posteriors in that projected parameter direction. Similar results were found in previous work \citep{y3-3x2ptkp}.

If we consider only the full-parameter difference between 3$\times$2pt and CMB, our result (1.8$\sigma$) is slightly higher than what was found in DES Y3 (1.5$\sigma$, Table III of \citep{y3-3x2ptkp}),  but both are still consistent. As such, we are able to combine the DES Y6 3$\times$2pt results with the CMB data, yielding the constraints in Table~\ref{tab:lcdm_param_constraint}. In detail, both the 3$\times$2pt results from DES and the CMB result have improved since our DES Y3 analysis \citep{y3-3x2ptkp} and slightly shifted: from Y3 to Y6 DES 3$\times$2pt, the constraining power improved by a factor of 2 in the $S_8-\Omega_{\rm m}$ plane, while $S_8$ shifted higher; from \textit{Planck} 2018 to the CMB combination used in this work (SPT+ACT+\textit{Planck}), the constraining power also improved and $S_8$ \rcwr{also} shifted higher. Figure~\ref{fig:lcdm_lin_desy136_cmb} shows this change. The net result is a slight increase in separation between 3$\times$2pt and CMB, even though the DES Y6 3$\times$2pt became \textit{more} consistent with \textit{Planck} 2018 than Y3, being now at 1.0$\sigma$ while it was at 1.5$\sigma$ in DES Y3. 

\begin{table*}
\setlength{\extrarowheight}{7pt}
\caption{\label{tab:lcdm_param_constraint} Summary of marginalized parameter constraints in \lcdm.  The mean and 68\% CL are provided for each cosmological parameter, followed by the maximum \textit{a posteriori} (MAP) value in parentheses, except for neutrino mass, for which the 95\% upper bound is given. \rcwr{For data combinations where we use normalizing flows to combine posterior samples directly (and without access to the full likelihood), we do not report MAP values as they cannot be reliably determined. } Parameters that are not significantly constrained are indicated by a dash. \rcwr{Except for DES Y1 and Y3, a}ll data have been re-analyzed with model and prior choices matching the DES Y6 3$\times$2pt analysis. ``DES all probes'' stands for the data combination used in Section~\ref{sec:des_probes}: DES Y6 3$\times$2pt, DES BAO, DES SN, DES CL. ``CMB'' stands for the CMB combination described in Section~\ref{sec:ext}: \textit{Planck} 2018, ACT DR6, SPT-3G DR1 {(only primary probes, we do not include CMB lensing)}. ``Ext. Low-$z$'' represents the combination of the most constraining low-redshift data described in Section~\ref{sec:ext}: DESI DR2 BAO, DES BAO (excl DESI), DES SN, SPT CL. We also define ``All Ext.'': CMB, Ext. Low-$z$.}
\begin{tabular}{lcccccccccc}
\hline
\hline
\textbf{$\Lambda$CDM} & $S_8$ & $\Omega_{\mathrm{m}}$ & $\sigma_8$ & $\Omega_{\mathrm{b}}\times 10^2$  & $n_{\mathrm{s}}$ & $h$ & $\sum m_{\nu}$ (eV)  & FoM$_{\sigma_8,\Omega_{\mathrm{m}}}$ \\ [0.2cm]
\hline
 \vspace{0.1cm}
{\textbf{DES Data}} &&&& \vspace{0.15cm}  \\
\hline
\multirow{2}{*}{3$\times$2pt} & $ 0.789^{+0.012}_{-0.012} $  & $ 0.333^{+0.023}_{-0.028} $  & $ 0.751^{+0.034}_{-0.036} $  & \rcwr{$4.58^{+0.75}_{-0.93}$}  & --  & --  & --  & \multirow{2}{*}{3907}    \\ [-0.2cm] 
 & (0.793)  & (0.316)  & (0.772)  & \rcwr{(4.16)}  & --  & --  & --  &   \\ 
\hline 3$\times$2pt  &  $ 0.789^{+0.011}_{-0.011} $  & $ 0.325^{+0.021}_{-0.025} $  & $ 0.759^{+0.032}_{-0.034} $  & \rcwr{$4.66^{+0.79}_{-0.97}$}  & --  & --  & --  & \multirow{2}{*}{4455}  \\ [-0.2cm] 
 (NL Bias)& (0.788)  & (0.330)  & (0.752)  & (4.65)  & --  & --  & --  &  \\
\hline \multirow{2}{*}{$\gamma_\mathrm{t}$+$w(\theta)$} & $ 0.743^{+0.023}_{-0.023} $  & $ 0.314^{+0.020}_{-0.029} $  & $ 0.729^{+0.043}_{-0.043} $  & --  & --  & --  & --  & \multirow{2}{*}{1833} \\ [-0.2cm] 
 & (0.765)  & (0.283)  & (0.787)  & --  & --  & --  & --  &   \\
\hline \multirow{2}{*}{$\xi_{\pm}$} & $ 0.783^{+0.019}_{-0.015} $  & $ 0.321^{+0.036}_{-0.047} $  & $ 0.763^{+0.053}_{-0.062} $  & --  & --  & --  & --  & \multirow{2}{*}{1450}   \\ [-0.2cm] 
 & (0.763)  & (0.377)  & (0.681)  & --  & --  & --  & --  &   \\

\hline \multirow{2}{*}{DES Y3 3$\times$2pt} & $ 0.776^{+0.017}_{-0.017} $  & $ 0.339^{+0.032}_{-0.031} $  & $ 0.733^{+0.039}_{-0.049} $  & --  & --  & --  & --  & \multirow{2}{*}{2068}  \\ [-0.2cm]
 & (0.776)  & (0.372)  & (0.696)  & --  & --  & --  & --  &   \\
 
 \hline\multirow{2}{*}{DES Y1 3$\times$2pt} & $ 0.747^{+0.027}_{-0.025} $ & $ 0.303^{+0.034}_{-0.041} $ & $ 0.747^{+0.052}_{-0.068} $ &--  & --  & -- & -- & \multirow{2}{*}{1085}   & \multirow{2}{*}{}  \\ [-0.2cm] 
 & (0.770) & (0.253) & (0.838) &--  & --  & --   \\

 \hline\multirow{2}{*}{ DES all probes} & $ 0.796^{+0.010}_{-0.011} $  & $ 0.320^{+0.011}_{-0.013} $ & $ 0.771^{+0.019}_{-0.020} $ & \rcwr{$4.05^{+0.35}_{-0.90}$} &--  & --  &--  & \multirow{2}{*}{8373}  \\ [-0.2cm] 
   & -- & -- & -- & -- & -- & -- & --   \\

 \hline \vspace{0.1cm}
{\textbf{External Data}} &&&& \vspace{0.15cm} \\
 \hline\multirow{2}{*}{ Ext. Low-$z$} & $ 0.801^{+0.019}_{-0.020} $ & $ 0.304^{+0.007}_{-0.008} $ & $ 0.796^{+0.019}_{-0.019} $  & \rcwr{$4.84^{+0.96}_{-0.42}$} & --  & --  & -- & \multirow{2}{*}{7426}  \\ [-0.2cm] 
 & -- & -- & -- & -- & -- & -- & --  \\
 
 \hline\multirow{2}{*}{ CMB}  & $ 0.836^{+0.012}_{-0.013} $  & $ 0.324^{+0.007}_{-0.012} $ & $ 0.805^{+0.014}_{-0.007} $ & $ \rcwr{5.04^{+0.07}_{-0.12}} $ & $ 0.970^{+0.004}_{-0.004} $ & $ 0.667^{+0.009}_{-0.006} $ & $ <0.24 $ & \multirow{2}{*}{8994} \\ [-0.2cm] 
  & (0.837) & (0.314) & (0.818) & \rcwr{(4.93)} & (0.971) & (0.675) & (95\% CL)   \\
  
 \hline \vspace{0.1cm}
{\textbf{Combined Data}} &&&& \vspace{0.15cm} \\
 \hline\multirow{2}{*}{ 3$\times$2pt + Ext. Low-$z$} & $ 0.799^{+0.009}_{-0.010} $ & $ 0.307^{+0.006}_{-0.006} $ & $ 0.789^{+0.012}_{-0.013} $ & \rcwr{$4.99^{+0.77}_{-0.27}$} & -- & -- & $<0.47$  & \multirow{2}{*}{17045}  \\ [-0.2cm] 
  & -- & -- & -- & -- & -- & -- & (95\% CL)  \\
  
 \hline\multirow{2}{*}{3$\times$2pt + CMB} & $ 0.811^{+0.008}_{-0.008} $ & $ 0.313^{+0.006}_{-0.009} $ & $ 0.795^{+0.013}_{-0.008} $ & $ \rcwr{4.95^{+0.06}_{-0.10}} $ & $ 0.974^{+0.003}_{-0.003} $ & $ 0.675^{+0.007}_{-0.004} $ & <0.22 & \multirow{2}{*}{16598}  \\ [-0.2cm] 
  & (0.813) & (0.305) & (0.807) & \rcwr{(4.85)} & (0.974) & (0.681) & (95\% CL)   \\

\hline\multirow{2}{*}{ DES all probes + CMB}  & $ 0.815^{+0.008}_{-0.007} $ & $ 0.314^{+0.006}_{-0.007} $ & $ 0.797^{+0.012}_{-0.007} $ & \rcwr{$ 4.95^{+0.05}_{-0.08}$}  & $ 0.973^{+0.003}_{-0.003} $ & $ 0.674^{+0.006}_{-0.004} $ & $<0.21$ & \multirow{2}{*}{20590}  \\ [-0.2cm] 
  & -- & -- & -- & -- & -- & -- & (95\% CL)  \\ 

\hline\multirow{2}{*}{  3$\times$2pt + All Ext.} & $ 0.806^{+0.006}_{-0.007} $ & $ 0.302^{+0.003}_{-0.003} $ & $ 0.804^{+0.007}_{-0.006} $  & \rcwr{$ 4.83^{+0.03}_{-0.04} $} & $ 0.976^{+0.003}_{-0.003} $ & $ 0.683^{+0.003}_{-0.002} $ & $<0.14$ & \multirow{2}{*}{51955}  \\ [-0.2cm] 
  & -- & -- & -- & -- & -- & -- & (95\% CL)  \\ 

\hline \hline
\end{tabular}
\end{table*}

\begin{table*}
\setlength{\extrarowheight}{7pt}
%\vspace{2in}
\caption{\label{tab:postw} Summary of marginalized parameter constraints in \wcdm. The combination used for ``Ext. Low-$z$'', ``CMB'' and ``All Ext.'' is the same as in Table~\ref{tab:lcdm_param_constraint}, {while ``DES all probes'' does not include DES CL (see text for details)}.}
\begin{tabular}{lccccccccc}
\hline
\hline
\textbf{$w$CDM} & $S_8$ & $\Omega_{\mathrm{m}}$ & $\sigma_8$ & $\Omega_{\mathrm{b}}\times 10^{2}$ & $n_{\mathrm{s}}$ & $h$ & $w$ & $\sum m_{\nu}$ (eV) & FoM$_{\sigma_8,\Omega_{\mathrm{m}}}$  \\ [-0.1cm]
 &  &  &  &  &  &  & &   &  FoM$_{w,\Omega_{\mathrm{m}}}$ \\ [0.05cm]
\hline
 \vspace{0.1cm}
{\textbf{DES Data}} &&&& \vspace{0.15cm}  \\
\hline
\multirow{2}{*}{3$\times$2pt} & $0.781^{+0.021}_{-0.020} $  & $ 0.325^{+0.032}_{-0.035} $  & $ 0.753^{+0.035}_{-0.039} $  & \rcwr{$4.40^{+0.57}_{-1.19}$}  & --  & --  & $ -1.12^{+0.26}_{-0.20} $  & --  & 1584    \\ 
[-0.2cm] 
 & (0.778)  & (0.329)  & (0.743)  & (4.08)  & --  & --  & ($-1.12$)  & -- & 168  \\
 \hline 3$\times$2pt  & $0.778^{+0.020}_{-0.018}$  & $0.317^{+0.031}_{-0.032}$  & $0.760^{+0.034}_{-0.040}$  & \rcwr{$4.34^{+0.51}_{-1.18}$}  & --  & --  & $-1.13^{+0.24}_{-0.19}$  & --  & 1790    \\ [-0.2cm] 
 (NL Bias)& (0.798)  & (0.324)  & (0.768)  & (4.51)  & --  & --  & ($-0.89$)  & -- & 194 \\
 \hline DES Y3 3$\times$2pt  & $ 0.775^{+0.026}_{-0.024} $  & $ 0.352^{+0.035}_{-0.041} $  & $ 0.719^{+0.037}_{-0.044} $  & --  & --  & -- & $ -0.98^{+0.32}_{-0.20} $  & --  & 1124  \\ [-0.2cm]  
 & (0.779)  & (0.340)  & (0.731)  & --  & --  & --  & ($-1.04$) & -- & 116  \\ 
 \hline DES Y1 3$\times$2pt  & $0.782^{+0.036}_{-0.024}$  & $0.284^{+0.033}_{-0.030}$  & --  & --  & --  & -- & $-0.82^{+0.21}_{-0.20}$  & --  & --  \\ [-0.2cm] 
 & --  & --  & --  & --  & --  & --  & --  & -- & -- \\ 
 
 \hline DES all probes (-CL)  & $ 0.797^{+0.013}_{-0.012} $  & $ 0.317^{+0.013}_{-0.014} $  & $ 0.776^{+0.025}_{-0.025} $  & \rcwr{$4.39^{+0.63}_{-0.85}$} & --  & -- & $ -0.956^{+0.044}_{-0.041} $  & --  & 6584 \\ [-0.2cm] 

 & (0.802)  & (0.307)  & (0.792)  & (4.61)  & --  & --  & ($-0.936$) & -- & 1955   \\

 \hline \vspace{0.1cm}
 
{\textbf{External Data}} &&&& \vspace{0.15cm} \\
\hline
Ext. Low-$z$ & $ 0.794^{+0.020}_{-0.019} $  & $ 0.296^{+0.008}_{-0.008} $  & $ 0.799^{+0.018}_{-0.018} $  & \rcwr{$5.42^{+0.86}_{-0.55}$}  & --  & --  & $ -0.913^{+0.037}_{-0.034} $   & --  & 6709    \\ [-0.2cm] 
 & -- & -- & -- &  -- & --  &  -- & -- & -- &  3655    \\
\hline CMB   & $ 0.812^{+0.015}_{-0.022} $  & $ 0.268^{+0.011}_{-0.044} $  & $ 0.863^{+0.051}_{-0.023} $  & \rcwr{$4.17^{+0.15}_{-0.68}$}  & $ 0.970^{+0.003}_{-0.003} $  & --  & $ -1.238^{+0.094}_{-0.192} $   & $<0.28$  & 2181     \\ [-0.2cm] 

& (0.796)  & (0.225)  & (0.919)  & --  & (0.972)  & --  & ($-$1.376)  &(95\% CL) & 524 \\
\hline \vspace{0.1cm}
{\textbf{Combined Data}} &&&& \vspace{0.15cm} \\
 \hline 3$\times$2pt + Ext. Low-$z$  & $ 0.803^{+0.009}_{-0.010} $  & $ 0.308^{+0.006}_{-0.007} $  & $ 0.793^{+0.013}_{-0.013} $  & \rcwr{$5.30^{+0.72}_{-0.30}$}  & $ 0.974^{+0.024}_{-0.009} $  & --  & $ -0.962^{+0.031}_{-0.029} $ & --  & 16046 \\ [-0.2cm] 
  & --  &  -- &  -- &  -- & -- &  -- & -- & -- & 5139    \\
\hline3$\times$2pt & $ 0.801^{+0.011}_{-0.011} $  & $ 0.280^{+0.017}_{-0.025} $  & $ 0.831^{+0.031}_{-0.023} $  & \rcwr{$4.40^{+0.28}_{-0.39}$}  & $ 0.972^{+0.003}_{-0.003} $  & $ 0.716^{+0.029}_{-0.026} $  & $ -1.159^{+0.100}_{-0.101} $  & $<0.31$ & 5150 \\ [-0.2cm] 
+ CMB & (0.806)  & (0.274)  & (0.844)  & (4.34)  & (0.973) & (0.720)  & ( $-1.139$)  & (95\% CL) & 990  \\

\hline DES all probes {(-CL)} & $ 0.812^{+0.009}_{-0.007} $ & $ 0.317^{+0.007}_{-0.007} $ & $ 0.790^{+0.012}_{-0.010} $ & $ \rcwr{5.03^{+0.10}_{-0.11}} $ & $ 0.974^{+0.003}_{-0.003} $ & $ 0.669^{+0.007}_{-0.007} $  & $ -0.978^{+0.025}_{-0.027} $  & $<0.23$  & 17422  \\ [-0.2cm] 
+ CMB  & -- & -- & -- & -- & -- & -- & -- & (95\% CL)  & 5449 \\
\hline 3$\times$2pt + All Ext.  & $ 0.807^{+0.007}_{-0.007} $  & $ 0.305^{+0.004}_{-0.005} $  & $ 0.801^{+0.008}_{-0.008} $  & $ \rcwr{4.91^{+0.07}_{-0.08}} $  & $ 0.977^{+0.003}_{-0.003} $  & $ 0.679^{+0.005}_{-0.005} $  & $ -0.981^{+0.021}_{-0.022} $  & $ <0.13 $  & 32033  \\ [-0.2cm] 
& -- & -- &  -- &  -- & -- &  -- & --  &  (95\% CL)   & 12970\\

\hline
\hline
\end{tabular}
\end{table*}

\begin{table}
    \centering
    \caption{\label{tab:consistency} Consistency of various independent data sets, in units of $\sigma$, as introduced in Sec.~\ref{sec:tension_metrics}, with a derived probability-to-exceed shown in parentheses. In the table below, ``DES all probes'' stands for the data combination used in Section~\ref{sec:des_probes}: DES Y6 3$\times$2pt, DES BAO, DES SN, DES CL; ``CMB'' stands for the CMB combination described in Section~\ref{sec:ext}: \textit{Planck} 2018, ACT DR6, SPT-3G DR1,  primary without lensing; ``Ext. Low-$z$'' represents the combination of the most constraining low-redshift data described in Section~\ref{sec:ext}: DESI DR2 BAO, DES BAO (excl. DESI), DES SN, SPT CL.}
    \begin{tabular}{l l  c }
    \hline
  
   Dataset 1 & Dataset 2 & Parameter  \\
             &           & difference \\
    \hline
      \hline
    $\Lambda$\textbf{CDM} & & \\
    
    \textit{DES internal consistency} \\
    DES 3$\times$2pt & DES SN + BAO & 0.35 (0.725)  \\
    DES 3$\times$2pt & DES CL & 2.0 (0.042) \\
    DES 3$\times$2pt + CL & DES SN + BAO  & 0.31 (0.753) \\
    DES 3$\times$2pt + SN + BAO & DES CL  & 0.51 (0.61)  \\

    \hline
    \textit{DES consistency with external data} \\
    DES 3$\times$2pt & Planck   & 1.0 (0.319)  \\
    DES 3$\times$2pt & CMB   & 1.8 (0.08)\\
    DES 3$\times$2pt + CL & CMB   & 2.0 (0.042) \\
    DES 3$\times$2pt + SN + BAO & CMB   & 2.5 (0.012) \\
    DES all probes & CMB   & 2.8 (0.006) \\
    DES 3$\times$2pt & Ext. Low-$z$  & 0.2 (0.843)  \\
    DES 3$\times$2pt + Ext. Low-$z$ & CMB   & 2.3 (0.021)\\
    \hline
    \hline
    $w$\textbf{CDM} & & \\
    
    DES 3$\times$2pt & CMB   & 1.3 (0.186) \\
    DES 3$\times$2pt & DES SN + BAO & 0.34 (0.73)\\
    DES 3$\times$2pt + SN + BAO  & CMB   & 2.5 (0.012) \\
    DES 3$\times$2pt & Ext. Low-$z$  & 1.1 (0.284) \\
    DES 3$\times$2pt + Ext. Low-$z$ & CMB   & 2.4 (0.015) \\
    \hline

\end{tabular}

\end{table}

\subsection{wCDM}
\label{sec:wcdm}

We test a model with one additional free parameter, the dark energy equation of state parameter $w$, assumed to be constant in time. In this \wcdm~model, the dark energy density evolves with redshift as $\rho_{DE} \sim (1+z)^{3(1+w)}$. We find
\begin{align}
w&=-1.12^{+0.26}_{-0.20} \quad (-1.12) \notag \\
\Omega_{\rm m} &= 0.325^{+0.032}_{-0.035}  \quad 
(0.329)\\
\rcwr{S_8} &= \rcwr{0.781^{+0.021}_{-0.020} \quad (0.778)}.\notag
\end{align}
Our constraints show no evidence for departure from a $\Lambda$CDM model, e.g. are consistent with $w=-1$. When using a nonlinear galaxy bias model, we find $w=-1.13^{+0.24}_{-0.19} \ (-0.89)$, $\Omega_{\rm m} = 0.317^{+0.031}_{-0.032} \ (0.324)$ \rcwr{and $S_8 = 0.778^{+0.020}_{-0.018} \ (0.798)$},
very similar to the linear bias results, and still within $1\sigma$ of $\Lambda$CDM.  
The $\Lambda$CDM constraint on $S_8$ of $0.789^{+0.012}_{-0.012} \ (0.793)$ is  significantly weakened to  $0.781^{+0.021}_{-0.020} \ (0.778)$ in $w$CDM.

We further assess whether the data prefer a $w$CDM universe over a $\Lambda$CDM one using the criterion described in Section~\ref{sec:model_comparison}. We obtain $\Delta \chi^2=-0.82$ (0.9$\sigma$) as defined in Equation~\ref{eq:deltachi2}, corresponding to no  significant preference for $w$CDM. %\footnote{\journal{}{After submission we found that the $\beta_{ij}$ factors had been initialized  with different input nuisance parameters for each worker, spanning the priors. We checked that this produces a negligible impact on the posteriors (shifts of less than 0.02$\sigma$ in the mean and less than 1\% in the width of the marginalized constraints on cosmological and nuisance parameters), and decided not to update the chains. The maximum a posteriori (MAP) estimates are more sensitive, since each minimizer run used a slightly different likelihood normalization and point-mass template, shifting the best-fit $\chi^2$ by $\mathcal{O}(1)$. Re-evaluating and re-optimizing the MAP searches for 3$\times$2pt with a common fixed $\beta_{ij}$ changes the $w$CDM versus $\Lambda$CDM difference from $\Delta\chi^2=-0.82$ ($0.9\sigma$) to $\Delta\chi^2\approx-0.1$ ($0.3\sigma$),  resulting in even less preference for $w$CDM. We do not update the quoted MAP values or $\Delta\chi^2$, since these estimates come from a reduced re-optimization of the existing searches rather than a full rerun, and the conclusion is unchanged}}
.  

\begin{figure*}[t]
    \centering
    \includegraphics[width=\columnwidth]{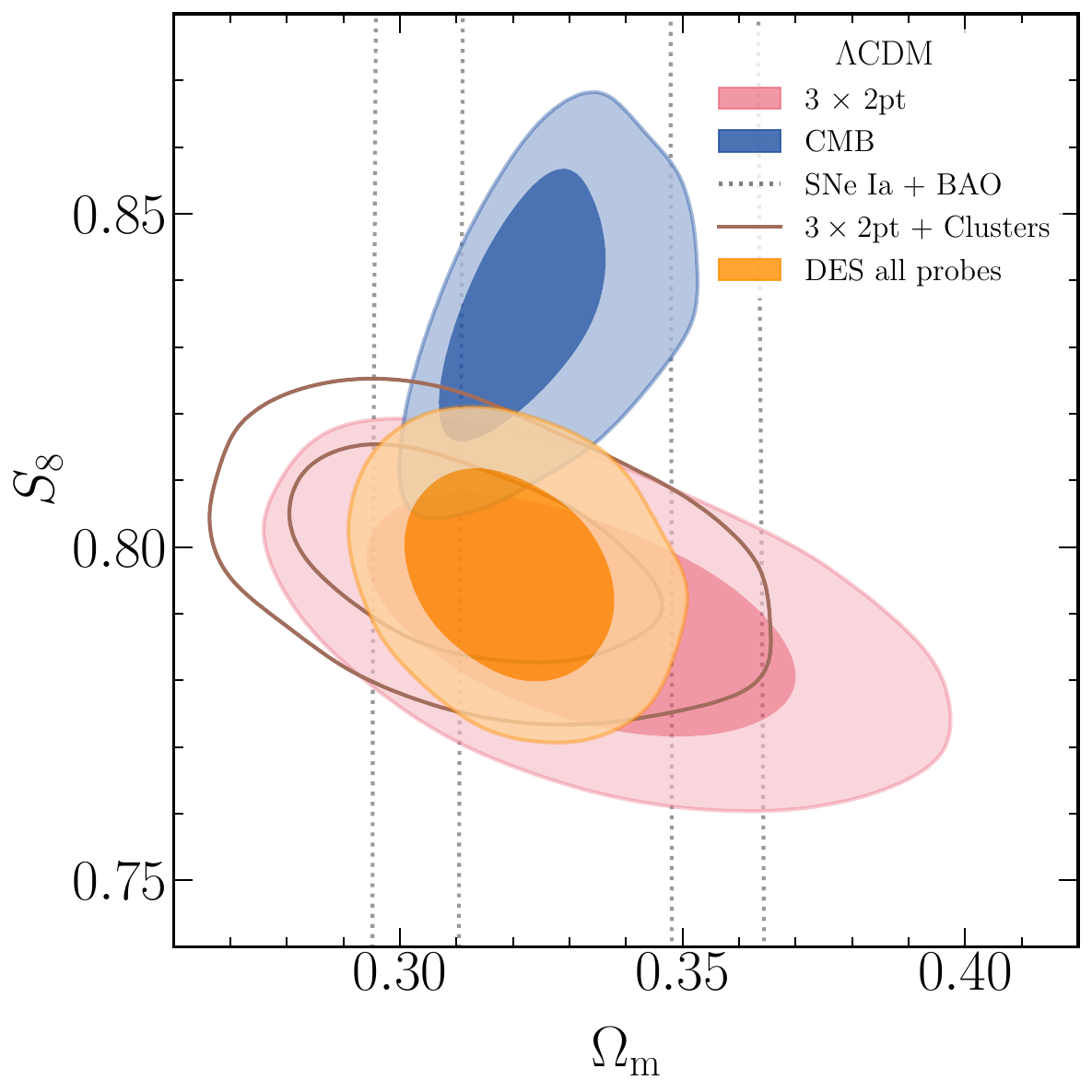}
    \includegraphics[width=\columnwidth]{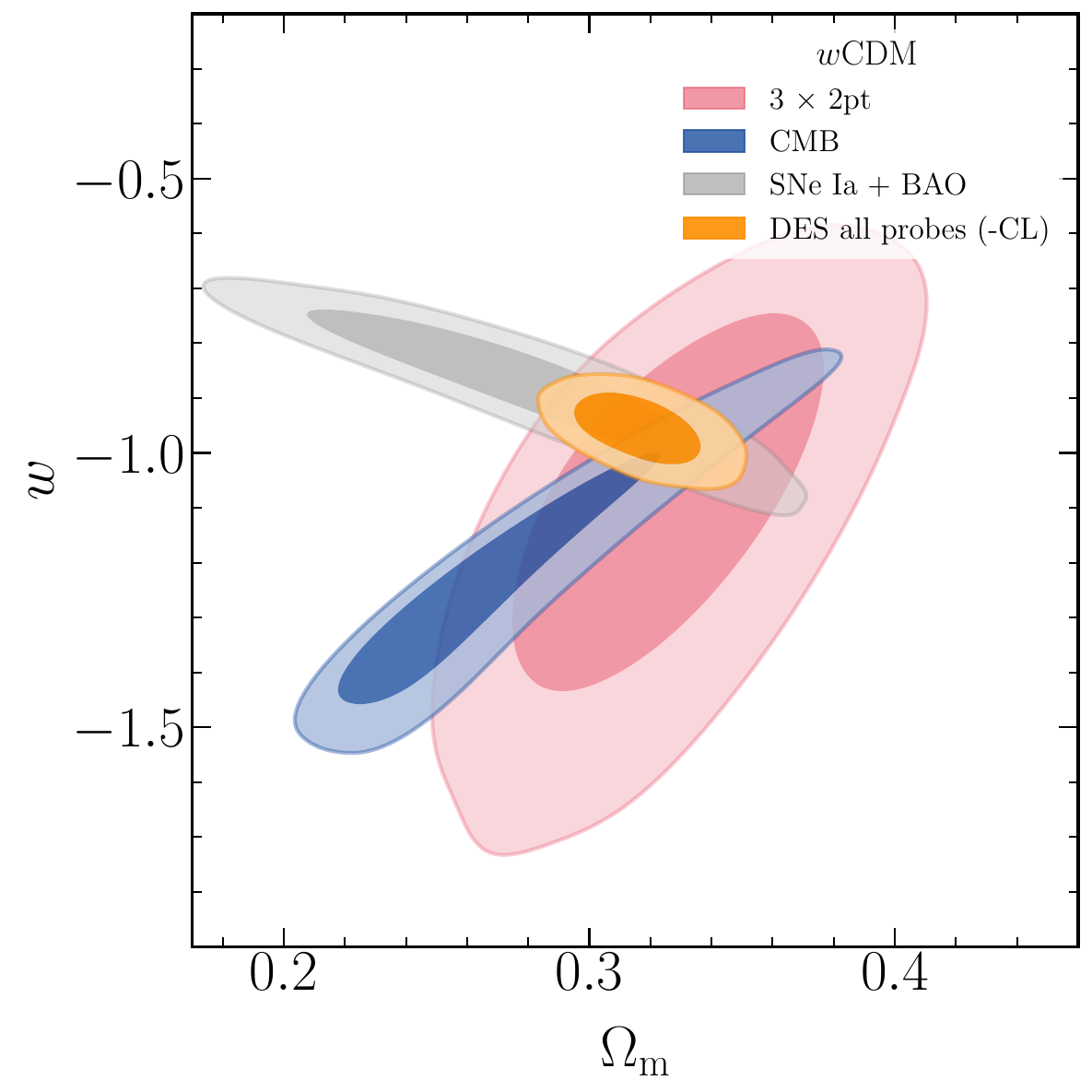}
    \caption{Combination of all the DES cosmological probes in the \lcdm (left panel) and $w$CDM (right panel) models. We show marginalized constraints on parameters from large-scale structure and weak lensing (3$\times$2pt; pink), type Ia supernovae combined with BAO (black dotted/gray), galaxy cluster clustering and number counts combined with 3$\times$2pt (brown). The combination of DES Y6 3$\times$2pt, Clusters, SNe Ia, and BAO is shown in orange (All DES). Blue contours correspond to CMB constraints from \textit{Planck}, ACT and SPT combined. The combination of DES probes is consistent with CMB at the 2.8$\sigma$ level in \lcdm and 2.5$\sigma$ in \wcdm\  in full parameter space.
    }
    \label{fig:all_des}
\end{figure*}

We compare the consistency of our results with those from the CMB in the $w$CDM model in the right panel of Figure~\ref{fig:all_des}, which shows posteriors of $3\times2$pt and CMB in the $w-\Omega_{\rm m}$ plane, which lay on top of each other. However the agreement in $S_8$--$\Omega_{\rm m}$ is similar as in $\Lambda$CDM. In the full parameter space, both datasets are consistent at the $1.3\sigma$ level. Projected in $S_8$ the parameter difference is $1.0\sigma$, while in the $w-S_8$ plane, we observe a higher difference with respect to the CMB, reaching $2.1\sigma$.

\section{Combination with other DES data}\label{sec:des_probes}

In this section we present cosmological constraints from combining the latest results from the four main cosmological probes in DES: the recently recalibrated Type Ia supernovae (DES SN), baryon acoustic oscillation (DES BAO), galaxy clusters (DES CL), and the 3$\times$2pt results from this work (DES Y6 3$\times$2pt). Since its conception, \rcwr{the Dark Energy Survey} was designed to combine these four complementary cosmological probes \rcwr{through its coherent and self-calibrating low-redshift cosmology program}. We describe below the likelihoods we use for these other three probes.

\begin{itemize}

\item \textit{DES BAO:} DES measured the Baryonic Acoustic Oscillation (BAO) feature from the clustering of galaxies using the full survey data in \cite{y6-BAOkp}, using a sample optimized for BAO in \cite{y6-BAO-sample}. This sample consists of approximately 16 million red and bright galaxies over $4,273$ deg$^2$ and a redshift range of $0.6<z<1.2$, with an effective redshift of $z_{\rm eff}=0.85$. The BAO analysis combines three different estimators into a consensus measurement of the ratio of the angular distance ($D_M$) to the sound horizon scale ($r_{\rm d}$) of $D_M(z_{\rm eff})/r_{\rm d}=19.51\pm0.41$. This 2.1\% measurement is the tightest BAO constraint from a photometric galaxy survey and competes in precision with Stage-III spectroscopic surveys. Interestingly, this measurement is in $\sim2\sigma$ tension with the \textit{Planck} 2018 best-fit -- \lcdm prediction of $D_M(z=0.85)/r_{\rm d}=20.39$. 

\rcwr{We note that the DES BAO measurement is not fully independent of DESI BAO, as \rcwr{the two} surveys share $\sim 1,000$ deg$^2$. In a followup work \citep{Mena26}, we reanalyze the DES BAO in the area that does not overlap with DESI, using the angular correlation function, finding $D_M(z=0.85)/r_{\rm d}=19.74\pm0.60$. This new measurement, although looser, is fully independent of DESI. We will hence use this when combining DES BAO with DESI BAO in the next section, and denote this result as ``DES BAO (excl. DESI)''.}

\item \textit{DES SN:} The DES supernovae type Ia (SNe Ia) dataset used here is the ``DES-Dovekie'' SN Ia sample \citep{Popovic2025a}, which is a reanalysis of the original DES-SN5YR \cite{DESY5SN2024, Sanchez2024, Vincenzi2024} SN Ia light curves with updated photometric cross-calibration, among other improvements \citep{Popovic2025a}. This new sample contains 1623 likely SNe Ia from the full DES supernova survey, combined with approximately 200 SNe from historical low-redshift samples. The pipeline and sample building papers are described in \citep{Popovic2025a}. \rcwr{These data were used to constrain cosmology assuming a time-evolving dark energy equation of state in \citep{Popovic2025b}, in combination with CMB and
DESI BAO, finding a reduced tension with $\Lambda$CDM ($3.2\sigma$) vs. the same data combination using the original DES-SN5YR data ($4.2\sigma$).}

\item \textit{DES CL:} The latest galaxy cluster constraints are based on \textsc{redMaPPer} clusters \citep{Rykoff2014} from the Y3 data. This red-sequence-based optical cluster finder extracted $\sim16,000$ galaxy clusters in the DES Y3 footprint above richness of 20 and in the redshift range $0.2<z<0.85$. In DES Y1 \cite{Costanzi2021}, it was found that \textsc{redMaPPer} clusters suffer from line-of-sight projection effects, where interloper galaxies contaminate the richness estimation by 20-40\% and bias the lensing results. This effect was later studied in \citep{Wu2022, Zhang2023, Zhou2024, Lee2025}. \rcwr{In DES Y3, we took the alternative approach of discarding the small-scale information that suffers from projection effects. Three additional two-point functions that include cluster density $\delta_{c}$ were added to the standard 3$\times$2pt analysis: $\langle \delta_{c} \delta_{c}\rangle$, $\langle \delta_{c} \delta_{g}\rangle$, $\langle \delta_{c} \gamma_\mathrm{t}\rangle$.} This methodology was first developed in DES Y1 \citep{To2021,y6-cardinal} and applied to DES Y3 data in \citep{DES2025_clusters}. It was found that adding galaxy clusters mainly helps in constraining $\Omega_{\rm m}$ {(by $\sim 25\%$)} and that clusters prefer a slightly lower value of $\Omega_{\rm m}$. 

One important note is that in \citep{DES2025_clusters}, the galaxy samples used in the 3$\times$2pt block were the same as those used for cross-correlation with clusters. This means that several shared nuisance parameters, in particular the mass-observable relation, can be self-calibrated. Here, however, the Y3 source and lens samples are different from the Y6 ones. This means that the combination does not invoke self-calibration. In addition, to be conservative, we only use the part of the data vector in \cite{DES2025_clusters} that does not overlap with the Y6 3$\times$2pt data vector.\footnote{Specifically, the cluster (CL) likelihood we use to combine with Y6 $3\times2$pt includes Y3 $\langle \delta_{c} \delta_{g}\rangle$, $\langle \delta_{c} \gamma_\mathrm{t}\rangle$ and number counts (N), and ignores Y3 clustering $\langle \delta_{g} \delta_{g}\rangle$.} 
\rcwr{Both of these factors result in less constraining power gain when combining with clusters compared to what is shown in \citep{DES2025_clusters}, but we nevertheless found that there is an impact from including the cluster information} ($\sim 16\%$ tighter $\Omega_{\rm m}$ constrain). In addition, we assume the correlation of the Y6 $3\times2$pt and the Y3 cluster likelihood is small (the later being dominated by shape/shot noise) and we use normalising flows to combine them (see Section~\ref{sec:sampling}).

In an upcoming paper \citep{y6-clusters}, we will present a complete self-calibrating analysis \rcwr{that updates the cross-correlations between the Y3 cluster sample and the Y6 lens and source samples.}

\end{itemize}

In the left panel of Figure~\ref{fig:all_des}, we show the individual and combined constraints for $\Lambda$CDM from the four DES probes: DES Y6 3$\times$2pt, DES BAO, DES SN and DES CL. We find
\begin{align}
S_8 = 0.794^{+0.009}_{-0.012}, \notag \\
\Omega_{\rm m} = 0.322^{+0.012}_{-0.011}, \notag \\
\sigma_{8} = 0.766^{+0.017}_{-0.022}.
\end{align}

Combining the four probes results in a 1.3\% constraint on $S_8$, which is comparable in precision to the constraint from the primary CMB. The FoM$_{\sigma_8,\Omega_{\mathrm{m}}}$ of the four DES probes and primary CMB are also very similar, 8721 and 8994, respectively. We note that these are the first cosmological constraints combining all four dark energy probes from a single experiment, an important milestone for Stage-III dark energy experiments. Comparing between the different probes, we see that $S_8$ is primarily constrained by 3$\times$2pt, SN+BAO places a strong constraint on $\Omega_{\rm m}$, and clusters contribute to a small shift and tightening in the $\Omega_{\rm m}$ direction. The addition of the other probes to $3\times 2$pt pulls the $\Omega_b$ constraint against the lower prior edge. All four probes appear fully consistent in this parameter space, and we confirm this fact quantitatively in Table~\ref{tab:consistency}, {where different combinations of DES probes are always consistent to within $\lesssim 2\sigma$}.

Comparing the four DES probes and the CMB constraints, we see that in the $S_8-\Omega_{\rm m}$ plane the 2$\sigma$ contours overlap. Quantitatively, we find a $2.8\sigma$ parameter differences in the full parameter space between the combination of all DES probes and the CMB. Compared to the case with only 3$\times$2pt in Section~\ref{sec:linear_gal}, it is interesting that the addition of DES BAO, SN and CL increased the strength of differences with CMB. We further break down the different components and calculate the consistency between the different DES probes in Table~\ref{tab:consistency}. We find that 3$\times$2pt+SN+BAO gives 2.5$\sigma$, while clusters increase the tension slightly. \rcwr{It is not clear that there is a specific DES probe that drives the overall difference with the CMB.}

When combining all DES probes with the primary CMB, we find further improvements in all parameters, with $S_8=0.815^{+0.008}_{-0.007}$, $\Omega_{\rm m}=0.314^{+0.006}_{-0.007}$. The FoM$_{\sigma_8,\Omega_{\mathrm{m}}}$ increases to 20590, which is a more than 2$\times$ gain over the CMB FoM$_{\sigma_8,\Omega_{\mathrm{m}}}$. 

For $w$CDM, we combine three of the four probes: DES Y6 3$\times$2pt, DES BAO and DES SN. As the DES clusters analysis of \citep{DES2025_clusters} did not test the pipeline in $w$CDM, we choose to be conservative and omit it in this combination. In a forthcoming paper we will present dark energy constraints including DES clusters \citep{y6-clusters}. The combined constraints are shown in the right panel of Figure~\ref{fig:all_des}. We find $w = -0.956^{+0.044}_{-0.041}$, $\Omega_{\rm m} = 0.317^{+0.013}_{-0.014}$ and $S_8 = 0.797^{+0.013}_{-0.012}$. Similar to before, we find no evidence for departure from a cosmological constant of $w=-1$, now with a constraint on $w$ tighter than $5\%$. \rcwr{This is more than twice as constraining as the equivalent} constraint in DES Y3 at $w=-0.84^{+0.11}_{-0.10}$, where we combined the mid-survey (DES Year 3) versions of the 3$\times$2pt \citep{y3-3x2ptkp}, SN \citep{DES2019Y3SN}, and BAO \citep{DES2022Y3BAO}. \rcwr{We also do not find a significant preference of $w$CDM over $\Lambda$CDM (0.5$\sigma$) when doing model comparison via $\Delta \chi^2.$}

Comparing the different probes, \rcwr{degeneracy-breaking is effective between 3$\times$2pt and SN+BAO} -- the degeneracy directions in the $w-\Omega_{\rm m}$ plane for the two sets of contours are orthogonal, resulting in the tight constraint when combining. The FoM$_{w,\Omega_{\mathrm{m}}}$ goes from 168 to 1955 when 3$\times$2pt is combined with SN+BAO, a more than 10$\times$ gain. Compared to the primary CMB constraints, we find that the parameter difference in the full {$w$CDM} parameter space is $2.5\sigma$, consistent with the $\Lambda$CDM case.

\section{Combination with External Probes}\label{sec:ext}

In this section we present results from comparing and combining the 3$\times$2pt constraints from this work with other external datasets to achieve the tightest cosmological constraints to date. In particular, we use the following external likelihoods:

\begin{itemize}
\item \textit{CMB}: We use a combination of TT, EE and TE primary CMB power spectrum measurements from \textit{Planck} 2018 \cite{Planck2020_cosmo}, Atacama Cosmology Telescope (ACT)-DR6 \cite{Louis2025} and South Pole Telescope (SPT)-3G DR1 \cite{Camphuis2025} experiments. We combine the following likelihoods:
\begin{itemize}
	\item \textit{Planck} 2018 (PR3 combination -- \texttt{Plik} likelihood with Legacy maps) \textit{lite} likelihood with $2 < \ell_{\rm TT} < 1000$ for TT spectra (including two low-$\ell$ TT bins as described in \cite{Prince2019}) and $30 < \ell_{\rm TE, EE} < 600$ for TE and EE spectra. When used with these cuts and in combination with ground-based CMB data, this is expected to be entirely consistent with the more recent `PR4' \textit{Planck} result from \texttt{CamSpec} likelihood with \texttt{NPIPE} maps \cite{Jense2025}.
    \item \textit{Planck} 2018 (\texttt{simall}) likelihood with $2 < \ell_{\rm EE} < 30$ for low-$\ell$ polarization information, necessary to constrain the optical depth to reionisation $\tau$. We note that other analyses have used the \textit{Planck} \texttt{SRoll2} low-$\ell$ EE likelihood \cite{Delouis2019,Pagano2020}, which disfavors lower values of $\tau$ and various authors (e.g., \cite{GarciaQuintero2025,Sailer2025}) have pointed out that this choice can affect (although not to high significance) the level of consistency of combined data sets with \lcdm.
	\item ACT-DR6 \textit{lite} likelihood with $ 600 < \ell_{\rm TT} < 8500$ for TT spectra and $ 600 < \ell_{\rm TE, EE} < 8500$ for TE and EE spectra \cite{Louis2025}.
	\item SPT-3G DR1 \textit{lite} likelihood with \texttt{candl} \cite{Balkenhol2024} \texttt{SPT3G\_D1\_TnE} data using $400 < \ell_{\rm TT} < 3000$ for TT spectra and $400 < \ell_{\rm TE, EE} < 4000$ for TE and EE spectra.
\end{itemize}
We follow the official ACT analysis of \cite{Louis2025} in ignoring the covariance between ACT-DR6 and \textit{Planck} in the overlapping multipole range, which was not found to affect results. Likelihoods referred to as \textit{lite} include a pre-marginalisation over nuisance parameters which model microwave foregrounds. Residual uncertainties are modeled using amplitude parameters $A_{\rm CMB}$ (which is shared across ACT and \textit{Planck} likelihoods so that $A_{\rm ACT} = A_{\rm Planck}$ as in \cite{Louis2025}), $T_{\rm cal}$ (for SPT-3G), and a polarization calibration parameter $P_{\rm ACT}$ (in the ACT part of the likelihood) and $E_{\rm cal}$ (in the SPT-3G part of the likelihood).

We do not combine the DES constraints with the CMB lensing likelihood from these experiments, as the DES 3$\times$2pt is not independent of the CMB lensing information. A proper combination will need to take into account the cross-correlation \citep{DESSPT2023}.

\item \textit{DESI BAO:}
We use BAO measurements from the Dark Energy Spectroscopic Instrument (DESI) Data Release 2 \cite{desi-dr2} covering an effective redshift range $0.295 < z < 2.330$. More explicitly, we include isotropic BAO scaling parameters measured for the BGS sample, and both isotropic and Alcock-Paczynski scaling parameters for the LRG1, LRG2, LRG3+ELG1, ELG2, QSO and Lyman-$\alpha$ samples.

\item \textit{SPT CL:} We use the constraints derived from cluster abundance with SPT described in \citep{Bocquet2024}. The constraint uses a cluster sample constructed from a combination of SPT-SZ, SPTpol ECS and SPTpol 500d data, which comprises {of} 1,005 clusters in the redshift range  $0.25<z<1.78$. The cluster masses are calibrated via a combination of DES Y3 and HST weak lensing data. 

\item \textit{BBN:} Big Bang Nucleosynthesis (BBN) theory predicts the abundance of light elements in the early Universe, such as deuterium and helium, as well as their relation to the baryon-to-photon ratio. Therefore, the observational determination of the primordial deuterium abundance and the helium fraction can be used to compute the physical baryon density parameter today,
$\Omega_b\, h^2$. We employ a BBN $\Omega_b\, h^2 = 0.02218 \pm 0.00055$ constraint from a recent analysis \citep{2024BBN} that recalculates the predictions while marginalizing over uncertainties in reaction rates in the standard $\Lambda$CDM model. We mainly use this likelihood to derive constraints on the Hubble parameter in Sec.~\ref{sec:hubble}. 

\end{itemize}

\begin{figure*}
    \centering
    \includegraphics[width=0.45\linewidth]{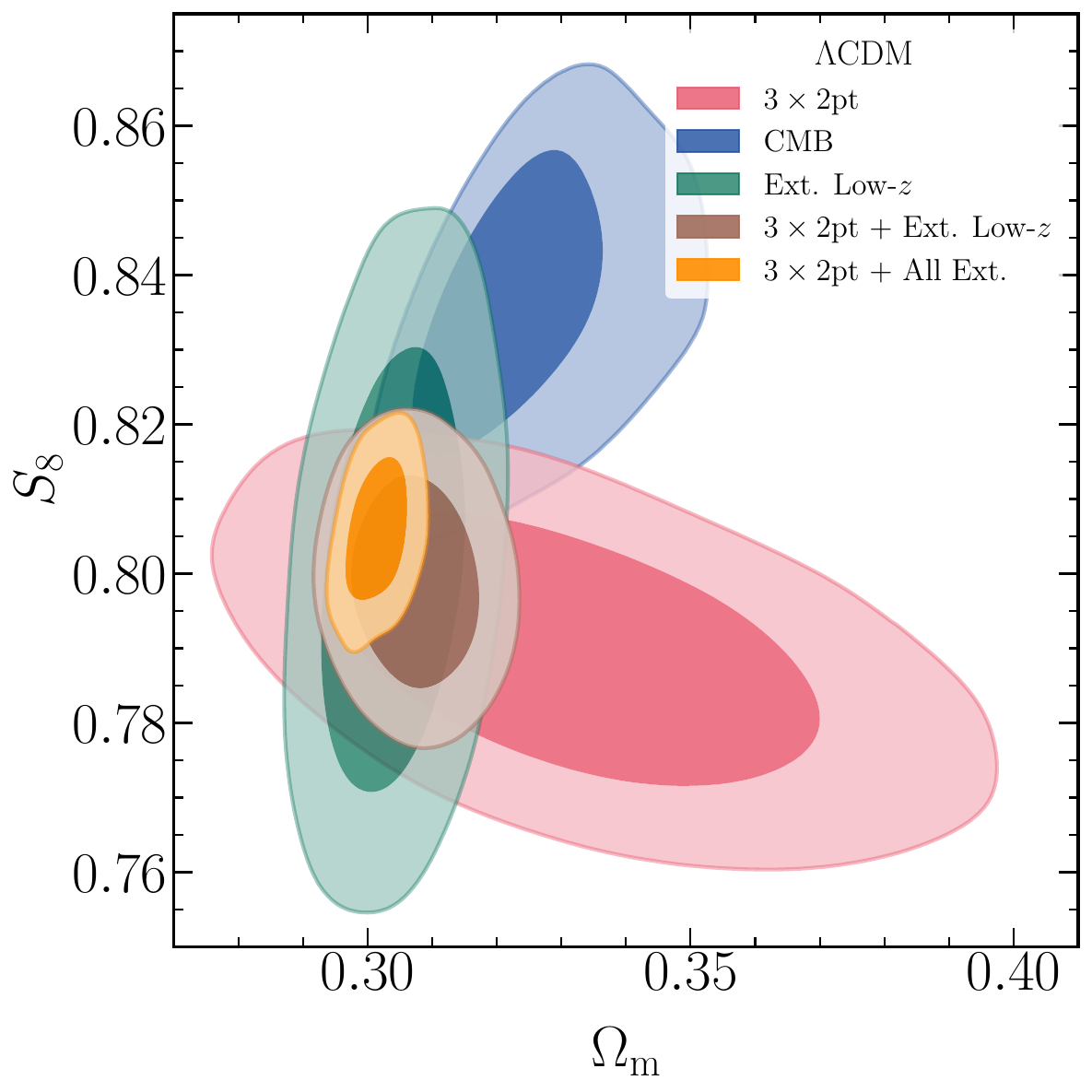}
\includegraphics[width=0.45\linewidth]{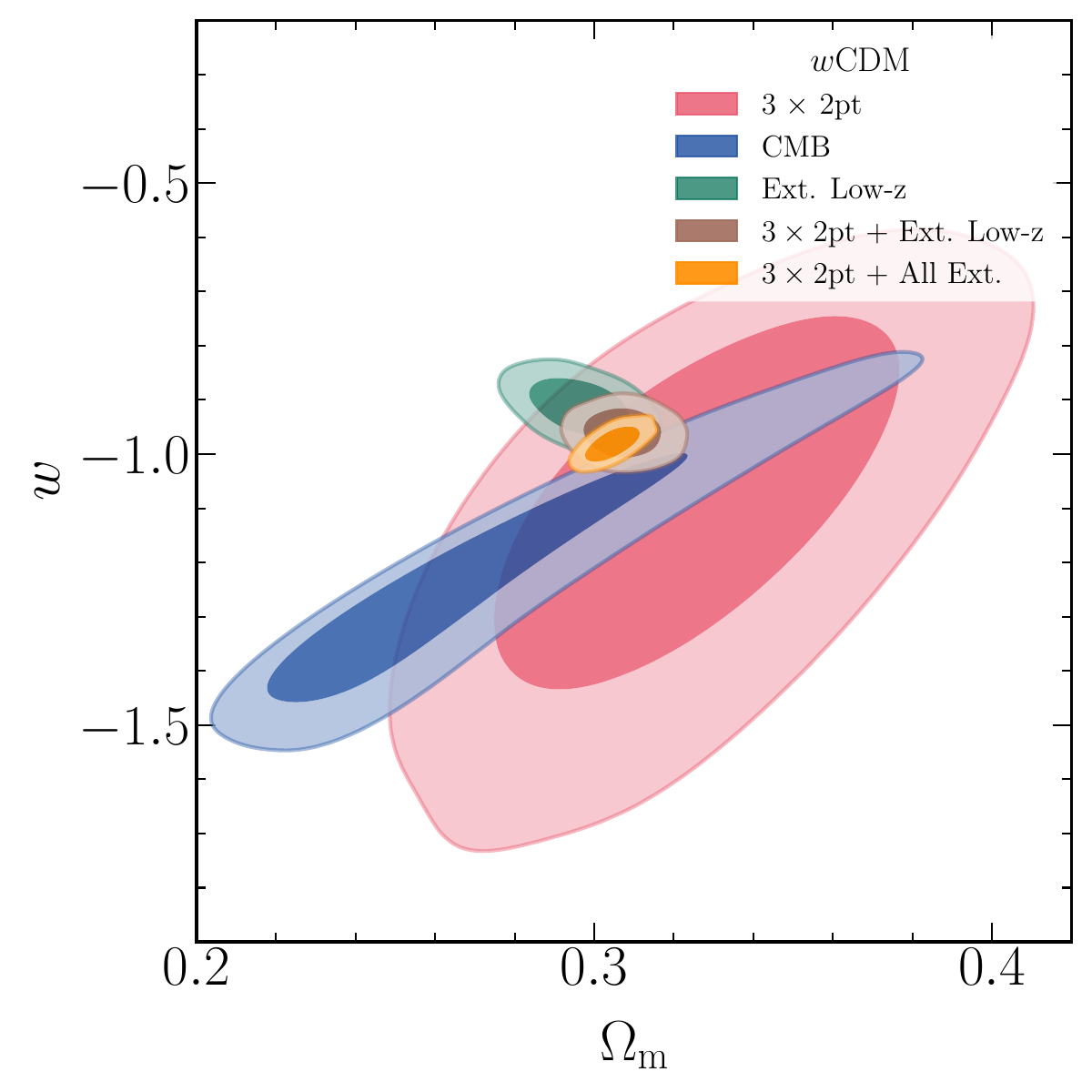}
    \caption{$\Lambda$CDM (left) and $w$CDM (right) constraints for DES Y6 3$\times$2pt combined with other low redshift external probes. \journal{}{Blue contours correspond to CMB constraints from \textit{Planck}, ACT and SPT combined.} Combining DES Y6 3$\times$2pt, external low-redshift probes and the CMB gives the orange contours.  }
    \label{fig:extlcdm}
\end{figure*}

In addition to the above external likelihoods, when combining with supernova data, we use the DES SN described in Section~\ref{sec:des_probes} as that is the most constraining supernova dataset to date. We \rcwr{also combine} with DES BAO (excl. DESI), which is the DES BAO excluding objects that overlap with DESI, as described in the previous section.

\subsection{Early vs. late universe}

We now combine our DES Y6 3$\times$2pt results with the most constraining combination of internal and external cosmological probes. Specifically, we first look at the combination we call ``Ext. Low-$z$'', which consists of, from the list above,  DESI BAO, DES Y6 BAO (excl. DESI), DES SN and SPT CL. Figure~\ref{fig:extlcdm} plots the results of combining our 3$\times$2pt results with Ext. Low-$z$, for $\Lambda$CDM and $w$CDM. The constraints are listed in Table~\ref{tab:lcdm_param_constraint}.  

We find the tightest constraints from the low-$z$ universe in $\Lambda$CDM to be $S_8 = 0.799^{+0.009}_{-0.010}$, $\Omega_{\rm m} = 0.308^{+0.006}_{-0.006}$ and $\sigma_{8} = 0.789^{+0.012}_{-0.013}$.

In particular, adding 3$\times$2pt to Ext. Low-$z$ results in a 2.4$\times$ gain in the FoM$_{\sigma_8,\Omega_{\mathrm{m}}}.$ 

From Figure~\ref{fig:extlcdm} it can be clearly seen that, as expected, 3$\times$2pt contributes mostly in constraining the structure growth, or $S_8$, {while Ext. Low-$z$ provides strong constraining power in $\Omega_{\rm m}$, primarily driven by the BAO and SN measurements, as seen in Figure~\ref{fig:omegam}.}
The external constraint on $\Omega_{\rm m},$ as well as the FoM$_{\sigma_8,\Omega_{\mathrm{m}}},$ are about 2$\times$ better here than the DES-only low-redshift constraints, mostly due to the addition of DESI BAO.    

The right panel of Figure~\ref{fig:extlcdm} shows the $w$CDM constraints combining all these datasets. In this model we find $w=-0.962^{+0.031}_{-0.029}$ and $\Omega_{\rm m} = 0.308^{+0.006}_{-0.007}$.

This is a 3\% constraint on $w$, and is still consistent with the cosmological constant ($w=-1$). Overall, the DES 3$\times$2pt mostly impacts the constraint on $S_8$, but also moves $\Omega_{\rm m}$ to slightly higher values compared to the Ext. Low-$z$ probes, increasing the consistency with the primary CMB in the $w-\Omega_{\rm m}$ plane. 

In Figure~\ref{fig:extlcdm} we also overlay the primary CMB constraints for $\Lambda$CDM and $w$CDM. The comparison with 3$\times$2pt+ Ext. Low-$z$ represents our best understanding of the two models from the late and early universe. We find a 2.3$\sigma$ difference between the 3$\times$2pt + Ext. Low-$z$ results and the CMB when considering the full parameter space for $\Lambda$CDM, and 2.4$\sigma$ for $w$CDM. 
While these numbers do not indicate a significant tension between the low- and high-redshift universe, it is worth noting that they are all higher than the values seen in DES Y3 \citep{y3-3x2ptkp}. As discussed in Section~\ref{sec:linear_gal}, this is partially driven by the combined CMB constraints.   

\subsection{Joint cosmological constraints in $\Lambda$CDM and $w$CDM}

We now combine DES Y6 3$\times$2pt + Ext. Low-$z$ and CMB to obtain the final joint constraint. This produces the tightest $\Lambda$CDM joint constraint from all major cosmic surveys to date. We show the $S_8-\Omega_{\rm m}$ projection of this constraint in the left panel of Figure~\ref{fig:extlcdm}, but will examine the other projections in the sections below. It is worth noting that there is nontrivial degeneracy breaking when combining such a complex set of data in this high-dimensional space. 

We find the tightest constraints available in $\Lambda$CDM to be
\begin{align}
S_8 = 0.806^{+0.006}_{-0.007}, \notag \\
\Omega_{\rm m} = 0.302^{+0.003}_{-0.003}, \notag \\
\sigma_{8} = 0.804^{+0.007}_{-0.006} .
\end{align}
This combination has a FoM$_{\sigma_8,\Omega_{\mathrm{m}}}$ of 51955, 7.4 times larger than 3$\times$2pt+~Ext. Low-$z$ alone, and a factor of 1.5 larger than the \rcwr{previous} most constraining combination we presented four years ago with DES Y3 \citep{y3-3x2ptkp}. In that work, we combined the DES Y3 3$\times$2pt with eBOSS BAO and RSD, Pantheon SN and CMB constraints from \textit{Planck} 2018. The subsequent upgrades in each of the physical probes together contribute to this gain. Interestingly, this combination also results in $S_8$ shifting lower by about 0.6$\sigma$ compared to \rcwr{our result in Y3}. Given that the CMB constraints shifted to higher $S_8$ values from \textit{Planck} to the joint \textit{Planck}+ACT+SPT combination that is used in this work, this suggests that the low-redshift constraints now have substantially greater impact on the joint constraints.  

Since we are considering a flat universe, a constraint on $\Omega_{\rm m}$ translates directly into a constraint on the dark-energy density parameter, giving $\Omega_{\Lambda} = 0.698^{+0.003}_{-0.003}$ and $\Omega_{\Lambda}h^{2} = 0.326^{+0.004}_{-0.004} $. In physical units this translates to $\rho_{\Lambda} = 6.120^{+0.075}_{-0.069}\times 10^{-30}\,\text{g/}\text{cm}^{3}$.
The value of the cosmological constant itself is  $\Lambda = 8\pi G\rho_{\Lambda} = 1.027^{+0.013}_{-0.012} \,  \times 10^{-35}\,\text{s}^{-2}$ or in distance units $\Lambda/c^{2} = 1.14^{+0.014}_{-0.028} \,  \times 10^{-56}\,\text{cm}^{-2}$.

In $w$CDM we show the best constraining combination in the right panel of Figure~\ref{fig:extlcdm} for the $w-\Omega_{\rm m}$ plane. We find:
\begin{align}
w=-0.981^{+0.021}_{-0.022}, \notag \\
\Omega_{\rm m} = 0.305^{+0.004}_{-0.005},
\end{align}
a 2\% constraint on $w$ that is consistent with the cosmological constant ($w=-1$). 

\subsection{Comparison of lensing probes}

\label{sec:lensing_probes}

\begin{figure}
    \centering
    \includegraphics[width=0.48\textwidth]{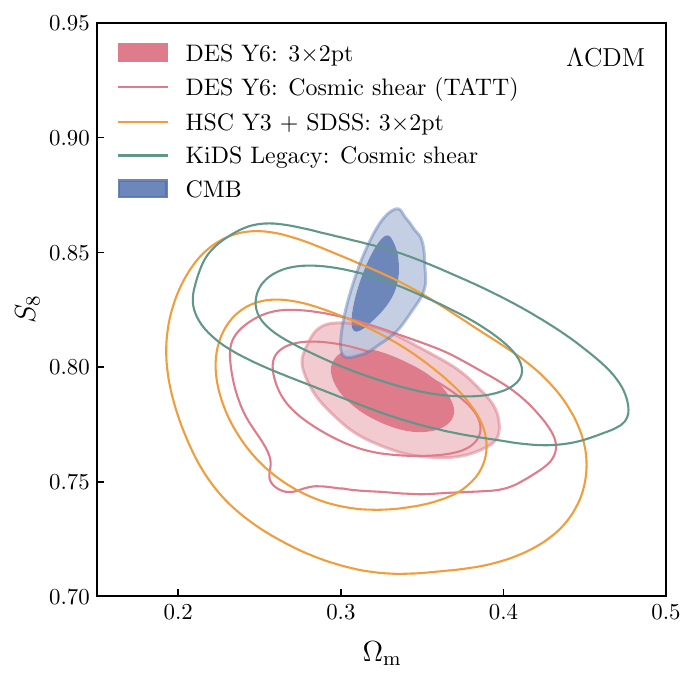}
     \caption{Comparison of $S_8-\Omega_{\rm m}$ for the main Stage-III lensing surveys. Here we select the results from each survey that is most comparable to either our 3$\times$2pt or cosmic shear analysis. For KiDS, we show the KiDS-Legacy cosmic shear constraints \citep{Wright2025} (green). For HSC, we show the 3$\times$2pt constaints from \citep{Zhang2025} (orange). \journal{}{Blue contours correspond to CMB constraints from \textit{Planck}, ACT and SPT combined.}}   
    \label{fig:3x2-lensing}
\end{figure}

In this section we compare our 3$\times$2pt results with results from other Stage-III lensing analyses: the Kilo-Degree Survey \citep[KiDS,][]{Kuijken2015} and the Hyper Suprime-Cam Subaru Strategic Program \citep[HCS-SSP,][]{Aihara2017}. A quantitative evaluation of the consistency of the three Stage-III results is not straightforward because of the numerous differences in model choices, scale cuts, and analysis techniques, and multiple results from each collaboration.
We plot here results that are representative of each collaboration's constraints on structure growth. 

For KiDS, the most recent 3$\times$2pt results come from \citep{Heymans2020}, where the lensing data from the 1000 degree$^2$ KiDs survey were combined with spectroscopic data from BOSS and 2dFlens to form the 3$\times$2pt data vector. Since then, KiDS made significant updates to their shear catalog, resulting in the KiDS-Legacy cosmic shear results \citep{Wright2025}. Given the update, we show the KiDS-Legacy cosmic shear constraints in Figure~\ref{fig:3x2-lensing} (green), and compare it to both the DES Y6 3$\times$2pt and the cosmic shear results (pink solid and open, respectively). We find that the constraining powers of KiDS-Legacy and DES Y6 from cosmic shear are similar, and the constraints are consistent: $S_8 = 0.815^{+0.016}_{-0.021}$ (KiDS) and $S_8 = 0.783^{+0.019}_{-0.015}$ (DES). DES 3$\times$2pt provides tighter constraints primarily in the $\Omega_{\rm m}$ direction but also in $S_8$. It is also interesting to observe from Figure~\ref{fig:3x2-lensing} that the degeneracy direction in this parameter plane is similar, as expected from the similar redshift range of the shear catalogs.

For HSC Y3, three versions of the 3$\times$2pt analysis were carried out. All of them use the SDSS DR11 spectroscopic galaxies as the lens sample, as opposed to DES's photometric lenses. In \citep{Sugiyama2023} (referred to as the large-scale analysis), the authors computed the matter power spectrum using a standard Boltzmann code (similar to DES's use of CAMB), but employed a halo model to describe both nonlinear structure formation and galaxy bias-contrasting with DES Y6's perturbative bias models combined with conservative scale cuts to avoid halo-model systematics. In \citep{Miyatake2023}, an emulator trained on simulations was used to model smaller scales, yielding tighter constraints in $\Omega_{\rm m}$. However, the latest analysis of \citep{Zhang2025}  is most similar to this work in three key respects: it employs tomographic binning of the source sample (rather than a single redshift bin), applies conservative scale cuts determined by theory uncertainties (similar to \citepalias{y6-methods}), and uses simple perturbative bias models without relying on halo model predictions for small-scale clustering. We show their constraint in Figure~\ref{fig:3x2-lensing} in orange. Due to the smaller survey area (416 deg$^2$),  HSC Y3 constraints are substantially wider, $S_8=0.780^{+0.029}_{-0.030}$. %$S_8=0.804^{+0.051}_{-0.051}$. 
They are also consistent with the DES results. 
We note that in Figure~\ref{fig:3x2-lensing} we plot directly the published chains from each survey, instead of a matched re-analysis of their data. We leave a more detailed comparison with unified analysis choices \citep[e.g.][]{Chang2019, Longley2023, DESKIDS2023, Porredon2025} for future investigation. 

In summary, the final DES $3\times2$pt and \rcwr{cosmic shear} analyses yield cosmological constraints in the $S_8-\Omega_{\rm m}$ plane that are consistent with the latest results from the other Stage-III weak-lensing surveys. \rcwr{Overall, the marginalized posteriors in this parameter space from all the weak lensing surveys are also consistent with the CMB, though the inferred $S_8$ from lensing remains systematically slightly lower than that inferred from CMB measurements. KiDS-Legacy finds the highest $S_8$ values, while DES $3\times2$pt gives the tightest $S_8$ constraint with the highest significance of deviation from CMB, but still below $3\sigma.$}

\subsection{Constraints on the total amount of matter}\label{sec:matter}

The possibility of dynamical dark energy (DDE) has attracted considerable attention following recent measurements, with several studies attributing this preference to underlying $\Omega_{\rm m}$ tensions between probes \citep[e.g.][]{Tang2025}. Initial hints of tension within $\Lambda$CDM emerged from the combination of DES SN with BOSS BAO \citep{DESY5SN2024}, which showed significant $\Omega_{\rm m}$ disagreement. The DESI DR1 \citep{desi-dr1} and DR2 BAO measurements \citep{desi-dr2}, when combined with type Ia supernovae data and CMB constraints, strengthened this evidence, yielding indications for DDE within the $w_0$--$w_a$CDM framework at the $\sim 3\sigma$ level. These analyses found SN measurements yielded higher $\Omega_{\rm m}$ values than CMB, which in turn yielded higher values than BAO. The recent recalibration of DES SN \citep{Popovic2025a} finds a $\sim$1$\sigma$ lower $\Omega_{\rm m}$ value, though still higher than BAO predictions. 

\begin{figure}
    \centering
    \includegraphics[width=0.44\textwidth]{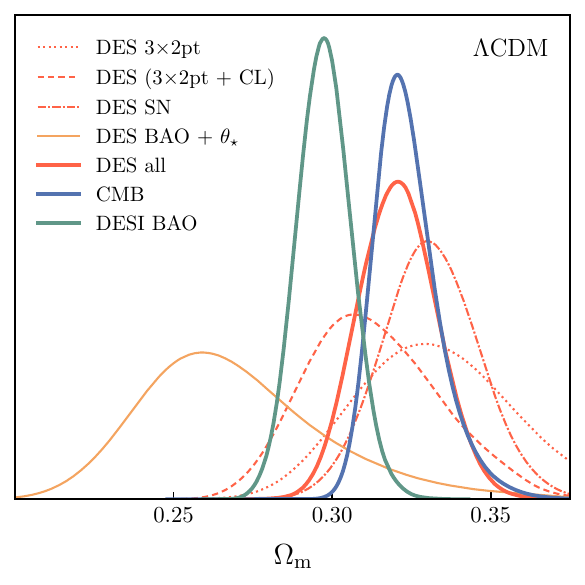}
    \caption{Marginalized constraints on the total amount of matter $\Omega_{\rm m}$ in the $\Lambda$CDM model from different DES probes and their combination, compared to CMB and DESI BAO. ``DES all'' includes Y6 3$\times$2pt, SN, BAO, and Clusters. We obtain excellent agreement between ``DES all'' and CMB. }
    \label{fig:omegam}
\end{figure}

In Figure~\ref{fig:omegam}, we compare the marginalized $\Omega_{\rm m}$ constraints inferred from DES probes -- including 3$\times$2pt, SN, BAO and CL -- and assess their level of agreement with DESI BAO and CMB constraints. Our DES $3\times2$pt, SN and CL results are in excellent agreement with the CMB, while  DES BAO+$\theta_\star$ yields a lower result.\footnote{\rcwr{For the DES BAO only case, we apply the $\theta_\star$ prior to obtain an $\Omega_{\rm m}$ constraint. $\theta_\star$ is the angular scale of the acoustic peak, which isolates the geometric/background information from the CMB. We incorporate this via a Gaussian likelihood taken from the \textit{Planck} 2018 temperature and polarization data \cite{Planck2020_cosmo}, with $100\theta_\star = 1.04109 \pm 0.00030\,$.}} All combined and without the $\theta_\star$ prior, the DES probes \rcwr{find very good agreement with} the CMB and \rcwr{are at 1.7$\sigma$ difference} from DESI BAO.
A more rigorous and complete comparison of geometric vs.~growth probes will be explored in the forthcoming paper \citep{y6-extensions}.

\subsection{Neutrino mass}

\begin{figure}
    %\centering
    \raggedleft
    \includegraphics[width=\linewidth]{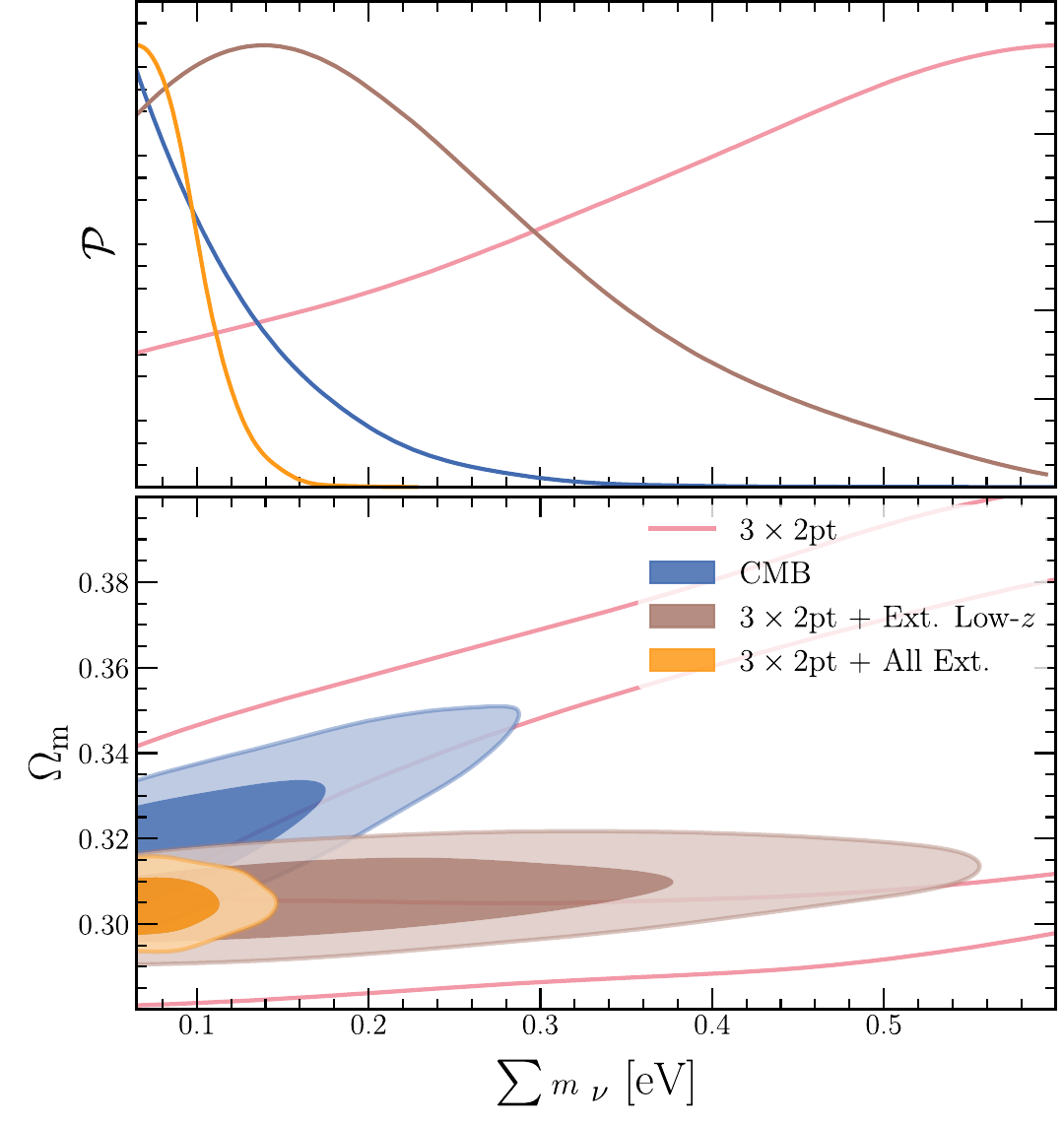}
    \caption{Constraints on the sum of neutrino mass coming from the DES Y6 3$\times$2pt and combination with other low- and high-redshift external probes. We note that all chains presented here contain a top-hat prior of $\sum m_\nu \in [0.06-0.6]$ eV.}
    \label{fig:Y3-mnu_vs_om}
\end{figure}

Figure \ref{fig:Y3-mnu_vs_om} shows the marginalized constraints on the sum of neutrino masses $\sum m_\nu$ and the matter density $\Omega_{\rm m}$ in flat $\Lambda$CDM, from our fiducial 3$\times$2pt, CMB, 3$\times$2pt+Ext. Low-$z$ and 3$\times$2pt+All Ext. As in Y3 \citep{y3-3x2ptkp}, the $3\times2$pt analysis alone does not provide a meaningful constraint on $\sum m_\nu$, and the marginalized posterior remains bounded by the adopted prior range of 0.06-0.6 eV.  
Our measurement of $\Omega_\mathrm{m}$ peaks in a region that is consistent with the CMB value, unlike the BAO case where $\Omega_\mathrm{m}$ is significantly lower than the CMB (Figure~\ref{fig:omegam}). This lower preference for $\Omega_{\rm m}$  drives the joint neutrino constraints (orange contour in Figure~\ref{fig:Y3-mnu_vs_om}) toward the lower boundary \citep{Elbers_neutrino_DESIfullshape_2025,Chebat_freq_neut_2025}, due to the degeneracy between $\Omega_{\rm m}$ and $\sum m_\nu$.

The CMB {(as defined in Section~\ref{sec:ext})} alone yields an upper limit of $\sum m_\nu < 0.24$ eV (95$\%$ CL), while the inclusion of DES $3\times2$pt slightly tightens this constraint to $\sum m_\nu < 0.22$ eV. When combining DES $3\times2$pt with the external low-redshift (Ext Low-$z$) probes, we obtain an upper bound of $\sum m_\nu < 0.47$ eV.
When we further combine this with the CMB, we find $\sum m_\nu < 0.14$ eV, with an improvement due to the different $\Omega_{\rm m}$--$\sum m_\nu$ degeneracies of high-$z$ and low-$z$ probes (cf. Figure \ref{fig:Y3-mnu_vs_om}). This {upper} limit is noticeably weaker than the one reported by the DESI Collaboration using DESI DR2 combined with the CMB (without lensing), $\sum m_\nu < 0.067$ eV  \citep[95$\%$ CL, ][]{desi-dr2}. The main reason is that we adopt physical priors that include the lower bound of 0.06 eV imposed by neutrino oscillation experiments \citep{review_particle_physics}, instead of the commonly used minimal mass of zero. {Yet, the trend in our combined result remains consistent with external constraints, as the posterior in the $\Lambda$CDM framework consistently peaks at the lower prior boundary \citep{Elbers_neutrino_DESIfullshape_2025,Chebat_freq_neut_2025}.}

All the marginalized neutrino constraints in $\Lambda$CDM are reported in Table \ref{tab:lcdm_param_constraint}. 

\subsection{Constraints on Hubble parameter}\label{sec:hubble}

\begin{figure}
    \centering
    \includegraphics[width=0.48\textwidth]{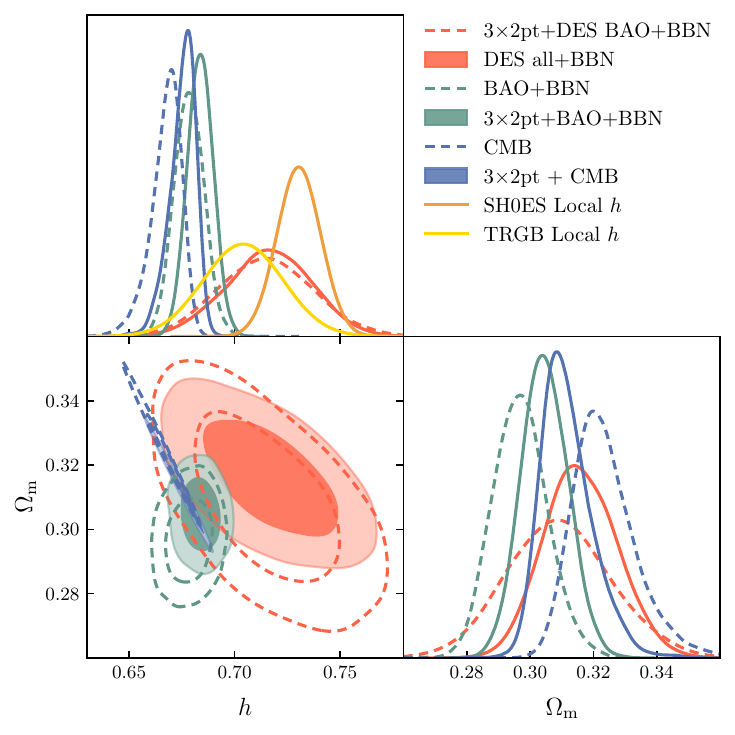}
    \caption{Marginalized constraints on the Hubble parameter $h$ and the total amount of matter $\Omega_{\rm m}$ in the $\Lambda$CDM model
    compared to  local measurements of $h$. The ``BAO'' label denotes the combination of DESI DR2 BAO and DES BAO measurements, with overlapping regions removed from the DES dataset. } 
    \label{fig:h0_omegam}
\end{figure}

Next we consider the Hubble parameter evaluated at the present time, $H_0 = 100h \text{ km s}^{-1} \text{ Mpc}^{-1}$, motivated by significant tensions observed between different measurements.

Multiple direct local measurements prefer a higher value of the expansion \rcwr{rate} \rcwr{than that inferred from the primary CMB}, most prominently the $h = 0.7304 \pm 0.0104$ constraint from the SH0ES team \citep{Sh0es2022} based on an astronomical distance ladder using Cepheid variables, and $h = 0.704 \pm 0.019$ \citep{Freedman2025} using the Tip of the Red Giant Branch (TRGB) method \citep{Freedman2025}.  

In the $\Lambda$CDM model, some of these local measurements stand in contrast to constraints from the CMB by \textit{Planck} + ACT + SPT, which prefer $h = 0.667^{+0.009}_{-0.006}$, a 4.6$\sigma$ tension in the 1D $h$ projection compared to the SH0ES result, while consistent at 1.6$\sigma$ with the TRGB result. \rcwr{This value comes from reanalyzing the relevant likelihood with our priors, including varying neutrino masses, from Table~\ref{tab:lcdm_param_constraint} and our sampler.} 

In Figure~\ref{fig:h0_omegam} we explore what  DES 3$\times$2pt  adds to this picture. While 3$\times$2pt provides only weak direct constraints on $h$, its strong constraining power on other cosmological parameters, especially $\Omega_{\rm m}$, helps break parameter degeneracies when combined with the CMB. When we combine 3$\times$2pt with CMB data, we find $h = 0.675^{+0.007}_{-0.004}$, which is both more constraining and slightly higher than the CMB-only value. This upward shift arises primarily because 3$\times$2pt prefers a lower $\Omega_{\rm m}$, which in turn pushes $h$ upward through the degeneracy between these parameters. Due to this shift to higher $h$, since the uncertainties are smaller, the tension with local Universe measurements remains practically unchanged from the CMB-only case.

We can also constrain the value of $h$ independently of CMB data using a combination of BAO and BBN constraints on $\Omega_{\rm b} h^2$, which yields $h = 0.678 \pm 0.007$ when using DESI DR2 BAO + DES BAO (excl. DESI). When adding DES $3\times2$pt measurements, we also find a slightly more constraining and higher value of $h = 0.684 \pm 0.006$.  These constraints are tighter than CMB-derived values, and are in $3.9\sigma$ tension with the local SH0ES measurement
and $1.0\sigma$ agreement with TRGB. 

For the first time, we can also obtain an independent measurement of $h$ just from DES data ($3\times2$pt+SN+BAO+CL) and BBN:  $0.715^{+0.021}_{-0.019}$, which is consistent with and as constraining as the most recent measurement reported by the TRGB method, \rcwr{ with a $2.3\sigma$ shift compared with CMB}. This is because DES $3\times2$pt constrains $\Omega_b$ to be slightly lower than the CMB-preferred value, which leads to higher $h$ through the BBN prior. While the DES-based constraint is slightly higher in $h$ when combined with BBN compared to the combination including DESI BAO, the two constraints are still consistent.

Strong lensing time-delay measurements can also constrain $H_0$, albeit with larger uncertainty at current precision levels. From the recent \textsc{TDCOSMO} results \citep{TDCOSMO2025}, we find {consistent results 
$\left( h=0.727^{+0.048}_{-0.036}\right)$} when reanalyzing their likelihood with our priors and sampler. When combined with 3$\times$2pt data, we obtain a tighter constraint on $H_0$ with a slightly lower value of \rcwr{$h=0.715 \pm 0.033$}, primarily due to breaking degeneracies in the $\Omega_{\rm m}$ direction. \rcwr{This result is consistent} with the fiducial value reported by \citet{TDCOSMO2025} of $h=0.716^{+0.039}_{-0.033}$, obtained when combining the TDCOSMO-2025 sample with $\Omega_{\rm m}$ constraints from the Pantheon+ SN Ia dataset.

\section{Conclusions}\label{sec:conclusion}

In this paper we present the $\Lambda$CDM and $w$CDM constraints from the DES Y6 3$\times$2pt analysis, combining three two-point functions of galaxy clustering and galaxy weak lensing. This result combines the full six-year DES data of 140 million weak lensing source galaxies and 9 million lens galaxies, covering roughly 4,300 deg$^2$. It rests atop 18 other supporting papers that describe components that went into this analysis, as well as the prior 3$\times$2pt analyses from DES in Year 1 \citep{y1-keypaper} and Year 3 \citep{y3-3x2ptkp}.

We carry out extensive testing of our data and methodology in this paper, and in the 18 supporting papers, to ensure the robustness of the results. In addition, the combination of the three two-point functions naturally provides a self-calibration mechanism to safeguard against a number of systematic effects.

Assuming $\Lambda$CDM cosmology and linear galaxy bias, the DES Y6 3$\times$2pt analysis yields $S_8 = 0.789^{+0.012}_{-0.012}$ and $\Omega_{\rm m} = 0.333^{+0.023}_{-0.028}$. This is a factor of two increase in constraining power compared to DES Y3, and rivals state-of-the-art CMB constraints in the $S_8$ direction, which measures the clustering amplitude of cosmic structure. Comparing with the latest CMB constraints from temperature and polarization based on a combination of \rcwr{SPT-3G DR1, ACT-DR6, }and \textit{Planck} 2018, {we find a full parameter space difference of 1.8$\sigma$. When projected onto $S_8$, i.e., focusing on growth of structure, the difference increases to 2.6$\sigma$, with DES preferring a lower value}. 
\rcwr{The full parameter difference reduces to 1.0$\sigma$ when considering \textit{Planck} 2018 only, due both to \textit{Planck} 2018 preferring lower $S_8$ values on its own than when combined with other CMB data, and to decreased uncertainties from the combined CMB dataset. The full parameter difference between DES Y3 3$\times$2pt and \textit{Planck} 2018 is 1.5$\sigma$}.

We then combine, for the first time, the four dark energy probes that were proposed at the inception of DES \citep{DES2005}: galaxy clusters, BAO, SN Ia and 3$\times$2pt. While we perform this combination in a sub-optimal way because the cluster analysis uses a different source population than the 3$\times$2pt analysis, we still find improvement in the cosmological constraints, especially in $\Omega_{\rm m},$ when combining all probes compared to 3$\times$2pt. We find $S_8 = 0.794^{+0.009}_{-0.012}$ and $\Omega_{\rm m} = 0.322^{+0.012}_{-0.011}$. This DES-only constraint is  consistent at 2.8$\sigma$ with the primary CMB constraints in the full parameter space. When combined with the CMB primary results, we find $S_8=0.815^{+0.008}_{-0.007}$ and $\Omega_{\rm m}=0.314^{+0.006}_{-0.007}$.

We also combine with a collection of external low-$z$ probes (we refer to as Ext. Low-$z$), which include DESI BAO DR2, DES BAO (minus DESI overlap), DES SN, and SPT cluster abundances. We find further improvements in the cosmological constraints, though the $S_8$ constraint is still dominated by DES 3$\times$2pt. This combination is consistent with the CMB at the 2.3$\sigma$ level, allowing us to combine them and achieve final constraints of 0.8\% in $S_8$ and 1\% in $\Omega_{\rm m}$.  

We also explore $w$CDM constraints for 3$\times$2pt only, all DES probes, and combinations with exernal data. We find that marginalized $w$ posteriors from all combinations are consistent with $-1$. DES alone constrains $w$ with better than $5\%$ precision, while combining DES 3$\times$2pt + Ext. Low-$z$ + CMB results in  $2\%$ precision in $w$ that is still consistent with $-1$. We find no \rcwr{significant} preference for $w$CDM over $\Lambda$CDM from the 3$\times$2pt data. Notably, the 2--3$\sigma$ differences between low-redshift and CMB datasets persist comparably in both cosmological models, with full-parameter differences of 2.3$\sigma$ for 3$\times$2pt + Ext. Low-$z$ (2.8$\sigma$ for all DES probes) in $\Lambda$CDM and 2.4$\sigma$ (2.5$\sigma$) in $w$CDM. \rcwr{This minimal extension to evolving dark energy density does not decrease the parameter differences observed in $\Lambda$CDM. We refer to \citep{y6-extensions} for exploration of alternative cosmological scenarios.}

We discuss how the new 3$\times$2pt constraints from DES impact other $\Lambda$CDM tensions in cosmology today. DES data with a BBN prior  constrain $H_0$ to $0.715^{+0.021}_{-0.019}$,  very similar in mean and uncertainty to the TRGB local $H_0$ constraint. We also find that DES 3$\times$2pt does not  significantly impact the upper bound of neutrino mass derived from other probes. Combining DES 3$\times$2pt with Ext. Low-$z$ and CMB yields $\sum m_\nu < 0.14$ eV. We caution that we have placed a sharp prior at $\sum m_\nu = 0.06-0.6$ eV, which removes the parameter space of unphysical negative neutrino masses and reflects the current bounds from neutrino oscillation experiments. 

\rcwr{In the larger context of weak lensing experiments, there has been a historic trend across measurements to prefer a slightly lower $S_8$ than the CMB-predicted value.
The full-survey results from DES and e.g.~KiDS \citep{Wright2025} have tended to find slightly higher $S_8$ values than before, while remaining in excellent agreement with our previous DES analyses.
At the same time, the joint CMB constraints from \textit{Planck} 2018, SPT-3G DR1, and ACT-DR6 also shrank asymmetrically to slightly higher predictions of $S_8$ vs.~\textit{Planck} alone, leaving the DES-CMB consistency at $2.8 \sigma$.
This leaves the overall state of the field in a similar place as it was in previous years: we see no substantial evidence for a discrepancy with the CMB in either $S_8$ or the full parameter space, with various estimates of significance across weak lensing experiments still ranging from $\sim 1\sigma$ to $\lesssim 3 \sigma$. }

The first cosmology analysis using two-point functions in DES was performed in 2016 \citep{DES2016shearSV} using cosmic shear from Science Verification data, alongside with first clustering and galaxy-galaxy lensing measurements \cite{2016MNRAS.455.4301C,2017MNRAS.465.4204C}. Over the past 10 years, we have processed more data, developed more sophisticated tools, invented new methodologies and trained a new generation of young scientists. This paper uses $\sim$30 times the area, $\sim$45 times the number of weak lensing galaxies, a much more optimized clustering sample, and a substantially much more complex and robust analysis framework compared to those first papers. We expect this first analysis using all 6 years of DES data for large-scale structure cosmology to be followed by more creative uses of the Y6 data going forward.

DES is a precursor survey to the Vera C. Rubin Observatory Legacy Survey of Space and Time (LSST), the Euclid mission, and the Nancy Grace Roman Space Telescope High Latitude Imaging Survey (HLIS). Part of the legacy of DES will be its traces in the planned cosmology analysis of these Stage-IV surveys (e.g. position and shear catalogs, redshift calibration, synthetic source injection, the blinding process and multi-stage analysis plan, the 3$\times$2pt pipeline, etc.). \rcwr{The journey of DES, together with other Stage-III galaxy imaging surveys, will continue with LSST, Euclid and Roman, opening the next golden era of galaxy imaging surveys.} 

\section{Data Availability}

The Y6KP data products in this analysis, including catalogues, data-vector measurements, masks and chains, are publicly available at \url{https://des.ncsa.illinois.edu/releases/y6a2}, see \cite{y6-das}. The DES Y6 Gold catalogue derived from the DES Data Release 2 (DR2) \cite{des-dr2} is publicly available at \url{https://des.ncsa.illinois.edu/releases} and described in \cite{y6-gold}. 

\section*{Acknowledgments}

{We would like to thank Angus H. Wright, Catherine Heymans and Rachel Mandelbaum for their input to Sec.~\ref{sec:lensing_probes}, and the KiDS and HSC collaborations for providing their sampled posteriors used therein. We thank Sebastian Bocquet and the SPT collaboration for sharing the SPT-clusters chains. \rcwr{We also thank the TDCOSMO collaboration for providing early access to their likelihood and to Martin Millon for his help implementing it in \textsc{CosmoSIS}.}}
\newline

{\bf Author contributions: } All authors contributed to this paper and/or carried out infrastructure work that made this analysis possible. Some highlighted contributions include

\begin{itemize}[leftmargin=*]
\itemsep0em
\item Source sample selection and calibrations: A.~Amon, K.~Bechtol, M.~R.~Becker, M.~Gatti, R.~A.~Gruendl, M.~Jarvis, F.~Menanteau, A.~Roodman, E.~S.~Rykoff, T.~Schutt, E.~Sheldon, M.~Yamamoto
\item Lens sample selection and calibrations: S.~Avila, M.~Crocce, J.~Elvin-Poole, E.~Legnani, S.~Lee, J.~Mena-Fern{\'a}ndez, A.~Porredon, W.~Riquelme, M.~Rodriguez-Monroy, I.~Sevilla-Noarbe, N.~Weaverdyck
\item Photometric redshifts: A.~Amon, A.~Alarcon, G.~M.~Bernstein, G.~Camacho-Ciurana, A.~Campos, C.~Chang, W.~d'Assignies, J.~de~Vicente, G.~Giannini, J.~Myles, J.~Prat, C.~Sanchez, C.-H.~To, 
L.~Toribio~San~Cipriano, M.~A.~Troxel, N.~Weaverdyck, B.~Yin
\item Simulations and derived calibrations: D.~Anbajagane, J.~Beas-Gonzalez, M.~R.~Becker, J.~DeRose, S.~Everett, E.~Legnani, S.~Mau, T.~Schutt, M.~Tabbutt, C.-H.~To, M.~Yamamoto, B.~Yanny
\item Data-vector measurements and cosmological inference: A.~Amon, F.~Andrade-Oliveira, J.~Blazek, G.~Campailla, J.~M.~Coloma-Nadal, W.~d'Assignies, A.~Fert{\'e}, G.~Giannini, M.~Jarvis, E.~Krause, E.~Legnani, J.~Muir, A.~Porredon, J.~Prat, M.~Raveri, S.~Samuroff, D.~Sanchez-Cid, N.~Weaverdyck, M.~Yamamoto
\item Paper writing and figures: A.~Alarc{\'o}n, A.~Amon, G.~M.~Bernstein, G.~Campailla, C.~Chang, M.~Crocce, W.~d'Assignies, A.~Drlica-Wagner, G.~Giannini, E.~Krause, J.~Muir, A.~Porred{\'o}n, J.~Prat, D.~Sanchez-Cid, I.~Sevilla-Noarbe, M.~A.~Troxel, N.~Weaverdyck, M.~Yamamoto, B.~Yin
\item Comments to manuscript: S.~Avila, G.~M.~Bernstein, S.~Bocquet, T.~Davis, T.~Diehl, S.~Dodelson, J.~Frieman, I.~Harrison, K.~Herner, O.~Lahav, R.~Kron, E.~Krause, R.~Miquel
\item Coordination and scientific management: M.~R.~Becker and M.~Crocce (Year Six Key Project Coordinators), C.~Chang and M.~A.~Troxel (Science Committee Chairs), A.~Alarc{\'o}n, A.~Amon, S.~Avila, J.~Blazek, A.~Fert{\'e}, G.~Giannini, A.~Porredon, J.~Prat, M.~Rodr{\'i}guez-Monroy, C.~Sanchez, D.~Sanchez-Cid, T.~Schutt, N.~Weaverdyck, M.~Yamamoto, B.~Yin (Working Group and Analysis Team leads)
\end{itemize}

The remaining authors have made contributions to this paper that include, but are not limited to, the construction of DECam and other aspects of collecting the data; data processing and calibration; developing broadly used methods, codes, and simulations; running the pipelines and validation tests; and promoting the science analysis. 
\newline

{\bf Funding and Institutional Support:} 

Funding for the DES Projects has been provided by the U.S. Department of Energy, the U.S. National Science Foundation, the Ministry of Science and Education of Spain, the Science and Technology Facilities Council of the United Kingdom, the Higher Education Funding Council for England, the National Center for Supercomputing Applications at the University of Illinois at Urbana-Champaign, the Kavli Institute of Cosmological Physics at the University of Chicago, the Center for Cosmology and Astro-Particle Physics at the Ohio State University, the Mitchell Institute for Fundamental Physics and Astronomy at Texas A\&M University, Financiadora de Estudos e Projetos, Funda{\c c}{\~a}o Carlos Chagas Filho de Amparo {\`a} Pesquisa do Estado do Rio de Janeiro, Conselho Nacional de Desenvolvimento Cient{\'i}fico e Tecnol{\'o}gico and the Minist{\'e}rio da Ci{\^e}ncia, Tecnologia e Inova{\c c}{\~a}o, the Deutsche Forschungsgemeinschaft and the Collaborating Institutions in the Dark Energy Survey. 

The Collaborating Institutions are Argonne National Laboratory, the University of California at Santa Cruz, the University of Cambridge, Centro de Investigaciones Energ{\'e}ticas, Medioambientales y Tecnol{\'o}gicas-Madrid, the University of Chicago, University College London, the DES-Brazil Consortium, the University of Edinburgh, the Eidgen{\"o}ssische Technische Hochschule (ETH) Z{\"u}rich, Fermi National Accelerator Laboratory, the University of Illinois at Urbana-Champaign, the Institut de Ci{\`e}ncies de l'Espai (IEEC/CSIC), the Institut de F{\'i}sica d'Altes Energies, Lawrence Berkeley National Laboratory, the Ludwig-Maximilians Universit{\"a}t M{\"u}nchen and the associated Excellence Cluster Universe, the University of Michigan, NSF NOIRLab, the University of Nottingham, The Ohio State University, the University of Pennsylvania, the University of Portsmouth, SLAC National Accelerator Laboratory, Stanford University, the University of Sussex, Texas A\&M University, and the OzDES Membership Consortium.

Based in part on observations at NSF Cerro Tololo Inter-American Observatory at NSF NOIRLab (NOIRLab Prop. ID 2012B-0001; PI: J. Frieman), which is managed by the Association of Universities for Research in Astronomy (AURA) under a cooperative agreement with the National Science Foundation.

The DES data management system is supported by the National Science Foundation under Grant Numbers AST-1138766 and AST-1536171. \rcwr{This work used Jetstream2 and OSN at Indiana University through allocation PHY240006: Dark Energy Survey from the Advanced
Cyberinfrastructure Coordination Ecosystem: Services $\&$ Support (ACCESS) program, which is supported by U.S. National Science Foundation grants
2138259, 2138286, 2138307, 2137603, and 2138296.}

The DES Spanish institutions are partially supported by MICINN/MICIU/AEI (/10.13039/501100011033) under grants PID2021-123012NB, PID2021-128989NB, PID2022-141079NB, PID2023-153229NA, PID2024-159420NB, PID2024-156844NB, 
CEX2020-001058-M, CEX2020-001007-S and CEX2024-001441-S, some of which include ERDF/FEDER funds from the European Union, and a grant by LaCaixa Foundation (ID 100010434) code LCF/BQ/PI23/11970028. IFAE is partially funded by the CERCA program of the Generalitat de Catalunya. We acknowledge the use of Spanish Supercomputing Network (RES) resources provided by the Barcelona Supercomputing Center (BSC) in MareNostrum 5 under allocations AECT-2024-3-0020, 2025-1-0045 and 2025-2-0046. 

We  acknowledge support from the Brazilian Instituto Nacional de Ci\^encia e Tecnologia (INCT) do e-Universo (CNPq grant 465376/2014-2).

This research used resources of the National Energy Research Scientific Computing Center (NERSC), a Department of Energy User Facility using NERSC award HEP-ERCAP0035553.

Some calculations were done using the University of Cincinnati's Advanced Research Computing center as well as the Symmetry cluster at Perimeter Institute. Research at Perimeter Institute is supported in part by the Government of Canada through the Department of Innovation, Science and Economic Development and by the Province of Ontario through the Ministry of Colleges and Universities.

This document was prepared by the DES Collaboration using the resources of the Fermi National Accelerator Laboratory (Fermilab), a U.S. Department of Energy, Office of Science, Office of High Energy Physics HEP User Facility. Fermilab is managed by Fermi Forward Discovery Group, LLC, acting under Contract No. 89243024CSC000002.

\appendix

\section{Unblinding investigations}
\label{sec:unblinding_details}

\subsection{Result from parameter-level unblinding tests}

The final stage of (parameter-level) unblinding included three main components that needed to be passed before fully unblinding:
\begin{itemize}
\item \textbf{Goodness of fit via $\Delta_{\rm PPD}$.} We calculated this separately for $\xi_{\pm}$, $2\times2$pt and 3$\times$2pt under a $\Lambda$CDM model. For $\xi_{\pm}$ we additionally looked at the values when assuming the NLA IA model instead of TATT. We also looked at $3\times2$pt under a $w$CDM model. For these $\Delta_{\rm PPD}$s, we targeted them to be greater than 0.01 to unblind. 

\item \textbf{Internal consistency via $\Delta_{\rm PPD}$.} From the 3$\times$2pt data vector, we can form 7 pair-wise consistency checks. We calculated all of them assuming a $\Lambda$CDM model. These are $P(2\times2 {\rm pt}|3\times2 {\rm pt})$, $P(\xi_{\pm}|3\times2 {\rm pt})$, $P(w|2\times2 {\rm pt})$, $P(\gamma_\mathrm{t}|2\times2 {\rm pt})$, $P(w|\xi_{\pm}+\gamma_\mathrm{t})$, $P(\gamma_\mathrm{t}|\xi_{\pm}+w)$, $P(\xi_{\pm}|2\times2 {\rm pt})$. We targeted $\Delta_{\rm PPD}>0.01$ for most of these to unblind. The exception was $P(w|\xi_{\pm}+\gamma_\mathrm{t})$: in our simulation tests \citep{y6-ppd}, we found that this suffers from significant prior-volume effects resulting in very small $\Delta_{\rm PPD}$ values even when the data are generated self-consistently. We therefore decided to unblind even if this $\Delta_{\rm PPD}$ is $<0.01.$
\item \textbf{Visual inspection of the nuisance parameters.} We examined the posterior of a subset of the nuisance parameters compared to their priors to see if they significantly push against the prior edges. Here we only looked at parameters where we have derived tight Gaussian priors from simulations or external data, including the source and lens redshift uncertainties ($u^\mathrm{l}$, $u^\mathrm{s}$), the source calibration uncertainties ($m$), and the lens magnification parameters ($\alpha$).

\end{itemize}

\begin{figure*}
    \centering
    \includegraphics[width=\linewidth]{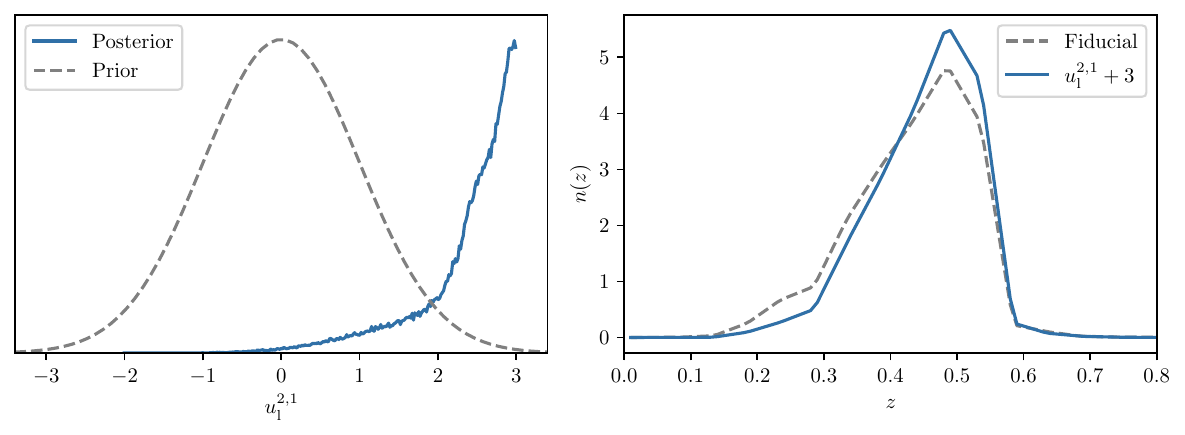}
    \caption{\label{fig:bin2}Left: Posterior distribution of the leading (first) redshift mode coefficient for \maglimpp\ bin 2, $u_\mathrm{l}^{2,1}$ (blue), compared to its prior (dashed grey). The posterior peaks toward the upper edge of the prior range. Right: Corresponding mean redshift distribution for \maglimpp\ bin 2, reconstructed from using the fiducial mode coefficients (dashed grey) and an artificially increased value of $u_\mathrm{l}^{2,1}$ (blue), chosen to match the posterior shift in the left panel. Increasing the $u_\mathrm{l}^{2,1}$ distribution narrows the bin 2 redshift distribution, enhancing the main peak and suppressing the low-redshift contribution around $z \approx 0.25$.}
\end{figure*}

The first column Table~\ref{tab:ppds} shows the $\Delta_{\rm PPD}$s from our parameter-level unblinding tests in our first attempt to unblind using the full 3$\times$2pt data vector assuming $\Lambda$CDM and linear galaxy bias---it was clear that many of the tests were failing. When further examining the posterior of the nuisance parameters, we found that for the first mode of lens bin 2, the posterior deviated significantly from the prior (see Figure \ref{fig:bin2}), and that bin contributed significantly to the internal tension. Also the posterior of the magnification coefficient for lens bin 6  deviated from its prior, albeit not as severely. A number of investigations were carried out (described in the next section) to isolate the root cause of this tension but were inconclusive. To be conservative, the team decided to remove lens bin 2 for unblinding and for our fiducial results but assess the impact of this decision on cosmology after unblinding. With the removal of lens bin two, the $\Delta_{\rm PPD}$ values are shown in the second column of Table~\ref{tab:ppds} -- we find acceptable $\Delta_{\rm PPD}$s for all goodness of fit and internal consistency tests. In turn, Table~\ref{tab:chi2}
shows the goodness of fit based on reduced $\chi^2$ (obtained at the MAP of the chains). Results are broadly consistent with those of Table~\ref{tab:ppds}, with $p$-values increasing above the threshold of 0.01 once lens bin 2 is removed. 

When assuming nonlinear galaxy bias as our model, we found similar trends in the $\Delta_{\rm PPD}$ values as listed in the third and forth columns of Table~\ref{tab:ppds}, except for $P(\gamma_\mathrm{t} | w + \xi)$, that remains below $1\%$, but for a small margin. As a result, we also remove lens bin 2 in our analysis with nonlinear galaxy bias.  

\begin{table}
    \centering
    \caption{$\Delta_{\rm PPD}$ values for the unblinding tests with and without lens bin 2. The ``No lens bin 2'' is the analysis that we unblind and use for all results in this paper, unless noted otherwise. Entries with ``-'' are dominated by prior volume effects and not considered.}
    \label{tab:ppds}
        \begin{tabular}{lllll}
        \hline \midrule
        & \multicolumn{2}{l}{\textbf{Linear galaxy bias}} & \multicolumn{2}{l}{\textbf{Nonlinear galaxy bias}} \\
        & All data & No lens bin 2& All data & No lens bin 2 \\ 
        \midrule
        \multicolumn{5}{l}{\textbf{Goodness of fit}} \\
        $\xi_{\pm}$                             & 0.162 & 0.153 & 0.153 & 0.153 \\
        $\xi_{\pm}$ (NLA)                       & 0.137 & 0.139 & 0.139 & 0.139 \\
        2$\times$2pt                            & 0.032 & 0.173 & 0.108 & 0.233 \\
        3$\times$2pt                            & 0.006 & 0.059 & 0.006 & 0.076 \\
        3$\times$2pt ($w$CDM)                   & 0.007 & 0.060 & 0.007 & 0.066 \\            
        \midrule 
        \multicolumn{5}{l}{\textbf{Internal consistency}} \\            
        $P(2\times2 {\rm pt}|3\times2 {\rm pt})$& 0.062 & 0.161 & 0.036 & 0.208 \\
        $P(\xi_{\pm}|3\times2 {\rm pt})$        & 0.141 & 0.137 & 0.136 & 0.132 \\
        $P(w|2\times2 {\rm pt})$                & 0.293 & 0.250 & 0.334 & 0.460 \\
        $P(\gamma_\mathrm{t}|2\times2 {\rm pt})$         & 0.045 & 0.181 & 0.256 & 0.312 \\
        $P(w|\xi_{\pm}+\gamma_\mathrm{t})$               & - & - & - & - \\
        $P(\gamma_\mathrm{t}|\xi_{\pm}+w)$               & 0.000 & 0.053 & 0.000 & 0.004 \\
        $P(\xi_{\pm}|2\times2 {\rm pt})$        & 0.004 & 0.049 & 0.007 & 0.051 \\            
        \midrule    \hline
        \end{tabular}

\end{table}

\begin{table}
    \centering
    \caption{Goodness of fit based on reduced $\chi^2$ and corresponding $p$-values for the case of linear galaxy bias (with and without bin 2). The overall conclusions are consistent with the goodness of fit results based on $\Delta_{\rm PPD}$ shown in Table~\ref{tab:ppds}}
    \label{tab:chi2}
        \begin{tabular}{lll}
        \hline \midrule
        & \multicolumn{1}{l}{\textbf{Linear galaxy bias}} \\
        & All data & \, \, \, No lens bin 2 \\ 
        \midrule    
        \multicolumn{3}{l}{$\chi^2/\nu$} \\
        $\xi_{\pm}$                             & 309.1 / 278.4 = 1.11 & \, \, \, 308.1 / 278.7 = 1.11  \\
        $\xi_{\pm}$ (NLA)                       & 299.7 / 263.4 = 1.14 & \, \, \, 301.2 / 263.5 = 1.14  \\
        2$\times$2pt                            & 402.0 / 363.5 = 1.11 & \, \, \, 333.2 / 307.3 = 1.08  \\
        3$\times$2pt                            & 737.0 / 644.9 = 1.14 & \, \, \, 654.3 / 588.4 = 1.11  \\
        3$\times$2pt ($w$CDM)                   & 733.7 / 644.7 = 1.14 & \, \, \, 653.4 / 588.2 = 1.11    \\            
        \midrule 
        \multicolumn{3}{l}{p-value ($\sigma$s)} \\            
        $\xi_{\pm}$                             & 0.099 (1.28) & \, \, \, 0.109 (1.23)  \\
        $\xi_{\pm}$ (NLA)                       & 0.062 (1.54) & \, \, \, 0.055 (1.6)  \\
        2$\times$2pt                            & 0.08 (1.41) & \, \, \, 0.148 (1.05) \\
        3$\times$2pt                            & 0.007 (2.47) & \, \, \, 0.031 (1.87) \\
        3$\times$2pt ($w$CDM)                   & 0.008 (2.39) & \, \, \, 0.032 (1.86)  \\            
        \midrule    \hline
        \end{tabular}
\end{table}

\subsection{Diagnostic tests for lens bin 2}

One of the primary hints for potential systematic effects in lens bin 2 is the posterior of the first mode $u_\mathrm{l}^{2,1}$ of the redshift uncertainty. When visualized in the redshift distribution space, the implication is that the data prefers a redshift distribution that is pushed to a higher value (see right panel in Figure~\ref{fig:bin2}). With these observations, we carried out the following investigations in the case of linear galaxy bias model. 

\subsubsection{Priors on redshift uncertainty}\label{sec:priors_red_app}

We explored several motivated approaches in widening the priors on the redshift uncertainty, in case we had somehow under-estimated the redshift uncertainty. This did not sufficiently resolve the problem, and we do not think this is a likely scenario as we have extensively tested the redshift pipelines as shown in \citepalias{y6-sourcepz, y6-lenspz, y6-wz}.   

First we increased the priors on WZ systematic parameters \citepalias[$\{s_{{\rm u}k}\}_{\rm u}$, cf. ][]{y6-wz}, and obtained new redshift modes. Next we looked at dropping the additional constraints coming from WZ entirely and only used SOMPZ to construct the modes. Finally, we increased the widths of the priors on the lens bin 2 mode amplitudes, $\{u_\mathrm{l}^{2,i}\}_i$, by a factor of 3, allowing for larger variations in the reconstructed $n(z)$ compared to the fiducial priors. For this test, we also increased the prior on the magnification coefficient $\alpha^6$ by a factor of 3. We found that:
\begin{itemize}
    \itemsep0em
    \item the lens bin 2 $n(z)$ and its variation, reconstructed from the larger-WZ-priors-redshift-modes were identical to the fiducial case, meaning that the WZ systematic function was already flexible enough in the first run;
    \item using SOMPZ only for the modes yielded a level of discrepancy similar to that obtained when using SOMPZ+WZ, with posteriors for lens bin 2 comparable to those shown in Figure~\ref{fig:bin2}, demonstrating that WZ was not the cause of the problem;
    \item with wider priors, the $3\times2$pt $\Delta_{\rm PPD}$ increases to 0.01, which lies at the threshold value, but remains significantly smaller than the $3\times2$pt $\Delta_{\rm PPD}$ obtained when excluding lens bin 2 (0.06); the posteriors for lens bin 2 from this wide-prior run are discrepant with the fiducial-prior results at more than the $5\sigma$ level.
\end{itemize}

\subsubsection{Parametrization of redshift uncertainty}

One of the new implementations in Y6 uses modal decomposition \citepalias{y6-nzmodes} to characterize the redshift uncertainties instead of the more \textit{ad hoc} ``shift and stretch'' parametrization used in Y3 and other literature. One concern is that once we have chosen a certain set of modes, the model is only capable of modeling $n(z)$s that are linear combinations of the modes, potentially excluding a true $n(z)$ that has a qualitatively different shape if it falls outside of our redshift error modeling and likelihood. One example is to examine the right-hand panel of Figure~\ref{fig:bin2}. This constrained approach would not be able to model the true $n(z)$ if it has significant support at redshift 0.7. 

To diagnose this possibility, we also ran the full set of PPD tests assuming a ``shift and stretch'' parametrization, where the priors on the shift and stretch are derived from the same ensemble of SOMPZ+WZ $n(z)$ realizations where we derive the principle modes from. This essentially creates a different basis set for modeling redshift errors. We found that this does not significantly improve the $3\times2$pt $\Delta_{\rm PPD}$ values (0.007), and the unblinding tests still failed. The posteriors on the lens bin 2 mean redshift are shifted by about $2\sigma$ from the prior mean ($\Delta\langle z\rangle \sim 0.015$), and the posterior of the stretch parameter is around 0.8, corresponding to a $5\sigma$ tension with the calibration prior. This test therefore rules out the new redshift marginalisation method (modes) as the cause of the problem.

\subsection{Covariance mismatch}
\label{sec:cov-missmatch}

The $\Delta_{\rm PPD}$ numbers are sensitive to the overall amplitude of the covariance matrix. A few percent change could result in the $\Delta_{\rm PPD}$ values changing from 10$^{-3}$ to 10$^{-2}$. There are a few known minor inconsistencies in our covariance matrix modeling that we tested to ensure that these are not the reason for the unacceptable $\Delta_{\rm PPD}$s. 

First, the original version of \textsc{CosmoCov} used in this work did not support \textsc{HMCode2020}, so we instead used \textsc{HaloFit} for modeling the nonlinear power spectrum. We tested that this effect results in a $<2.5\%$ difference in some elements of the diagonal of the covariance, which should be a negligible effect. Second, we assumed no IA in the covariance matrix. The validity of this assumption to first order depends on the best-fit IA constraints, which we use to generate the final covariance matrix. However fiducial IA values led to negligible changes in $\Delta \chi^2$, hence we conclude this should not cause low $\Delta_{\rm PPD}$s.

After unblinding, we updated the covariance to include baryonic feedback using \textsc{HMCode2020}, fixing $\Theta_{\rm AGN}$ to our baseline value, and IA. To assess the impact of potential missing components in the fiducial covariance, we recomputed it at the baseline best-fit cosmology, galaxy bias, and intrinsic alignment parameters, adopting the NLA model. The resulting posteriors showed no significant shifts and only negligible changes in constraining power, confirming that the fiducial covariance was sufficiently accurate and stable for the reported cosmological constraints.

\subsection{Magnification priors}\label{sec:app_mag6}

Similar to the first mode of the lens bin 2, the posterior of the magnification parameter for lens bin 6, $\alpha^6$, pushed against the priors. It was not as severe as $u_\mathrm{l}^{2,1}$, but since magnification could be degenerate with cosmological parameters, we check whether the $\Delta_{\rm PPD}$ numbers improve significantly when we relax the $\alpha^6$ prior by a factor of 3 (we also relax the priors on $u_\mathrm{l}^{2,1}$, cf. Appendix \ref{sec:priors_red_app}). 
We found that the $\Delta_{\rm PPD}$ increased to 0.01, which is larger than the initial value (0.006), but remained small, indicating persistent tension. Then we ran a chain with the original magnification priors, but removing lens bin 2, and obtained a $\Delta_{\rm PPD}$ of 0.059. We concluded that despite an apparent discrepancy between the magnification prior and the data posterior, internal consistency is acceptable.

\subsection{Discussion}

Given all the tests above, we were not able to unambiguously identify the root cause of the internal inconsistency in our data aside from the fact that it is related to the lens bin 2. We conclude that there is an unknown systematic effect that is likely associated with the lens bin 2. \rcwr{The particular redshift range that lens bin 2 covers has posed long-standing challenges related to the DES filter transitions and color-redshift degeneracies, and this could be at play here. These issues can be exacerbated by sample selection that directly depends on \photoz's in the wide-field survey, as \maglimpp\ does.\footnote{The source sample does not directly rely on \photoz's for sample selection in the same way, and may be more robust to these issues because of this.}  It is possible that we are only now reaching a level of precision in our measurements where this internal discrepancy can be detected at the significance we see here. It will continue to be a substantial challenge for future photometric galaxy lensing surveys.} 
To be conservative, we removed all elements of the 3$\times$2pt data-vector that involve lens bin 2 before proceeding with further blinding tests and analyses.  

After unblinding, we also examined the parameter constraints when we include lens bin 2 in our data vector. The best-fit cosmological parameters with and without lens bin 2 do not differ significantly, nor does the constraining power. Figure~\ref{fig:robustness}  shows our fiducial 3$\times$2pt constraints excluding lens bin 2, compared with the case where lens bin 2 is used. We find small shifts in $\Omega_{\rm m}$ and $\sigma_8$ at the 0.4$\sigma$ level, and negligible change in $S_8$. Restoring lens bin 2 improves the overall constraining power by $\lesssim10\%$ in terms of FoM$_{\sigma_8,\Omega_{\mathrm{m}}}$, driven by $\Omega_{\rm m}$, and shows negligible improvement in $S_8$ errors. This is a reassuring result, suggesting that the unknown systematic effect that is contributing to the internal tension does not  impact our cosmological constraints or conclusions. This is also a testimony of the robustness of the 3$\times$2pt combination. We will leave further discussion of the impact of lens bin 2 to the 2$\times$2pt combination in \citepalias{y6-2x2pt}.  

\section{Prior-volume effects} 
\label{sec:projection}

\begin{table}
\centering
\caption{Prior-volume effects assessment for $\Lambda$CDM. Pull statistics (truth$-$mean)/$\sigma$ for simulated runs with input parameters at the MAP values from 3$\times$2pt $\Lambda$CDM analysis, for the fiducial 3$\times$2pt linear bias analysis and the subsets 2$\times$2pt and cosmic shear (CS), as well for the nonlinear (NL) galaxy bias case. }
\label{tab:proj_effects_lcdm}
\begin{tabular}{lcccccc}
\hline
Par & $3\times2$pt & $2\times2$pt & CS TATT & CS NLA & $3\times2$pt NL & $2\times2$pt NL \\
\hline
$S_8$ & $+$0.17 & $+$0.23 & $+$0.10 & $-$0.07 & $+$0.50 & $+$0.35 \\
$\Omega_{\rm m}$ & $-$0.32 & $-$0.37 & $-$0.14 & $-$0.31 & $-$0.64 & $-$0.84 \\
$\sigma_8$ & $+$0.26 & $+$0.33 & $+$0.06 & $+$0.17 & $+$0.65 & $+$0.76 \\

\hline
\end{tabular}
\end{table}

\begin{table}
\centering
\caption{Prior-volume effect assessment for $w$CDM. Pull statistics (truth$-$mean)/$\sigma$ for simulated runs with input parameters at the MAP values from the 3$\times$2pt $\Lambda$CDM analysis. }
\label{tab:proj_effects_wcdm}
\begin{tabular}{lccc}
\hline
Parameter & $3\times2$pt & $3\times2$pt NL \\
\hline
$S_8$ & $+$0.43 & $+$0.83 \\
$\Omega_{\rm m}$ & $-$0.05 & $-$0.15 \\
$\sigma_8$ & $+$0.22 & $+$0.54 \\
$w$ & $+$0.40 & $+$0.63 \\
\hline
\end{tabular}
\end{table}
Projection or (prior-)volume effects refer to systematic biases in marginalized parameter posteriors away from the peak of the full-dimensional posterior. This is a model-dependent effect that arises when changes in the parameter volume contribution to the marginal posterior density dominate over changes in the goodness-of-fit. These volume effects tend to be stronger in cases where the likelihood depends nonlinearly on the parameters. When the data are very constraining, the likelihood drops quickly from the maximum-likelihood point and volume effects become negligible; conversely, they can be exacerbated for models with poorly-constrained parameters (see e.g. \citep{Hadzhiyska_2023}).

Prior-volume effects complicated our scale cut validation procedure, as described in \citepalias{y6-methods}, making it difficult to distinguish between genuine systematic contamination and projection artifacts when testing different scale cut choices. These effects created counterintuitive behavior, where applying stricter scale cuts did not always improve validation performance. In some cases, removing more data points actually increased parameter shifts due to enhanced projection effects resulting from reduced constraining power. We assessed these effects by running chains with known input and comparing the posterior means to the true input values, quantified as: (truth$-$mean)/$\sigma$, {where $\sigma$ is the posterior standard deviation}. For this assessment, we first used the ``DV v6'' data vector employed throughout \citepalias{y6-methods} and found significant shifts of $>0.5\sigma$ in all parameters of interest, often exceeding $1\sigma$ (see more details in \citepalias{y6-methods}). Consequently, we could no longer rely on $\Delta S_8$ as a validation metric for $2\times2$pt and $3\times2$pt analyses, and instead only used 2D validation metrics in the $(\Omega_{\rm m}, S_8)$ plane that were less susceptible to these artifacts, eventually using the $S_8$ metric only for cosmic shear.

In view of this, we agreed on a procedure to address such effects when presenting the results after unblinding. First, we would evaluate these effects during the scale cut process described in \citepalias{y6-methods}, as we found they were impacting our constrains. After unblinding, we had agreed to generate a noiseless simulated data vector at the MAP values\footnote{Specifically, we generate a simulated input data vector using the best-fit parameters from the 3$\times$2pt $\Lambda$CDM analysis, including cosmology, galaxy bias, intrinsic alignments, and lens magnification. Redshift distributions and multiplicative shear are fixed to their fiducial values from Table~\ref{tab:params}. We adopt the same priors as in the data chains, except for lens magnification and intrinsic alignment parameters with Gaussian priors, which are shifted to match the input values.} from the real data chains, obtained with a minimizer following the procedure described in Appendix E of \citepalias{y6-methods}, and then, with this data vector as input, to run a chain with a setup identical to the data analysis to quantify prior-volume effects under the same conditions, including identical  covariance.  We agreed before unblinding that parameters showing prior-volume effects larger than 1$\sigma$ in the simulated data validation should be presented with caveats and would not be highlighted as key results in the abstract or summary tables, though they may be shown in plots alongside their MAP values with clear discussion of the impact. 

We present the results of these tests in Table~\ref{tab:proj_effects_lcdm} for $\Lambda$CDM and Table~\ref{tab:proj_effects_wcdm} for $w$CDM. Under the $\Lambda$CDM model, the fiducial linear bias 3$\times$2pt posterior exhibits minor prior-volume effects of less than \rcwr{$0.5\sigma$} for all parameters of interest. Cosmic shear and 2$\times$2pt analyses show similar behavior. Across all test configurations, linear galaxy bias models exhibit substantially smaller prior-volume effects than nonlinear bias models. For the nonlinear bias model, prior-volume effects are more pronounced, exceeding $0.5\sigma$ for most parameters but remaining below $1\sigma$ in all cases. For $w$CDM, the fiducial linear bias model presents minor biases, though larger than in $\Lambda$CDM, while the $w$CDM nonlinear bias model exhibits the largest biases, with a $0.83\sigma$ bias in $S_8$. 

\section{Robustness of the DES Y6 3$\times$2pt results}
\label{sec:robustness}

\begin{figure*}
    \centering
    \includegraphics[width=0.7\linewidth]{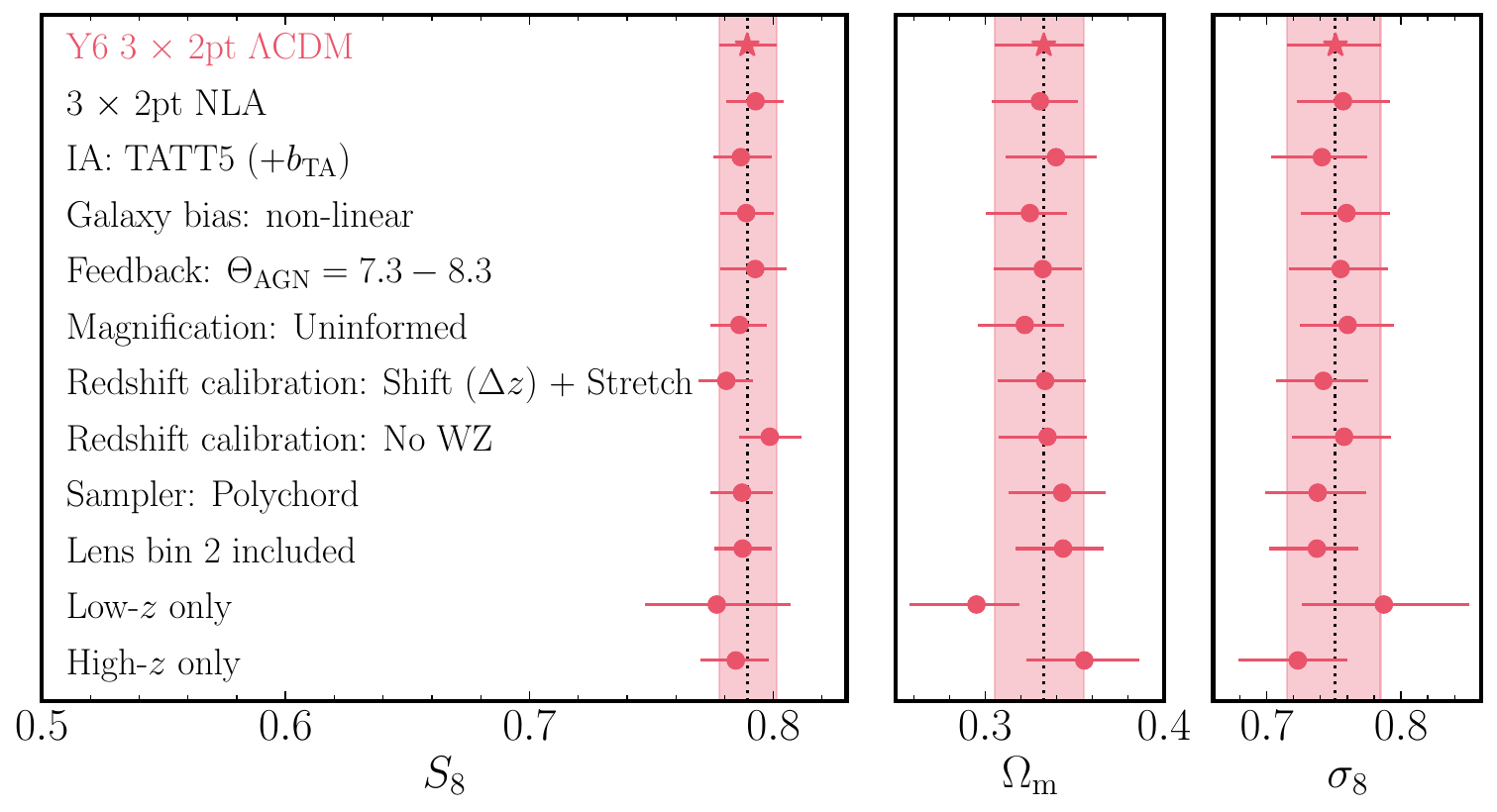}
    \caption{\label{fig:robustness} Constraints on $S_8$, $\Omega_{\rm m}$ and $\sigma_8$ in $\Lambda$CDM from the fiducial 3$\times$2pt results, compared to a number of robustness tests performed in Section~\ref{sec:robustness}. All constraints are shown as marginalized 1$\sigma$ constraints. From top to bottom we show our tests for astrophysical effects (IA, galaxy bias, baryonic feedback, magnification), different sampling approaches (changing samplers, different approaches to sample redshift uncertainty), and including/excluding part of the data vector (including lens bin 2, splitting the data by redshift).}
\end{figure*}

In this appendix, we examine the robustness of our constraints by altering a number of analysis choices and comparing the resulting cosmological constraints with our fiducial result. We summarize the exercise in Figure~\ref{fig:robustness} and discuss each test briefly below. The first set of tests concerns the choice of a certain model in our pipeline
\begin{itemize}
\itemsep0em
\item \textit{IA:} We use a 4-parameter TATT model as the fiducial IA model, but also look at the results using 1) the Nonlinear Linear Alignment model \citep[NLA,][]{Hirata2004,Bridle2007} and 2) a 5-parameter TATT model, where we in addition free $b_{TA}$. For 1), adopting the NLA IA model shifts $S_8$ higher by 0.24$\sigma$ with no change in the constraining power to $0.793^{+0.012}_{-0.012}$. For 2), we get an $S_8$ constraint of $0.787^{+0.012}_{-0.011}$, which is a 0.12$\sigma$ shift from our fiducial analysis. Our results therefore appear to be insensitive to the choice of the IA model.  
\item \textit{Galaxy bias:} As already discussed in Section~\ref{sec:nonlinear_gal}, we find that our results are not sensitive to changing the galaxy bias model from linear to nonlinear. 
\item \textit{Baryonic feedback:} Our fiducial analysis chooses to apply conservative scale cuts and not marginalize over baryonic feedback, but remove sufficient small-scale data so that the bias from fixing the modeled level of baryonic feedback is acceptable (see Section~\ref{sec:scale_cuts}). The parameter that controls the amplitude of the remaining baryonic effect is the sub-grid AGN heating temperature $T_{\rm AGN}$. We fix $\Theta_{\rm AGN} \equiv  {\rm log_{10}} (T_{\rm AGN}/{\rm K})$ to 7.7 in our fiducial analysis. We check the impact of freeing $\Theta_{\rm AGN}$ between 7.3 and 8.3 given the same scale cuts. 
We find the $S_8$ constraint changes to $0.793^{+0.013}_{-0.014}$, a $10\%$ less constraining power and a shift of $S_8$ higher by 0.3$\sigma$. 

\item \textit{Magnification:} Our fiducial analysis derives the \rcwr{Gaussian} priors on the lens magnification parameters through synthetic source injection, as described in \citepalias{y6-magnification}. \rcwr{This differs from the Y3 analysis, in which the magnification parameters were fixed to their estimated values. To assess the impact of this choice, we repeat the analysis with the magnification parameters fixed and find no noticeable change in the inferred value of $S_8$. Adopting uninformative flat priors instead of the fiducial priors leads to a shift of $-0.27\sigma$ in $S_8$.}   
\end{itemize}

The second set of tests concerns the different sampling strategies used in the analysis
\begin{itemize}
\itemsep0em
\item \textit{MCMC sampler:} All of the results in the paper use the \textsc{Nautilus} sample. We check for our fiducial result, if we switch to high-accuracy settings of the \textsc{PolyChord} nested sampling algorithm (the sampler validated in Y3 \citep{y3-samplers}), we find $S_8$ constraints for the fiducial 3$\times$2pt analysis to be $0.787^{+0.013}_{-0.013}$, which is consistent with what we derive from \textsc{Nautilus}.    

\item \textit{Photometric redshift:} One of the new implementations in DES Y6 is that we have taken a more flexible form for parametrizing the redshift uncertainty based on \citepalias{y6-nzmodes}, where we sample over principal modes of the redshift distribution uncertainties. We check that if we revert back to the more \textit{ad hoc} method of shifting and stretching a fixed mean $n(z)$, we get $S_8$ constraint of $0.781^{+0.011}_{-0.011}$. This shifts $S_8$ by about 0.5$\sigma.$ Tests in \citepalias{y6-nzmodes} show that the $n(z)$ modes and their amplitude priors better capture the uncertainties in the redshift compared to the shift-and-stretch parameters.  Our result here demonstrates that the redshift marginalization scheme is indeed important. 

We also run a chain with redshift modes constructed without the clustering information from \citepalias{y6-wz}, obtaining $S_8=0.798 \pm 0.013$. Thus adding WZ information produces a shift about 0.7$\sigma$ in $S_8$, with an uncertainty decrease around 10$\%$. This level of shift is larger than expected from the validation chains \citepalias{y6-sourcepz,y6-wz}, but the no-WZ and WZ marginalized $S_8$ contours remain consistent \rcwr{within  $1\sigma$.} This highlights the strong sensitivity of $S_8$ to redshift calibration \citep[such sensitivity is evident in the KiDS legacy results, e.g. Figures G1 and I1;][]{Wright2025}, and underscores the need for accurate, precise, and redundant redshift-calibration methods.

\end{itemize}

The third set of tests look at how including/excluding part of the data vector changes the results 
\begin{itemize}
\itemsep0em
\item \textit{Including lens bin 2:} We discussed in detail the motivation and investigation associated with the removal of lens bin 2 in Appendix~\ref{sec:unblinding_details}. Here we check the impact of including lens bin 2 back in the data vector (ignoring the internal inconsistency). We find that with lens bin 2, we have $S_8=0.787^{+0.012}_{-0.012}$, $\Omega_{\rm m}=0.344^{+0.023}_{-0.026}$, $\sigma_8=0.737^{+0.031}_{-0.035}$. The FoM$_{\sigma_8,\Omega_{\mathrm{m}}}$ is slightly higher at 4286 due to the additional statistical power, but the constraints are effectively unchanged (the $S_8$ constraints shift by 0.12$\sigma$). This result is reassuring, and suggests that if there is some unknown systematic effect that is causing lens bin 2 to be inconsistent with the rest of the data vector, the effect of that systematic effect is not coupled with our main cosmology results.

\item \textit{High-$z$/Low-$z$:} We split the data into halves of high and low redshift. The low-$z$ half consists of the 3$\times$2pt data vector formed by source bin 1 and 2 and lens bin 1 and 3, while the high-$z$ half contains the remaining bins. We find that the high-$z$ subset of the data gives $S_8=0.785^{+0.014}_{-0.014}$ and a FoM$_{\sigma_8,\Omega_{\mathrm{m}}}$ of 2793, while the low-$z$ subset gives $S_8=0.777^{+0.030}_{-0.029}$ and a FoM$_{\sigma_8,\Omega_{\mathrm{m}}}$ of 1154. Both subsets shift to slightly lower $S_8$ values, but all within 0.5$\sigma$. The high-$z$ subset dominates the constraining power both due to the higher lensing signal and the fact that we have removed lens bin 2.
\end{itemize}

\section{Limiting uncertainties in DES Y6 3$\times$2pt constraints}
\label{sec:breakdown}
The cosmological constraints from the DES Y6 3$\times$2\,pt analysis are shaped not only by the statistical power of the data but by the treatment of systematic uncertainties that affect both the modeling and the calibration of the observables. This appendix provides estimates of how different systematic effects contribute to the Y6 3$\times$2\,pt 
constraining power. As summarized in Section~\ref{sec:model}, we treat these systematic uncertainties through a combination of parametric modeling of systematic effects and data/scale cuts, which exclude measurements affected by systematics not included in the Y6 model. The latter are used to effectively mitigate several effects in the non-linear regime, as accurate models for these effects are ongoing research projects. Scale cuts, however, come at the cost of removing a substantial fraction of the signal-to-noise ratio from the data vector.

\textit{Impact of small-scale systematics}: As detailed in Section~\ref{sec:scale_cuts} and \citepalias{y6-methods,y6-1x2pt}, key uncertainties in interpreting data points below the scale cuts are baryonic feedback effects on the matter power spectrum, IA in the non-perturbative regime, non-linear galaxy bias, and non-linear matter clustering. 

For the cosmic shear data vector, our imposed scale cuts reduce the SNR from \journal{}{60} to 43, and for galaxy clustering (linear bias, without lens bin 2) our scale cuts reduce the SNR from 233 to 86.

Furthermore, point mass marginalization mitigates the non-local effect of the 1-halo term in galaxy-galaxy lensing above the scale cut. 
The combination of the scale cuts and point mass marginalization reduce the data vector S/N from 143 to 39 for galaxy-galaxy lensing (without lens bin 2), and from 302 to 95 for the fiducial (TATT, linear galaxy bias, no lens bin 2) 3$\times$2\,pt analysis. While it is not straightforward to quantify the contributions of individual systematic effects, nor to convert this S/N difference to cosmological constraining power, a factor three difference in S/N clearly indicates the promise of accurate small-scale modeling. For reference, Y6 represents a factor of 26\% improvement in S/N over the Y3 3$\times$2\,pt.

{\textit{Impact of parameterized systematics}}:
To assess how different parameterized systematic effects currently limit the cosmological precision of our analysis, we evaluate a suite of alternative analyses in which one systematic effect at a time is fixed to its best-fit model, thereby isolating how each systematic broadens the posterior on the $S_8$ parameter. Specifically, we quantify the impact of the uncertainty in our models for intrinsic alignment ($\sim 3\%$), redshift calibration ($\sim 14\%$) and magnification ($<1\%$).

The dominant term in this category arises from uncertainties in the redshift distributions of both source and lens samples. Intrinsic alignment, magnification and shear uncertainties have a negligible impact on the posterior. Although these effects do not set the overall precision floor, the comparison shows that observational calibration has become an important contributor at Y6 statistical depth, as opposed to what we found in Y3, where redshift uncertainty was negligible. Improvements in redshift calibration, for example, via deeper spectroscopic validation samples \citep{y6-sourcepz}, would reduce this component of the systematic error. 

\section{Full-parameter posterior for the DES Y6 3$\times$2pt analysis}
\label{sec:full_param}

In the main body of the article, we focused on the marginalized contours in the $S_8$-$\Omega_{\rm m}$ plane, although our analysis marginalizes over 50 nuisance parameters. Here, we make explicit some of the relevant parameter dependencies.

\begin{figure*}
    \centering
    \includegraphics[width=0.8\linewidth]{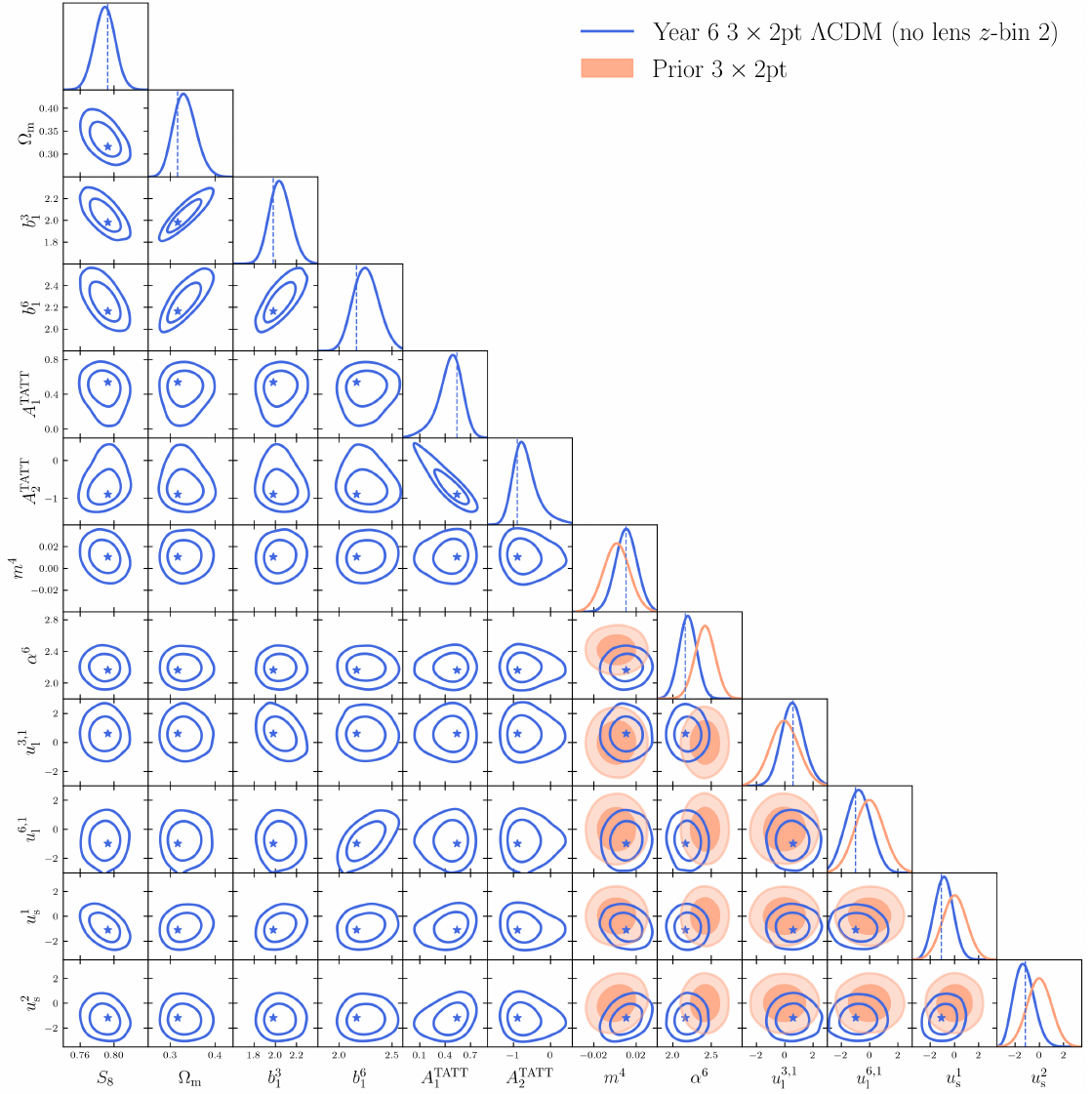}
    \caption{\label{fig:full_param} Cosmological and nuisance parameter posteriors for the $3\times2$pt linear galaxy bias $\Lambda$CDM case. We show only a subset of ten nuisance parameters out of the full set of 50. The star symbols correspond to the MAP values from minimizer runs. We also display the gaussian priors in the case of calibrated parameters. All other parameters have uninformative flat priors.}
    \label{fig:fullparam_post}
\end{figure*}

In Figure \ref{fig:full_param}, we show the posteriors of $S_8$ and $\Omega_{\rm m}$ along with a subset of the nuisance parameters that are marginalized over, for the linear $\Lambda$CDM chain. Among the nuisance parameters with uninformative flat priors, we chose to show the galaxy biases of lens bins 3 and 6, as well as the IA amplitudes. Among the nuisance parameters with calibrated Gaussian priors, we show the source bin 4 multiplicative shear bias parameter $m^4$, the lens bin 6 magnification coefficient $\alpha^6$, the first redshift modes of lens bins 3 and 6, and the first two redshift modes of the source bins. We deliberately selected parameters that illustrate differences between priors and posteriors, as well as correlations.

Below, we list the findings from the full nuisance-parameter posteriors that are not shown in this figure:
\begin{itemize}
\itemsep0em
\item The magnification coefficients are uncorrelated with all other parameters.
\item The second and third redshift modes of each lens bin $i$, $u_{\rm l}^{i,1/2}$, are uncorrelated with all other parameters.
\item The IA slopes $\alpha_{1/2}^{\rm TATT}$ are uncorrelated with all other parameters.
\item The shear bias coefficients $m^{2}$ and $m^{3}$ are uncorrelated with cosmology, while $m^{1}$ shows a weak correlation. We also observe correlations between the source redshift modes and the shear bias coefficients.
\end{itemize}

Below, we list the findings from the nuisance-parameter posteriors shown in Figure \ref{fig:full_param}:
\begin{itemize}
\itemsep0em
\item Cosmology and galaxy biases are correlated.
\item Each (linear) bias $b_1^i$ is correlated with the first mode 
of the corresponding lens bin $i$, but the lens redshift modes do not show a clear correlation with cosmology.
\item Some source redshift modes are correlated with $S_8$, the strongest being the first mode.
\item There is some correlation between IA amplitudes and cosmology.
\item There are weak correlations between IA amplitudes and source redshift modes, but not with lens redshift modes.
\item A small correlation is observed between $m^4$ and $S_8$.
\end{itemize}

Investigation of the full nuisance-parameter posteriors reveals the following about the calibrated priors:
\begin{itemize}
\itemsep0em
\item The posterior of $\alpha^6$ is in tension with the prior (cf. Appendix \ref{sec:app_mag6}).
\item The data are informative for the first redshift modes of the lens bins and the first three source redshift modes, as well as for the shear bias coefficient $m^4$. No significant tension is observed between priors and posteriors for these parameters.
\end{itemize}

\section{Changes since Y3}
\label{app:changes_since_y3}
This appendix lists major differences in data, measurements, models, and analysis methods between the current publication and the DES Y3 3$\times$2pt analysis reported in \citep{y3-3x2ptkp}.  See the cited supporting papers for further information.

\subsection{Images to catalogs}
\begin{itemize}
\itemsep0em
\item The Y6 analysis is based on 63,888 DECam exposures in the $griz$ bands (DR2) \citepalias{y6-gold} as compared to 30,319 exposures in $griz$ used in Y3 (DR1) \citep{des-dr1}. 
\item Enhanced star-galaxy separation and masking methods  \citepalias{y6-gold}.
\item Color-dependent PSF models \citepalias{y6-piff}.
\item Use \mdet\ in place of \mcal\ for shear estimation \citepalias{y6-metadetect}.
\item Increased simulation volume and realism for measuring blending effects on shear, and new parameterization of the blending corrections \citepalias{y6-imagesims}.
\item Deeper source catalog adding $\sim 3$ galaxies per arcmin$^2$ thanks to deeper imaging data in Y6. \citepalias{y6-metadetect}
\item Enhanced masking, including areas ill-suited for systematic weighting \citepalias{y6-mask}
\item Improved selection, including redshift-bin conditioned star-galaxy separation and SOM-based culling of redshift interlopers  \citepalias{y6-maglim}.
\item Higher resolution systematic galaxy weights with more realistic modeling \citepalias{y6-maglim}.
\item From the outset, two differentiated weighting schemes for clustering systematics (ENET and ISD) \citepalias{y6-maglim}.
\end{itemize}

\subsection{Summary statistics and redshifts}

\begin{itemize}
\itemsep0em
\item Larger area ($100\%$ of DES-wide in Y6 as opposed to $20\%$ in Y3) and modified injection schemes for better accuracy on quantities measured using \textsc{Balrog} source injections \citepalias{y6-balrog,y6-magnification}.
\item Use of $g$-band for source galaxy photometric redshifts instead of just $riz$ \citepalias{y6-piff,y6-metadetect}.
\item Enlarged spectroscopic calibration sample \citepalias{y6-sourcepz}.
\item Different formulation of systematic-error model, and use of importance sampling after SOMPZ instead of joint sampling \citepalias{y6-wz}.
\item Compression of $n(z)$ samples' variation into mode coefficients replaces shift-and-stretch approximation to $n(z)$ variation \citepalias{y6-nzmodes,y6-sourcepz,y6-lenspz,y6-wz}.
\item Shear-ratio likelihood from small scales not used in the 3$\times$2pt combination, but rather as an independent test of the redshift distributions and multiplicative shear biases for the highest redshift bins \citepalias{y6-gglens}.
\end{itemize}

\subsection{Models}
\begin{itemize}
\itemsep0em 
\item \textsc{HMCode2020} with fixed  $\Theta_{\rm AGN}=7.7$, instead  of \textsc{Halofit} without baryons for the nonlinear matter power spectra \citepalias{y6-methods}.
\item Scale cuts for linear galaxy bias case  $w(\theta)$ is 9 Mpc/$h$ instead of 8 Mpc/$h$ and $\xi_{+/-}$ are defined considering extreme baryonic feedback suppression as described by \textsc{Bahamas 8.0} \citepalias{y6-methods} ($\gamma_\mathrm{t}$ is the same). 
\item Gaussian priors on magnification coefficients, instead of fixed \citepalias{y6-magnification}.
\item Gaussian priors $\mathcal{N}(0.0, 3.0)$ truncated to $[-5,5]$ on the intrinsic alignment redshift evolution parameters $\eta_1$ and $\eta_2$ (instead of flat as in Y3),  $b_{\mathrm{TA}}$ fixed to unity (instead of the flat priors used in Y3) and the pivot redshift of $z_0=0.3$ instead of $z_0=0.62$ \citepalias{y6-methods}.
\item Correlated priors for multiplicative shear bias and source redshift distribution calibration mode \citepalias{y6-methods}. 
\item Nonzero central values for multiplicative shear bias in covariance matrix \citepalias{y6-methods}. 
\item As in Y3, the default covariance computation uses \citep{Takahashi2012} power spectrum model, while signal modelinng has been updated to 
\textsc{HMCode2020} with baryonic feedback~\citepalias{y6-methods}.
\item Corrections to the analytic shape/shot noise \citep{Troxel2018} computed from random catalogs in configuration space to account for the increased small-scale masking of the Y6 data as opposed to mask power spectrum (Y3).
\item Numerical debiasing of (minor) $w(\theta)$ weights over-correction  \citepalias{y6-maglim}.
\end{itemize}

\subsection{Inference techniques}
\begin{itemize}
\itemsep0em
\item Updated PPD methodology employs a Gaussian mixture model to compute an internal consistency metric which is interpretable without calibration and is more numerically stable than the Y3 implementation~\citepalias{y6-ppd}.  
\item \textsc{Nautilus} as fiducial MCMC nested sampler algorithm, instead of \textsc{Polychord} \citepalias{y6-methods}.
\item Thorough assessment of projection effects.
\item Inclusion of internal consistency via $\Delta_{\rm PPD}$ as condition for unblinding
\item {Using normalizing flow for combining independent chains.}
\end{itemize}

\bibliography{refs_short, des_short, des_y1kp_short, y3kp, y6kp}

\end{document}